\documentclass[10pt,journal]{IEEEtran}

\usepackage[T1]{fontenc}
\usepackage{lmodern}
\usepackage{microtype}

\usepackage{amsmath,amssymb,amsthm,mathtools}
\usepackage{bm}
\usepackage{bbm}
\usepackage{braket}

\usepackage{graphicx}
\usepackage{subcaption}
\usepackage{booktabs}
\usepackage{multirow}
\usepackage{array}
\usepackage{float}

\usepackage{tikz}
\usetikzlibrary{positioning,fit,calc,arrows.meta}

\usepackage{algorithm}
\usepackage[noend]{algpseudocode}
\usepackage{dblfloatfix}

\makeatletter
\renewcommand{\fps@figure}{!htbp}
\makeatother

\usepackage{cite}
\usepackage{xcolor}
\usepackage{quantikz}
\definecolor{paperSingleGate}{HTML}{DCEAF7}
\definecolor{paperTwoGate}{HTML}{F3D7B5}
\definecolor{paperMultiGate}{HTML}{E6D6F2}
\definecolor{paperMeasurement}{HTML}{D9EAD3}
\tikzset{
    publication wire/.style={draw=black,line width=0.5pt},
    publication gate/.style={draw=black,fill=paperSingleGate,
        rounded corners=0pt,line width=0.5pt,align=center,
        minimum height=6mm,inner ysep=0pt},
    publication block/.style={publication gate,fill=paperMultiGate},
    publication meter/.style={meter,draw=black,fill=paperMeasurement,
        line width=0.5pt,rounded corners=0pt,
        minimum width=8mm,minimum height=6mm,inner sep=0pt}
}

\usepackage{hyperref}
\usepackage[capitalize,noabbrev,nameinlink]{cleveref}
\usepackage{balance}
\usepackage{comment}

\usepackage{titlesec}

\titleformat{\section}
  {\centering\bfseries\large}
  {\thesection}
  {0.7em}
  {}

\titleformat{\subsection}
  {\centering\bfseries\normalsize}
  {\thesubsection}
  {0.7em}
  {}

\titleformat{\subsubsection}
  {\centering\bfseries\normalsize}
  {\thesubsubsection}
  {0.7em}
  {}

\titlespacing{\section}{0pt}{1em}{1em}
\titlespacing{\subsection}{0pt}{1em}{1em}
\titlespacing{\subsubsection}{0pt}{1em}{1em}

\newcommand{\E}{\mathbb{E}}
\newcommand{\Var}{\operatorname{Var}}

\newcommand{\poly}{\operatorname{poly}}
\newcommand{\Unif}{\operatorname{Unif}}

\newcommand{\rank}{\operatorname{rank}}

\graphicspath{{figs/}{Figures/}}

\newcommand{\QSGref}{
\textsuperscript{
\raisebox{-0.15\height}{
\includegraphics[height=0.9em]{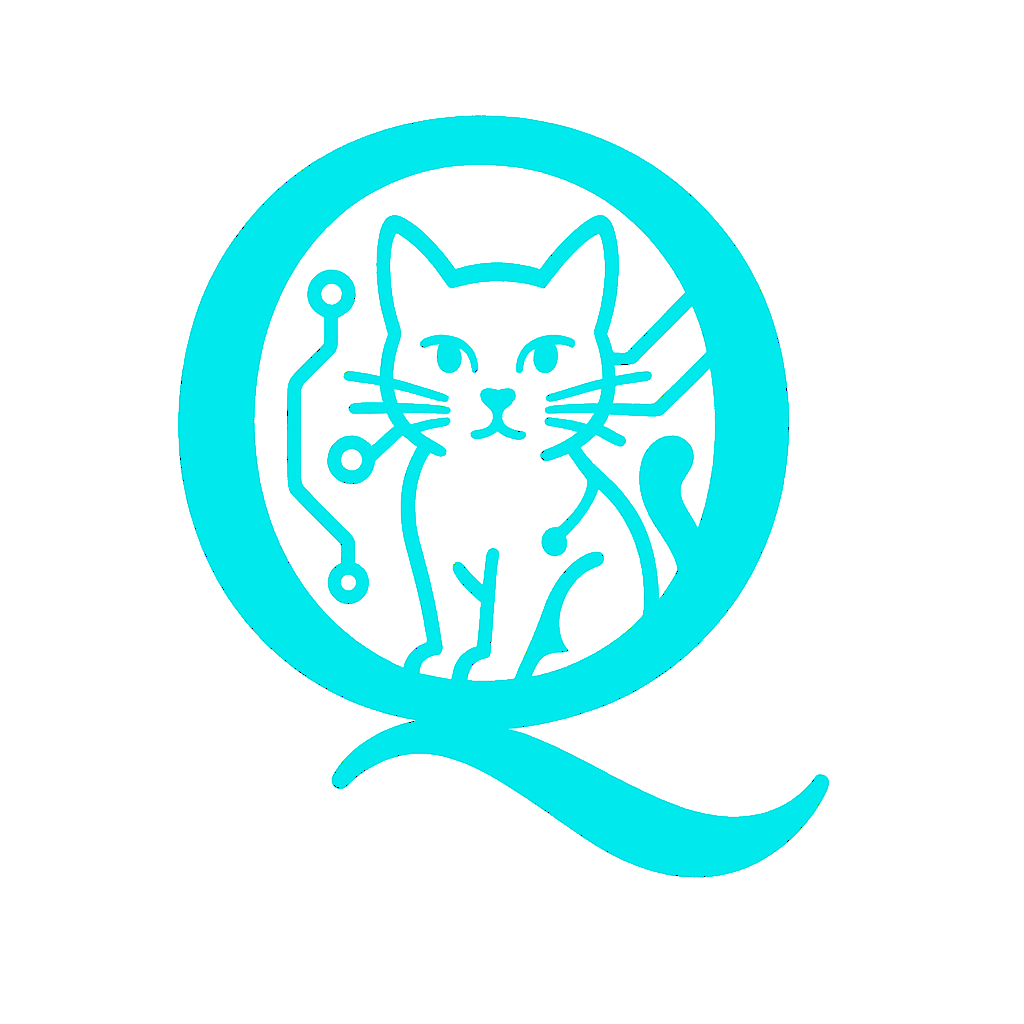}
}}}

\definecolor{paperorange}{rgb}{0.98,0.53,0.01}

\hypersetup{
    colorlinks=true,
    linkcolor=paperorange,
    citecolor=paperorange,
    urlcolor=paperorange,
    pdftitle={
        Trainability and Classical Simulability of
        Parameterized IQP-based Reinforcement Learning Policies
    },
    pdfauthor={André Sequeira et al.},
    pdfkeywords={
        quantum reinforcement learning,
        IQP circuits,
        barren plateaus,
        trainability,
        classical simulability
    }
}

\newtheorem{theorem}{Theorem}
\newtheorem{lemma}[theorem]{Lemma}
\newtheorem{proposition}[theorem]{Proposition}
\newtheorem{corollary}[theorem]{Corollary}
\newtheorem{backgroundproposition}{Proposition}
\newtheorem{mainproposition}[backgroundproposition]{Proposition}
\newtheorem*{embeddingcorollary}{Corollary (Trainability under neural Pauli embeddings)}

\theoremstyle{definition}
\newtheorem{definition}[theorem]{Definition}
\newtheorem*{policydefinition}{Definition}
\newtheorem*{rankdefinition}{Definition (Active-generator rank)}
\newtheorem*{lightconedefinition}{Definition (Backward interaction light cone)}

\theoremstyle{remark}

\title{
Trainability-Simulability Tradeoff in 
Parameterized IQP-based Reinforcement Learning Policies
}

\author{
\IEEEauthorblockN{
André Sequeira\QSGref
 }\\[0.8ex]
\IEEEauthorblockA{
\QSGref\ Quantum Software Engineering,
HASLab INESC TEC, Braga, Portugal
}
\\
\IEEEauthorblockA{
Corresponding author:
andre.m.sequeira@inesctec.pt
}
}

\begin{document}
% ================================================================

\maketitle
\thispagestyle{plain}
\pagestyle{plain}
% Keep paragraph and heading gaps natural on pages with large floats.
\raggedbottom

% ================================================================
% Abstract
% ================================================================

\begin{abstract}
Parameterized instantaneous quantum polynomial-time (IQP) circuits
generate distributions conjectured to be hard to sample classically.
As reinforcement learning (RL) policies, their $2^n$ measurement outcomes
are typically mapped to a smaller action space by a classical decoder.
This aggregation can mitigate initialization concentration while making
the policy classically simulable. We characterize this tradeoff through
the decoder's Fourier--Pauli expansion. Under independent uniform
initialization, we express policy variance as an exact sum over groups
of nonconstant Pauli components with identical active-generator
signatures. Each group contributes its squared aggregate coefficient
times $2^{-r}$, where $r$ is the binary rank of its active generators.
An inverse-polynomial aggregate coefficient with logarithmic rank
prevents exponential policy concentration. Sparse Pauli expansions and
contiguous action bins permit efficient classical additive estimation
of action probabilities. We introduce a majority decoder combining
protected low-weight components with exponentially large exact Fourier
support, preserving initialization nonconcentration while excluding
polynomial-size enumeration of the exact expansion. Numerical
experiments corroborate the variance laws. In ten-qubit CartPole
experiments, majority achieves higher returns than recursive parity and
the tested local readouts, while higher-weight Pauli sectors contribute
an increasing relative share of gradient magnitude during training.
A complementary 21-qubit block-chain policy outperforms an extensive
pool of fully connected classical neural networks in episode efficiency,
even though its bounded-width structure permits efficient classical
simulation. These results guide decoder design for quantum and
quantum-inspired RL, distinguishing policy nonconcentration, learning
performance, and classical simulability.
\end{abstract}


% ================================================================
\section{Introduction}
\label{sec:intro}
% ================================================================

Reinforcement learning (RL) provides a framework for sequential
decision-making in which an agent learns through interaction with an
environment. Rather than receiving a prescribed action for each
observation, the agent seeks a policy that maximizes cumulative reward,
accounting for the consequences of its decisions over time. This
formulation makes RL relevant to problems in control, robotics, and
resource allocation, where actions affect both future observations and
the opportunities for obtaining reward~\cite{sutton_reinforcement_2018}.
The policy representation is therefore a central modeling choice: it
determines which decision rules can be expressed and how they can be
learned from experience.

Parameterized quantum circuits (PQCs) offer a class of trainable models
for representing functions and probability distributions. Their
adjustable gates transform an encoded input into a quantum state whose
measurement statistics define the model output. Suitably constructed
PQC families possess universal function-approximation or
distribution-generation capabilities
~\cite{PerezSalinas2021UniversalApproximant,Barthe2025UniversalGenerativeModels},
motivating their investigation as learning architectures. In RL, this
construction yields a stochastic policy: the environment observation
is encoded in a circuit, an action is obtained from its measurement
outcomes, and the circuit parameters are optimized using the rewards
collected by the agent
~\cite{Chen2020DeepRL,Skolik2022QuantumAgents,Jerbi2021Parametrized,
Sequeira2023PolicyGradients}.

Instantaneous quantum polynomial-time (IQP) circuits provide a useful
test case. Their full output distributions can be hard to sample
classically under standard complexity assumptions
~\cite{BremnerJozsaShepherd2011,BremnerMontanaroShepherd2016,
BremnerMontanaroShepherd2017}, although selected expectation values
admit efficient randomized classical estimation~\cite{VanDenNest2011}.
This distinction supports classically trained IQP generative models
~\cite{RecioArmengolBowles2025,RecioArmengolAhmedBowles2025}.
It also creates a potential limitation for RL: coarse-graining a
difficult full-register sampling problem can produce an easily
simulated action distribution.

Connecting a quantum model to an environment requires specifying how
its measurement outcomes represent actions. An $n$-qubit circuit has
$2^n$ computational-basis outcomes, whereas the environment may offer
only $A\ll2^n$ discrete actions. Increasing the circuit width to enrich
the model does not require a corresponding increase in the action
space. A classical decoder therefore maps measured bit strings to
actions, assigning multiple outcomes to the same action
~\cite{Jerbi2021Parametrized,Meyer2023ActionDecoding}. The resulting
policy probability is the sum of the Born probabilities of all strings
assigned to that action.

This probability aggregation is more than an interface between the
circuit and the environment: it changes the quantities that must be
optimized and simulated. Each decoder defines action projectors with
specific Pauli components. Their structure can mitigate initialization
concentration, since observable locality plays a central role in
barren-plateau mechanisms~\cite{McClean2018Barren,Cerezo2021Cost},
including those of quantum policies~\cite{Sequeira2024Trainability}.
Yet structures that protect optimization signals can also make the
model classically accessible~\cite{Cerezo2025TrainabilitySimulability}.
This raises the central question of this work: how does the classical
decoder jointly determine policy initialization variance and the cost
of classical access to action probabilities?

The relevant access model must be specified. At deployment, sampling a
measurement string and decoding it produces an action. Likelihood-based
training additionally requires action probabilities: REINFORCE uses
their logarithmic derivatives, and proximal policy optimization (PPO)
uses probability ratios~\cite{Williams1992REINFORCE,Schulman2017PPO}.
Efficient additive estimation of all $A=\poly(n)$ action probabilities
therefore enables approximate classical action sampling, but does not
by itself guarantee efficient relative-error or score-function
estimation when an action probability is very small. Our simulation
claims concern the stated additive accuracy and sampling task, not an
unqualified efficient implementation of every policy-gradient update.

Our analysis uses the Walsh--Pauli expansion of each action projector.
For an individual diagonal Pauli observable under independent uniform
IQP angles, its variance is $2^{-r^u}$, where $r^u$ is the binary rank
of the active generator supports. This component identity also appears
in Theorem~3.8 of Shen \emph{et al.}~\cite{Shen2026IQPTrainability},
who study IQP Born-machine trainability; we give a self-contained,
independently obtained proof. Their graph examples likewise connect
rank to graph neighborhoods. The policy-level identity additionally
requires vanishing covariances between distinct Pauli expectations.
We establish this whenever the generator supports span $\mathbb F_2^n$;
independent local rotations on every qubit are sufficient but not
necessary. For arbitrary generator sets, we derive an exact grouped
variance law that accounts for coincident signatures and coefficient
cancellation. We use these identities to relate decoder coefficients
to policy concentration and classical estimation.

For repeated IQP-like circuits, the variance decomposition and
propagated-Pauli bounds are applications of the framework of Letcher,
Woerner, and Zoufal~\cite{Letcher2024Tight}, not new general variance
bounds. Their prescribed initial local layers and independent
initialization are essential. Alternating commuting blocks remove
global commutativity; their analysis nevertheless provides
decoder-dependent nonconcentration certificates. The contribution
here is to combine these ingredients with explicit action decoding,
neural encoding, and simulation criteria, rather than to claim either
the component rank identity or general Pauli-variance additivity as
a wholly new principle.

\noindent\textbf{Contributions.}
\begin{enumerate}
    \item \emph{Decoder-dependent initialization certificates.}
    We assemble exact IQP policy-variance laws and repeated-IQP
    lower bounds under explicit architecture and initialization
    assumptions. The individual-term IQP sum holds for full-rank
    generator support lists; otherwise, equal-signature terms must be
    grouped before squaring. A sufficiently weighted logarithmic-rank
    component, or a nonzero grouped coefficient, prevents exponential
    policy concentration. For graph
    IQPs with all independent local rotations, its rank is its
    structural backward light-cone size; other components may have
    extensive light cones. Neural encodings preserve these results
    at each fixed observation under the stated independent-offset
    assumptions.

    \item \emph{Action-level classical simulation.}
    We specialize expectation-value estimation to efficiently
    accessible polynomial-size Pauli expansions and to contiguous
    action intervals, including non-power-of-two action counts.
    Together with sparse-postprocessing results
    ~\cite{TakahashiMiyamotoKunihiro2026}, these criteria show why
    standard decoders can protect variance while making the
    coarse-grained policy classically accessible.

    \item \emph{Protected components with dense exact support.}
    Majority--modulo decoding combines inverse-polynomial low-weight
    coefficients with exponentially large exact Pauli support in
    the proved block-size regime. With a suitable low-rank
    component, it retains nonconcentration while excluding direct
    polynomial-size enumeration of the exact expansion. This is
    a representation-level design principle, not a classical
    hardness result; aggregate and approximate simulation remain
    separate questions.

    \item \emph{Mechanism tests and a quantum-inspired policy.}
    Initialization experiments through 16 qubits test the variance
    laws. Controlled CartPole studies compare majority with parity,
    valid local readouts, and MLPs; gradient-sector diagnostics
    measure relative participation rather than causal performance
    contributions. A connected, classically contractible 21-qubit
    policy achieves higher episode-based AUC than the tested MLPs
    under its original initialization, but this ordering reverses
    in the uniform-initialization follow-up. These results separate
    nonconcentration, initialization bias, learning performance,
    and classical runtime.
\end{enumerate}

\noindent\textbf{Organization.}
Section~\ref{sec:model} defines the circuits, policy interface, and
concentration criterion. Section~\ref{sec:main-results} collects the
main statements; Sections~\ref{sec:single-iqp-trainability}--\ref{sec:majority-modulo-construction}
develop the decoder-specific variance and simulation results.
Section~\ref{sec:experiments} presents the principal experiments and
Section~\ref{sec:discussion} discusses their scope. Proofs, secondary
CartPole studies, detailed gradient diagnostics, and circuit diagrams
are provided in the appendices.

% Kept as a separate macro so that the wide rank/light-cone comparison
% can enter the float queue immediately before the main results.
\long\def\RankLightconeFigure{%
\begin{figure*}[!t]
    \centering
    \resizebox{0.98\textwidth}{!}{%
    \begin{tikzpicture}[
        wire/.style={publication wire},
        measured wire/.style={line width=1.15pt, draw=blue!72!black},
        gate/.style={publication gate, minimum width=12mm,
                     minimum height=6mm, align=center, font=\scriptsize},
        active gate/.style={gate},
        inactive gate/.style={gate, draw=black!38, fill=black!4,
                              text=black!50},
        two gate/.style={active gate, fill=paperTwoGate, minimum width=14mm,
                         minimum height=13.5mm},
        three gate/.style={active gate, fill=paperMultiGate, minimum width=17mm,
                           minimum height=21mm},
        measure/.style={publication meter,
                        minimum width=8mm, minimum height=6mm,
                        align=center, font=\scriptsize},
        title/.style={font=\bfseries\small, align=center},
        note/.style={font=\scriptsize, align=center}
    ]
        % (a) Non-graph program: rank smaller than footprint.
        \begin{scope}[xshift=0cm]
            \node[title] at (2.70,1.55)
                {(a) Non-graph generators};
            \foreach \y/\q in {0.75/1,0/2,-0.75/3}{
                \node[font=\scriptsize, anchor=east] at (0.38,\y)
                    {$\lvert0\rangle_{\q}$};
                \draw[wire] (0.45,\y)--(5.50,\y);
            }
            \draw[measured wire] (0.45,0.75)--(5.50,0.75);
            \node[two gate] at (1.55,0.375)
                {$R_{X_1X_2}(\theta_1)$};
            \draw[publication wire] (3.55,0.75)--(3.55,-0.75);
            \node[gate, fill=paperTwoGate] at (3.55,0.75)
                {$R_{X_1X_3}(\theta_2)$};
            \node[gate, fill=paperTwoGate] at (3.55,-0.75)
                {$R_{X_1X_3}(\theta_2)$};
            \node[measure] at (5.12,0.75) {};
            \node[note] at (2.70,-1.35)
                {$Z_1:\ \{g_1,g_2\}=\{e_1+e_2,e_1+e_3\}$};
            \node[note] at (2.70,-1.78)
                {$r^{e_1}=2$; three qubits are touched};
        \end{scope}

        % (b) Local-and-edge graph IQP: rank equals light cone.
        \begin{scope}[xshift=6.25cm]
            \node[title] at (3.70,1.55)
                {(b) Local-and-edge graph IQP};
            \foreach \y/\q in {0.75/1,0/2,-0.75/3}{
                \node[font=\scriptsize, anchor=east] at (0.38,\y)
                    {$\lvert0\rangle_{\q}$};
                \draw[wire] (0.45,\y)--(7.35,\y);
            }
            \draw[measured wire] (0.45,0)--(7.35,0);
            \node[inactive gate, minimum width=9mm] at (1.25,0.75)
                {$R_X(\alpha_1)$};
            \node[active gate, minimum width=9mm] at (1.25,0)
                {$R_{X_2}(\alpha_2)$};
            \node[inactive gate, minimum width=9mm] at (1.25,-0.75)
                {$R_X(\alpha_3)$};
            \node[two gate] at (3.05,0.375)
                {$R_{X_1X_2}(\beta_{12})$};
            \node[two gate]
                at (5.25,-0.375)
                {$R_{X_2X_3}(\beta_{23})$};
            \node[measure] at (6.95,0) {};
            \node[note] at (3.70,-1.35)
                {$Z_2:\ \{e_2,e_1+e_2,e_2+e_3\}$};
            \node[note] at (3.70,-1.78)
                {$r^{e_2}=\Delta_G^{\theta}(e_2)=3$};
        \end{scope}

        % (c) Higher-body counterexample with every local generator.
        \begin{scope}[xshift=14.25cm]
            \node[title] at (2.90,1.55)
                {(c) Local plus higher-body generators};
            \foreach \y/\q in {0.75/1,0/2,-0.75/3}{
                \node[font=\scriptsize, anchor=east] at (0.38,\y)
                    {$\lvert0\rangle_{\q}$};
                \draw[wire] (0.45,\y)--(5.90,\y);
            }
            \draw[measured wire] (0.45,0.75)--(5.90,0.75);
            \node[active gate, minimum width=9mm] at (1.25,0.75)
                {$R_X(\alpha_1)$};
            \node[inactive gate, minimum width=9mm] at (1.25,0)
                {$R_X(\alpha_2)$};
            \node[inactive gate, minimum width=9mm] at (1.25,-0.75)
                {$R_X(\alpha_3)$};
            \node[three gate] at (3.35,0)
                {$R_{X_1X_2X_3}(\gamma)$};
            \node[measure] at (5.52,0.75) {};
            \node[note] at (2.90,-1.35)
                {$Z_1:\ \{g_1,g_2\}=\{e_1,e_1+e_2+e_3\}$};
            \node[note] at (2.90,-1.78)
                {$r^{e_1}=2$; three qubits are touched};
        \end{scope}
    \end{tikzpicture}}
    \caption{Active-generator rank in the three circuit examples.
    (a) For $Z_1$, the two non-graph generators are independent but
    jointly touch three qubits, so the active rank is two. (b) On the
    three-qubit path, the local generator $X_2$ and the incident edge
    generators span $e_1,e_2,e_3$; hence the active rank equals the
    three-qubit backward-light-cone size. (c) Local generators alone do
    not guarantee this equality for higher-body programs: although all
    $X_i$ rotations are present, $Z_1$ activates only $X_1$ among them,
    and $X_1X_2X_3$ supplies only one additional independent direction.
    Grey gates are inactive for the displayed observable. The linked
    boxes in (a) represent one nonadjacent $X_1X_3$ rotation, leaving
    qubit 2 untouched. In every panel, the green meter denotes
    computational-basis measurement of the indicated qubit.}
    \label{fig:rank-lightcone-generator-examples}
\end{figure*}
}

\long\def\PolicyInterfaceBackground{%
\subsection{Explicit quantum policies and the learning interface}
\label{subsec:background-interface}

A parameterized quantum policy embeds an environment state $s$ in an
$n$-qubit state $|\psi_\theta(s)\rangle$ and measures
$x\in\{0,1\}^n$.  Let $A$ denote the number of available actions,
indexed by $\mathcal A:=\{0,\ldots,A-1\}$.  Since a control task typically
has $A\ll2^n$ actions, a deterministic decoder
$f:\{0,1\}^n\to\mathcal A$ assigns each measured bit string to an
action. The corresponding action probabilities are introduced below
through decoder projectors, making the quantum observable seen by the
learning algorithm explicit.

Figure~\ref{fig:agent-environment-policy-interface} shows this interface.
At inference, one circuit execution produces a bit string and therefore
one action after applying $f$.  One shot is sufficient to act, but it
does not reveal $\pi_\theta(a|s)$. Likelihood-based
training therefore needs repeated-shot probability estimation, analytic
access, or an equivalent controlled gradient procedure.

This differs from ``train classically, deploy quantum'' generative
models, where an implicit objective such as MMD can be optimized without
pointwise probabilities.  In policy-gradient RL, the coarse-grained
probability used to act also enters the learning rule.  Its classical
cost therefore cannot be inferred from the hardness of sampling the full
quantum output distribution.
}

% ----------------------------------------------------------------
% The circuit-family overview is placed first so that it opens page 3.
\begin{figure*}[p]
    \centering
    \resizebox{0.92\textwidth}{!}{%
    \begin{tikzpicture}[
        wire/.style={publication wire},
        graph edge/.style={line width=1.35pt, draw=violet!72!black},
        vertex/.style={circle, draw=blue!65!black, fill=blue!13,
                       line width=0.9pt, minimum size=8mm,
                       inner sep=0pt, font=\small},
        gate/.style={publication gate,
                     minimum width=8mm, minimum height=6mm,
                     align=center, font=\small},
        block/.style={publication block,
                      minimum width=4.2cm, minimum height=3.5cm,
                      align=center, font=\small},
        measure/.style={publication meter, minimum width=8mm,
                        minimum height=6mm, align=center, font=\small},
        label/.style={font=\scriptsize, align=center}
    ]
        \node[font=\bfseries\small] at (2.25,2.35)
            {(a) Interaction graph $G=(V,E)$};
        \node[vertex] (v1) at (0.75,1.05) {$1$};
        \node[vertex] (v2) at (2.25,1.75) {$2$};
        \node[vertex] (v3) at (3.75,1.05) {$3$};
        \node[vertex] (v4) at (3.20,-0.60) {$4$};
        \node[vertex] (v5) at (1.30,-0.60) {$5$};
        \draw[graph edge] (v1)--(v2)--(v3)--(v4)--(v5)--(v1);
        \draw[graph edge] (v2)--(v5);
        \draw[graph edge] (v2)--(v4);
        \node[label, text=blue!60!black] at (2.25,-1.30)
            {vertex $i\ \longleftrightarrow\ e^{-i\alpha_iZ_i/2}$};
        \node[label, text=violet!72!black] at (2.25,-1.78)
            {edge $(i,j)\ \longleftrightarrow\ e^{-i\beta_{ij}Z_iZ_j/2}$};

        % The single-block circuit is slightly reduced relative to the graph.
        \begin{scope}[shift={(1.10,0.08)},scale=0.90,transform shape]
            \node[font=\bfseries\small] at (10.65,2.35)
                {(b) Graph-IQP circuit};
            \foreach \y/\q in {1.45/1,0.55/2,-0.55/n-1,-1.45/n}{
                \node[font=\scriptsize, anchor=east] at (6.05,\y)
                    {$\lvert0\rangle_{\q}$};
                \draw[wire] (6.10,\y)--(15.20,\y);
                \node[gate] at (6.85,\y) {$H$};
                \node[gate] at (13.35,\y) {$H$};
                \node[measure] at (14.45,\y) {};
            }
            \node[font=\small] at (5.72,0.00) {$\vdots$};
            \node[font=\small] at (15.55,0.00) {$\vdots$};
            \node[block] (dg) at (10.10,0.00)
                {$D_G(\boldsymbol\alpha,\boldsymbol\beta)$\\[0.35em]
                 $\displaystyle
                 \prod_{i\in V}e^{-i\alpha_iZ_i/2}
                 \prod_{(i,j)\in E}e^{-i\beta_{ij}Z_iZ_j/2}$\\[0.35em]
                 \scriptsize all gates commute};
            \node[label] at (10.10,-2.28)
                {$H^{\otimes n}D_GH^{\otimes n}
                 \equiv
                 \prod_{i\in V}R_{X_i}(\alpha_i)
                 \prod_{(i,j)\in E}R_{X_iX_j}(\beta_{ij})$};
        \end{scope}
    \end{tikzpicture}}

    \vspace{1.50em}

    \resizebox{0.92\textwidth}{!}{%
    \begin{tikzpicture}[
        wire/.style={publication wire},
        gate/.style={publication gate,
                     minimum width=7mm, minimum height=6mm,
                     align=center, font=\small},
        zblock/.style={publication block,
                      minimum width=1.85cm, minimum height=3.2cm,
                      align=center, font=\small},
        xblock/.style={publication block,
                      minimum width=1.85cm, minimum height=3.2cm,
                      align=center, font=\small},
        initial/.style={publication block, fill=cyan!22,
                       minimum width=1.75cm, minimum height=3.2cm,
                       align=center, font=\small},
        measure/.style={publication meter, minimum width=8mm,
                        minimum height=6mm, align=center, font=\small},
        note/.style={font=\scriptsize, align=center}
    ]
        \node[font=\bfseries\small] at (9.10,4.15)
            {(c) Native Hadamard--diagonal representation};
        \foreach \y/\q in {3.30/1,2.55/2,1.45/n-1,0.70/n}{
            \node[font=\scriptsize, anchor=east] at (0.85,\y)
                {$\lvert0\rangle_{\q}$};
            \draw[wire] (0.92,\y)--(18.00,\y);
            \node[gate] at (1.70,\y) {$H$};
            \node[gate] at (5.00,\y) {$H$};
            \node[gate] at (9.20,\y) {$H$};
            \node[gate] at (15.15,\y) {$H$};
            \node[measure] at (17.05,\y) {};
        }
        \node[font=\small] at (0.50,2.00) {$\vdots$};
        \node[font=\small] at (17.70,2.00) {$\vdots$};
        \node[zblock] at (3.30,2.00) {$D_Z^{(0)}$};
        \node[zblock, minimum width=2.25cm] at (7.05,2.00)
            {$D_Z^{(1)}$};
        \node[font=\large, fill=white, inner sep=2pt] at (11.05,2.00)
            {$\cdots$};
        \node[zblock, minimum width=2.25cm] at (13.10,2.00)
            {$D_Z^{(R)}$};
        \node[note] at (10.05,0.10)
            {$R$ alternating commuting blocks};

        \node[font=\bfseries\small] at (9.10,-0.75)
            {(d) Equivalent commuting-Pauli representation};
        \foreach \y/\q in {-1.60/1,-2.35/2,-3.45/n-1,-4.20/n}{
            \node[font=\scriptsize, anchor=east] at (0.85,\y)
                {$\lvert0\rangle_{\q}$};
            \draw[wire] (0.92,\y)--(18.00,\y);
            \node[measure] at (17.05,\y) {};
        }
        \node[font=\small] at (0.50,-2.90) {$\vdots$};
        \node[font=\small] at (17.70,-2.90) {$\vdots$};
        \node[initial] at (2.10,-2.90)
            {$D_X^{(0)}$\\[-0.10em]\scriptsize Initial layer};
        \node[zblock, minimum width=2.25cm] at (4.70,-2.90)
            {$D_Z^{(1)}$};
        \node[xblock, minimum width=2.25cm] at (7.55,-2.90)
            {$D_X^{(1)}$};
        \node[font=\large, fill=white, inner sep=2pt] at (9.65,-2.90)
            {$\cdots$};
        \node[zblock, minimum width=2.25cm] at (11.70,-2.90)
            {$D_Z^{(R)}$};
        \node[xblock, minimum width=2.25cm] at (14.55,-2.90)
            {$D_X^{(R)}$};
        \node[note] at (9.60,-4.78)
            {$R$ alternating commuting blocks};
    \end{tikzpicture}}

    \vspace{1.50em}

    \resizebox{0.82\textwidth}{!}{%
    \begin{tikzpicture}[
        wire/.style={publication wire},
        inactive wire/.style={publication wire,draw=black!35},
        support wire/.style={line width=0.8pt, draw=blue!70!black},
        cone wire/.style={line width=0.8pt, draw=orange!82!black},
        graph edge/.style={line width=1.2pt, draw=black!55},
        active graph edge/.style={line width=1.6pt, draw=orange!82!black},
        vertex/.style={circle, draw=black!55, fill=black!8,
                       line width=0.9pt, minimum size=7.5mm,
                       inner sep=0pt, font=\small},
        support vertex/.style={vertex, draw=blue!70!black, fill=blue!18,
                                line width=1.25pt},
        cone vertex/.style={vertex, draw=orange!82!black, fill=orange!20,
                             line width=1.25pt},
        gate/.style={publication gate, minimum width=7.5mm,
                     minimum height=6mm, align=center, font=\scriptsize},
        two gate/.style={publication gate, fill=paperTwoGate,
                         minimum width=16mm, minimum height=13.5mm,
                         align=center, font=\scriptsize},
        measure/.style={publication meter,
                        minimum width=8mm, minimum height=6mm,
                        align=center, font=\scriptsize},
        label/.style={font=\scriptsize, align=center}
    ]
        \node[font=\bfseries\small] at (2.60,2.10)
            {(e) Backward Heisenberg light cone};
        \node[vertex] (g1) at (0.40,0.25) {$1$};
        \node[cone vertex] (g2) at (1.50,1.00) {$2$};
        \node[support vertex] (g3) at (2.70,0.25) {$3$};
        \node[cone vertex] (g4) at (3.90,1.00) {$4$};
        \node[vertex] (g5) at (5.00,0.25) {$5$};
        \draw[graph edge] (g1)--(g2);
        \draw[active graph edge] (g2)--(g3);
        \draw[active graph edge] (g3)--(g4);
        \draw[graph edge] (g4)--(g5);
        \node[label, text=blue!70!black] at (2.70,-0.55)
            {measured support: qubit $3$};
        \node[label, text=orange!82!black] at (2.70,-0.98)
            {backward interaction light cone: qubits $2,3,4$};
        \node[label] at (2.70,-1.40)
            {$\Delta_G^{\theta}(u)=3$};

        \node[font=\bfseries\small] at (12.05,2.10)
            {(f) Equivalent commuting $X/XX$ IQP circuit};
        \foreach \y/\q in {1.50/1,0.75/2,0.00/3,-0.75/4,-1.50/5}{
            \node[font=\scriptsize, anchor=east] at (6.45,\y)
                {$\lvert0\rangle_{\q}$};
        }
        \draw[inactive wire] (6.50,1.50)--(17.85,1.50);
        \draw[cone wire] (6.50,0.75)--(17.85,0.75);
        \draw[support wire] (6.50,0.00)--(17.85,0.00);
        \draw[cone wire] (6.50,-0.75)--(17.85,-0.75);
        \draw[inactive wire] (6.50,-1.50)--(17.85,-1.50);
        \foreach \y/\a in {1.50/\alpha_1,0.75/\alpha_2,0.00/\alpha_3,-0.75/\alpha_4,-1.50/\alpha_5}{
            \node[gate] at (7.55,\y) {$R_X(\a)$};
        }
        \node[two gate, draw=black!45, fill=black!5] at (9.15,1.125)
            {$R_{XX}(\beta_{12})$};
        \node[two gate] at (11.15,0.375)
            {$R_{XX}(\beta_{23})$};
        \node[two gate] at (13.25,-0.375)
            {$R_{XX}(\beta_{34})$};
        \node[two gate, draw=black!45, fill=black!5] at (15.35,-1.125)
            {$R_{XX}(\beta_{45})$};
        \node[measure] at (17.20,0.00) {};
        \node[label] at (12.15,-2.03)
            {Blue: measured support; blue and orange: backward interaction light cone.};
    \end{tikzpicture}}
    \caption{IQP and repeated IQP-like Pauli architectures.
    (a) Vertices of an interaction graph select one-qubit $Z$ phases and
    edges select two-qubit $ZZ$ phases. (b) The resulting commuting
    graph-IQP block between Hadamards, equivalently the $X/XX$ circuit of
    Eq.~\eqref{eq:graph-iqp-x}. (c) Native repeated form, with Hadamards
    interleaving polynomial-size diagonal $Z$ programs. (d) Equivalent
    alternating $X$- and $Z$-type Pauli layers; their generators may be
    higher-body and need not define a graph. The shaded $D_X^{(0)}$
    auxiliary layer is used in the repeated trainability bounds, and
    setting its angles to zero recovers the standard repeated-IQP
    architecture. An IQP circuit obeys the exact rank law, which for a
    graph IQP reduces to backward-light-cone size. (e)--(f) illustrate
    this reduction for $Z^u=Z_3$ on a five-qubit chain: propagation
    reaches qubits $2$, $3$, and $4$, so
    $\Delta_G^{\theta}(u)=3$. In (b)--(d) and (f), the green meters denote
    computational-basis measurement; the single meter in (f) measures only
    qubit $3$.}
    \label{fig:single-block-graph-iqp}
    \label{fig:iqp-theoretical-circuits}
    \label{fig:iqp-interaction-lightcone}
    \vspace{1.5em}
\end{figure*}

% ================================================================
\section{Background}
\label{sec:model}
% ================================================================

\subsection{Parameterized IQP circuits}

An $n$-qubit parameterized IQP circuit is a polynomial-size product of
commuting $X$-type Pauli rotations, also called an $X$ program
~\cite{ShepherdBremner2009}:
\begin{equation}
    U_X(\boldsymbol\theta)
    =\prod_{\ell=1}^{m}e^{-i\theta_\ell P_\ell/2},
    \label{eq:iqp-general}
\end{equation}
where $m$ is polynomial in $n$ and the $P_\ell\in\{I,X\}^{\otimes n}$
commute. The equivalent diagonal $Z$ program is
$D_Z(\boldsymbol\theta):=H^{\otimes n}U_X(\boldsymbol\theta)H^{\otimes n}$,
so $U_X=H^{\otimes n}D_ZH^{\otimes n}$.
Unless stated otherwise, the input is $|0^n\rangle$ and the parameters
are initialized independently as $\theta_\ell\sim\Unif[-\pi,\pi]$.
For the binary-rank analysis in Sec.~\ref{sec:single-iqp-trainability},
we write $P_\ell=X^{g_\ell}$, with $g_\ell$ its binary support vector.
This algebraic restriction does not make the full IQP output
distribution generically easy to sample classically
~\cite{BremnerJozsaShepherd2011,BremnerMontanaroShepherd2016}.

Most results use a graph $G=(V,E)$ to select one- and two-qubit
generators:
\begin{align}
    U_G(\boldsymbol\alpha,\boldsymbol\beta)
    &=\prod_{i\in V}e^{-i\alpha_iX_i/2}
      \prod_{(i,j)\in E}e^{-i\beta_{ij}X_iX_j/2}
      \label{eq:graph-iqp-x}\\
    &=H^{\otimes n}D_G(\boldsymbol\alpha,\boldsymbol\beta)H^{\otimes n},
    \nonumber
\end{align}
where $D_G$ is obtained by replacing every $X$ in the first line by
$Z$.  Figure~\ref{fig:single-block-graph-iqp} shows how vertices select
one-qubit rotations and edges select two-qubit rotations.

Repeated copies of the same $X$ program do not enlarge the circuit
class: because all generators commute,
$U_X(\boldsymbol\theta)^L=U_X(L\boldsymbol\theta)$ (after collecting
the angles), and the result remains an IQP circuit. The useful repeated
construction instead uses the $Z$-program realization with Hadamards
between diagonal layers. Those basis changes conjugate alternating
layers into commuting $X$-type and $Z$-type Pauli programs, which do not
commute with one another and therefore increase expressivity.  The
generators within a layer need not be graph edges.  For polynomial-size
binary support lists $s_{r\ell}$ and $t_{r\ell}$, write
\begin{equation}
    D_Z^{(r)}=\prod_\ell e^{-i\theta_{r\ell}Z^{s_{r\ell}}/2},
    \qquad
    D_X^{(r)}=\prod_\ell e^{-i\varphi_{r\ell}X^{t_{r\ell}}/2}.
    \label{eq:general-repeated-pauli-layers}
\end{equation}
Every factor inside either product commutes with the other factors in
that layer.  The supports may have arbitrary size, subject only to a
polynomial total number of rotations.  For example, a layer may contain
all one-, two-, and three-qubit strings,
\begin{equation}
    D_Z^{(r)}=
    \prod_i R_{Z_i}
    \prod_{i<j}R_{Z_iZ_j}
    \prod_{i<j<k}R_{Z_iZ_jZ_k},
    \label{eq:one-two-three-body-repeated-example}
\end{equation}
with independent angles suppressed for readability; this example has
$O(n^3)$ rotations and is not specified by a pairwise interaction graph.
In this work we consider the repeated IQP-like architecture
\begin{equation}
    |\psi_R\rangle
    =D_X^{(R)}D_Z^{(R)}\cdots D_X^{(1)}D_Z^{(1)}
     D_X^{(0)}|0^n\rangle,
    \label{eq:repeated-iqp-background}
\end{equation}
where $D_X^{(0)}=\bigotimes_i e^{-i\phi_iX_i/2}$ is an auxiliary local
layer.  Setting $\boldsymbol\phi=0$ makes this layer the identity and
recovers the standard repeated-IQP circuit, so the augmentation does not
reduce expressivity.  Its role is to supply, together with the local
rotations in $D_Z^{(1)}$, the two orthogonal randomized layers required
for the Pauli-light-cone bounds of Ref.~\cite{Letcher2024Tight}.  We
therefore assume that $D_Z^{(1)}$ contains a trainable $Z_i$ rotation on
every qubit; all remaining generators in every layer may be arbitrary
Pauli strings of the indicated axis.

Each $D_X^{(r)}$ or $D_Z^{(r)}$ is internally commuting, but the complete
Hadamard-interleaved alternating circuit is no longer an IQP circuit. The
family is more expressive and, for $R=\poly(n)$, contains universal
alternating dynamics~\cite{Lloyd2018QAOA}. Figure
~\ref{fig:iqp-theoretical-circuits} shows the diagonal and Pauli-rotation
views used below.

% Define the learning-interface overview here and queue it after the
% tensor-network discussion so that it opens page 7.
\long\def\PolicyInterfaceFigure{%
\begin{figure*}[!t]
    \centering
    \resizebox{0.94\textwidth}{!}{%
    \begin{tikzpicture}[
        font=\small,
        flow/.style={-{Latex[length=2.2mm,width=1.45mm]},
                     line width=0.95pt,draw=black!72},
        feedback/.style={flow,rounded corners=4pt},
        wire/.style={line width=0.85pt,draw=black!68},
        gate/.style={draw=black!70,rounded corners=3pt,line width=0.8pt,
                     fill=blue!10,minimum width=6.6mm,
                     minimum height=6.6mm,inner sep=1pt,align=center},
        block/.style={draw=violet!65!black,rounded corners=4pt,
                      fill=violet!12,line width=0.9pt,
                      minimum width=2.45cm,minimum height=3.35cm,
                      align=center},
        measure/.style={draw=black!70,rounded corners=3pt,line width=0.8pt,
                        fill=orange!12,minimum width=7.5mm,
                        minimum height=6.6mm,inner sep=1pt,align=center},
        decoder/.style={draw=green!48!black,fill=green!10,
                        rounded corners=4pt,line width=0.9pt,
                        minimum width=1.75cm,minimum height=3.35cm,
                        align=center},
        action/.style={draw=orange!68!black,fill=orange!10,
                       rounded corners=3pt,line width=0.8pt,
                       minimum width=1.05cm,minimum height=0.34cm,
                       inner sep=0.5pt,align=center,font=\scriptsize},
        environment/.style={draw=black!65,fill=black!3,
                            rounded corners=5pt,line width=0.95pt,
                            minimum width=3.20cm,minimum height=0.68cm,
                            align=center},
        io/.style={font=\small,align=center},
        note/.style={font=\scriptsize,align=center,text=black!78}
    ]

        % Compact graph-IQP circuit, matching Fig. 1(b).
        \foreach \y/\q in {2.15/1,1.25/2,0.00/n-1,-0.90/n}{
            \node[font=\scriptsize,anchor=east] at (1.60,\y)
                {$|0\rangle_{\q}$};
            \draw[wire] (1.65,\y) -- (8.55,\y);
            \node[gate] at (2.30,\y) {$H$};
            \node[gate] at (6.70,\y) {$H$};
            \node[measure] (m\q) at (8.05,\y) {$\mathcal M_Z$};
        }
        \node[font=\small] at (1.33,0.625) {$\vdots$};
        \node[block] (dg) at (4.50,0.625)
            {$D_G(\boldsymbol\alpha,\boldsymbol\beta)$\\[0.18em]
             \scriptsize commuting $Z/ZZ$ block};

        % Explicit classical bit string produced by the measurements.
        \node[font=\scriptsize,anchor=west] at (8.68,2.15) {$x_1$};
        \node[font=\scriptsize,anchor=west] at (8.68,1.25) {$x_2$};
        \node[font=\small] at (8.87,0.625) {$\vdots$};
        \node[font=\scriptsize,anchor=west] at (8.68,0.00) {$x_{n-1}$};
        \node[font=\scriptsize,anchor=west] at (8.68,-0.90) {$x_n$};

        \node[decoder] (decoder) at (11.00,0.625)
            {$f(x)$\\[-0.08em]\scriptsize\bfseries Classical\\[-0.12em]
             \scriptsize\bfseries decoder};
        \draw[flow] (9.48,0.625) -- (decoder.west);
        \node[note] at (9.50,2.65) {measured bit string $x$};

        % The decoder selects one member of the discrete action set.
        \node[action] (a1) at (13.55,1.70) {$a_1$};
        \node[action] (a2) at (13.55,1.22) {$a_2$};
        \node[action] (a3) at (13.55,0.74) {$a_3$};
        \node[font=\scriptsize] (adots) at (13.55,0.17) {$\vdots$};
        \node[action] (aA) at (13.55,-0.40) {$a_A$};
        \node[draw=orange!55!black,rounded corners=5pt,line width=0.8pt,
              inner xsep=4pt,inner ysep=4pt,
              fit=(a1)(adots)(aA)] (actionset) {};
        \node[note,below=2pt of actionset]
            {$\pi_{\boldsymbol\theta}(\cdot\mid s_t)$};
        \draw[flow,line width=1.05pt,
              -{Latex[length=2.7mm,width=1.75mm]}] (decoder.east) --
            node[midway,above=2.5pt,note] {$a_i$}
            (actionset.west |- decoder.east);

        \node[io] (act) at (16.00,0.75)
            {\textbf{Action}\\[-0.05em]
             $A_t\sim\pi_{\boldsymbol\theta}(\cdot\mid s_t)$};
        \draw[flow] (actionset.east |- act.west) -- (act.west);

        % Agent--environment feedback loop.
        \node[environment] (env) at (7.55,-3.10)
            {\textbf{Environment}};
        \draw[feedback] (act.south) |- (env.east);

        \node[io] (obs) at (4.50,3.15)
            {\textbf{Observation}\\[-0.05em]$O_t$};
        \draw[flow] (obs.south) -- (dg.north);

        \node[io] (reward) at (4.50,-1.78)
            {\textbf{Reward}\\[-0.05em]$R_t$};
        \coordinate (envout) at (4.50,-3.10);
        \draw[line width=0.95pt,draw=black!72] (env.west) -- (envout);
        \fill[black!72] (envout) circle (1.2pt);
        \draw[feedback] (envout) -- (0.16,-3.10) -- (0.16,3.15)
            -- (obs.west);
        \draw[feedback] (envout) -- (reward.south);

        % Repeated shots estimate the coarse-grained action probabilities.
        \draw[feedback] (actionset.north) -- (13.55,3.62)
            -- node[above,note] {repeat for a number of shots}
            (8.05,3.62) -- (m1.north);
    \end{tikzpicture}%
    }
    \caption{Agent--environment interface for a parameterized quantum
    policy. The observation is embedded in the circuit, a measured bit
    string is mapped to an action by the classical decoder $f$, and the
    resulting action is applied to the environment. A single shot is
    sufficient to sample an action at inference. Training a conventional
    likelihood-based policy, however, requires explicit access to the
    coarse-grained action probabilities, obtained from repeated shots or
    analytic evaluation.}
    \label{fig:agent-environment-policy-interface}
\end{figure*}
}

% ----------------------------------------------------------------
\subsection{Weak simulation of IQP output distributions}
\label{subsec:background-weak-simulation}

For a fixed circuit instance, let
\begin{equation}
    p_{\theta}(x)
    :=
    \bigl|\langle x|U_X(\boldsymbol\theta)|0^n\rangle\bigr|^2
    \label{eq:iqp-output-distribution}
\end{equation}
denote the complete $n$-bit Born distribution.  When the angles depend
on an observation $s$, every statement below applies pointwise at fixed
$s$.  A classical
\emph{weak simulator} is a randomized classical algorithm that produces
samples from some distribution $q_{\theta}$ whose total-variation (TV)
distance from $p_{\theta}$ satisfies
\begin{equation}
    d_{\rm TV}(p_{\theta},q_{\theta})
    :=\frac12\sum_{x\in\{0,1\}^n}
      |p_{\theta}(x)-q_{\theta}(x)|
    \leq\epsilon
    \label{eq:weak-simulation-tv}
\end{equation}
in time polynomial in $n$ and $\epsilon^{-1}$. The task of interest
is \emph{sampling}: generating outcomes, rather than calculating their
individual probabilities. Multiplicative-error sampling instead
requires $p_{\theta}(x)/c\leq q_{\theta}(x)\leq c\,p_{\theta}(x)$
for every outcome, for a fixed factor $c\geq1$. Efficient classical
IQP sampling with sufficiently small multiplicative error is excluded
unless the polynomial hierarchy (PH) collapses
\cite{BremnerJozsaShepherd2011}. Such pointwise guarantees are also
generally unrealistic for noisy quantum devices, since they constrain
exponentially rare outcomes and require exact preservation of zeros;
this is not an in-principle obstruction for an ideal quantum device.
Additive-error sampling is therefore the physically relevant setting
for a quantum--classical separation. Under the standard average-case
hardness conjectures, an efficient classical sampler for general IQP
circuits at sufficiently small constant TV error would likewise
collapse PH~\cite{BremnerMontanaroShepherd2016,
MarshallAaronsonDunjko2024}.

The policy studied in this work is the pushforward of
$p_{\theta}$ through a decoder $f$.  For
$B_a=\{x:f(x)=a\}$,
\begin{equation}
    \pi_{\theta}(a)
    =\sum_{x\in B_a}p_{\theta}(x)
    \label{eq:decoded-policy-from-iqp-output}
\end{equation}
A quantum device can sample this action distribution by drawing one
bit string and returning $f(x)$. We are interested in whether this
decoded action-sampling task retains a classical simulation obstacle.
Likelihood-based training additionally requires numerical access to
the coarse-grained probability. Our classical-simulability results
concern \emph{additive action-probability estimation}: given
$a,\epsilon,$ and $\delta$, the
algorithm must output $\widehat\pi(a)$ such that
\begin{equation}
    \Pr\!\left[
       |\widehat\pi(a)-\pi_{\theta}(a)|\leq\epsilon
    \right]
    \geq1-\delta
    \label{eq:additive-policy-estimation-task}
\end{equation}
using resources polynomial in
$n,\epsilon^{-1},$ and $\log\delta^{-1}$. This task is distinct from
weak simulation of the full output distribution: full-distribution
sampling hardness neither proves nor rules out efficient additive
estimation after the decoder has aggregated many basis states.

% ----------------------------------------------------------------
\subsection{Additive estimation of IQP observables and policies}
\label{subsec:background-simulability}

The weak-sampling obstruction above does not imply that every statistic
of an IQP output distribution is hard to compute.  In particular, any
fixed diagonal Pauli expectation admits an efficient randomized
additive estimator implied by Theorem~3 of Ref.~\cite{VanDenNest2011},
stated below for completeness.

\begin{samepage}
\begin{backgroundproposition}[Van den Nest IQP expectation-value estimator \cite{VanDenNest2011}]
\label{prop:van-den-nest-iqp-estimator}
Let $U_X(\boldsymbol\theta)$ be a polynomial-size parameterized IQP
circuit and let $Z^u$ be any diagonal Pauli word.  For every
$\epsilon>0$ with $\epsilon^{-1}=\poly(n)$, there exists a classical
algorithm producing an unbiased estimator $\widehat L_u$ of
\begin{equation}
    L_u(\boldsymbol\theta)
    =
    \langle0^n|U_X^\dagger(\boldsymbol\theta)
    Z^uU_X(\boldsymbol\theta)|0^n\rangle
\end{equation}
such that $\operatorname{Std}[\widehat L_u]<\epsilon$, using
$\poly(n)$ time and space.
\end{backgroundproposition}
\end{samepage}

A self-contained derivation is given in
Appendix~\ref{app:van-den-nest-estimator}.

The same expectation can be expressed in the Heisenberg picture by
propagating $Z^u$ backwards through the circuit and evaluating
$U_X^\dagger Z^uU_X$ on $|0^n\rangle$. The support reached by this
evolution defines the observable's backward light cone. For IQP
circuits, however, mutual commutativity allows every gate that commutes
with the original $Z^u$ to cancel from the conjugation, irrespective of
its position in the circuit. Thus newly reached qubits do not activate
further interactions as they can in noncommuting circuits. For
local-and-edge graph-IQP models, support can extend only across edges
crossing the original observable's support boundary, motivating the
following definition.

\begin{lightconedefinition}
For a local-and-edge graph-IQP circuit $U_G(\boldsymbol\theta)$,
the backward interaction light cone of $Z^u$ is the union of the
supports of the Pauli terms that can occur in
$U_G^\dagger(\boldsymbol\theta)Z^uU_G(\boldsymbol\theta)$.  We denote
its number of qubits by $\Delta_G^{\theta}(u)$.  At generic parameter
values it consists of the support of $Z^u$ and its immediate graph
neighbours; exceptional angles can remove branches and shrink the
realized cone.  Repeated circuits require a different, realization-
dependent propagated-Pauli weight, defined in
Sec.~\ref{sec:repeated-trainability}.
\end{lightconedefinition}
This is an observable-local quantity, not a property of the entire
graph.  For a 1-local string $Z_i$ it is exactly
$1+\deg_G(i)$; for a general support it counts the support together with
the vertices reached by edges crossing its boundary.  Vertices outside
this set may have arbitrarily large degree without changing the
single-block light cone.
Figure~\ref{fig:iqp-interaction-lightcone}(e)--(f) illustrates the
definition for $Z^u=Z_3$ on a five-qubit chain. Its backward light cone contains
qubits $2$, $3$, and $4$, and therefore $\Delta_G^{\theta}(u)=3$.

Takahashi, Miyamoto, and Kunihiro show that efficient additive
estimation of all measured Pauli-$Z$ expectations also permits efficient
additive estimation after any polynomial-time computable Boolean
post-processing with a known polynomial bound on its exact Fourier
sparsity~\cite{TakahashiMiyamotoKunihiro2026}. Unlike direct termwise
aggregation, this result does not require the nonzero Fourier
coefficients or their locations to be supplied. For IQP circuits,
Proposition~\ref{prop:van-den-nest-iqp-estimator} provides the required
expectation estimator. Our sparse decoders have explicit expansions,
so the simpler termwise construction suffices; the distinction is
summarized in Appendix~\ref{app:sparse-postprocessing-explanation}.
Neither guarantee requires sampling the full circuit output.

\subsection{Tensor-network simulation and circuit-dependent tractability}
\label{subsec:background-tensor-networks}

The preceding estimators exploit the observable or decoder structure.
A complementary route exploits the circuit structure through
\emph{tensor networks}~\cite{MarkovShi2008}. Input states and gates are
represented by multidimensional arrays whose shared indices describe
the connecting wires. \emph{Contraction} means summing over these shared
indices. Pairing the circuit with its complex conjugate and imposing
measurement conditions produces a network for probabilities or
marginals, without constructing the full statevector.

Efficiency depends on the sizes of the intermediate tensors, not just
the number of qubits. For a network of $N_{\rm TN}$ bounded-size tensors
with binary indices, a supplied contraction order of width $w$---the
maximum number of open indices on an intermediate tensor, including
the inputs---has arithmetic cost
\[
    \poly(N_{\rm TN})\,2^{O(w)}.
\]
Thus logarithmic width gives a polynomial-time contraction; finding a
suitable order is a separate problem. Graph-theoretic treewidth bounds
provide sufficient structural guarantees~\cite{MarkovShi2008}.
For a chain representation, the dimensions of the indices joining
neighboring tensors are called \emph{bond dimensions}. Bounded bond
dimension permits efficient chain contraction even for entangled
states. Neither IQP commutativity nor bounded interaction degree alone
guarantees such a representation.

The simulation task must also include the readout. Efficient
contraction of one output probability does not by itself establish
efficient sampling or evaluation of an arbitrary decoded probability.
If the conditional marginals needed for sequential sampling can be
contracted efficiently, one can sample the full bit string and apply
any efficiently computable decoder. Repeating this procedure gives
additive estimates of its action probabilities. Alternatively, a
structured decoder can be incorporated directly into the network to
sum its accepted outcomes without enumerating them or its Pauli
expansion.

Both routes are relevant here. The one-layer path and ring circuits
used to verify the variance laws have bounded-width sampling networks.
The connected block-chain model of
Sec.~\ref{subsec:connected-iqp-mlp} also admits direct majority readout:
each inter-block ket bond has dimension at most two, and pairing
amplitudes with their conjugates gives dimension at most four.
Majority depends only on total Hamming weight, whose generating
polynomial yields the action probabilities in $O(n^2)$ arithmetic
operations, without bond truncation
(Appendix~\ref{app:connected-iqp-simulation}). These are
circuit-dependent guarantees. Exponential exact Pauli support can
obstruct termwise enumeration while leaving tensor-network simulation
efficient; it is not, by itself, a certificate of classical hardness.

For a concrete example, consider four qubits initialized in
$|0000\rangle$, with circuit
\begin{equation}
    U_4=H^{\otimes4}
    \left[\prod_{j=1}^{3}R_{Z_jZ_{j+1}}(\beta_j)\right]
    \left[\prod_{i=1}^{4}R_{Z_i}(\alpha_i)\right]
    H^{\otimes4}.
    \label{eq:background-tn-chain-example}
\end{equation}
The first Hadamards prepare $|+\rangle^{\otimes4}$; the diagonal
rotations couple only neighbors on the path $1$--$2$--$3$--$4$;
the final Hadamards precede computational-basis measurement.
Across the cut $\{1,\ldots,j\}\mid\{j+1,\ldots,4\}$, only one
gate crosses, and
\[
    R_{Z_jZ_{j+1}}(\beta_j)
    =\cos(\beta_j/2)I\otimes I
     -i\sin(\beta_j/2)Z_j\otimes Z_{j+1}.
\]
All diagonal gates commute, so this bridge can act first. Every
remaining operation acts entirely on one side of the cut and thus
preserves its Schmidt coefficients. The two bridge terms produce
orthogonal states on each side because $Z|+\rangle=|-\rangle$ and
$\langle+|-\rangle=0$. The Schmidt coefficients are therefore
$|\cos(\beta_j/2)|$ and $|\sin(\beta_j/2)|$, giving the exact minimal
state bond dimension
\[
    \chi_j=
    \begin{cases}
        2,&\beta_j\notin\pi\mathbb Z,\\
        1,&\beta_j\in\pi\mathbb Z.
    \end{cases}
\]
For example, $\beta_1=\beta_2=\beta_3=\pi/2$ gives two equal
Schmidt coefficients $1/\sqrt2$ across every chain cut, for any
local angles $\alpha_i$. The state is entangled, but all three bonds
still have dimension two. Pairing each bond with its conjugate gives
a probability-network bond of dimension at most $\chi_j^2=4$.

These bounds hold for every choice of angles and remain unchanged
when the path is extended to $n$ qubits. Expanding each bridge into
its two product terms constructs a matrix-product representation in
$O(n)$ arithmetic operations:
\[
    \langle x|U_n|0^n\rangle=A_1(x_1)\cdots A_n(x_n),
\]
where the internal matrices have dimensions at most $2\times2$;
the first and last are a row and a column, respectively. Pairing
each local matrix with its complex conjugate gives
\[
    \begin{aligned}
    T_j(x_j)&:=A_j(x_j)\otimes\overline{A_j(x_j)},\\
    p_\theta(x)&=T_1(x_1)\cdots T_n(x_n).
    \end{aligned}
\]
The internal transfer matrices are at most $4\times4$, so any
specified bit-string probability is evaluated in $O(n)$ arithmetic
operations. This is a tensor-network factorization, not generally a
product of independent single-bit probabilities: the contracted
indices retain correlations between qubits.

Marginals are equally tractable. To leave bit $j$ unobserved, replace
$T_j(x_j)$ by $T_j(0)+T_j(1)$; the matrix dimensions do not increase.
Thus any specified marginal event also costs $O(n)$. In particular,
prefix probabilities give the conditional probabilities
\[
    \begin{aligned}
    &\Pr[X_j=b\mid X_{<j}=x_{<j}]\\
    &\qquad=\frac{\Pr[X_{<j}=x_{<j},X_j=b]}
                 {\Pr[X_{<j}=x_{<j}]},
    \end{aligned}
\]
for $b\in\{0,1\}$ and prefixes of positive probability. Sampling these conditionals
successively produces a sample from the complete output distribution;
even recomputing each contraction costs only $O(n^2)$ per sample.
Consequently, every output distribution generated by this path-IQP
family, for any angles, admits efficient classical probability
evaluation and sampling. The contractions are exact in exact
arithmetic, with numerical error in finite-precision implementations.
The guarantee concerns a compact representation and access to the
distribution, not explicitly listing its $2^n$ probabilities.
No statevector or bond truncation is required.

More generally, this method is efficient whenever an exact
matrix-product representation can be constructed in polynomial time
for a chosen qubit ordering, with maximum state bond dimension
$\chi_{\max}:=\max_j\chi_j=\poly(n)$. Constant bond dimension is
sufficient but not necessary. A probability or specified marginal
can then be evaluated in $O(n\chi_{\max}^3)$ arithmetic operations
by contracting the ket and bra tensors without explicitly forming
the doubled transfer matrices; this is the standard open-boundary
matrix-product contraction cost for fixed local dimension
\cite[Sec.~4.2.1]{Schollwoeck2011MPS}. Recomputing a marginal
contraction at each of the $n$ sampling steps gives the conservative
upper bound $O(n^2\chi_{\max}^3)$ above, not the optimal sampling cost.
For a canonical matrix-product state, sequential perfect sampling
instead costs $O(n\chi_{\max}^2)$ per sample, after an
$O(n\chi_{\max}^3)$ canonicalization when needed
\cite[Secs.~IV and VI]{FerrisVidal2012PerfectSampling}.
For pairwise graph-IQP circuits, a concrete sufficient condition is
an efficiently available ordering with at most $c_j=O(\log n)$
interaction gates crossing each prefix cut: the two-term gate
decomposition gives $\chi_j\leq2^{c_j}=\poly(n)$ and an efficient
construction. This condition is sufficient, not necessary, since
dependencies between crossing gates can reduce the actual bond
dimension. For more general network geometries, polynomial individual
bond dimensions alone do not suffice; the intermediate tensors along
the chosen contraction order must also remain polynomial in size.
Thus efficient simulation depends on constructible low-width
representations, not on IQP commutativity or bounded graph degree
alone~\cite{MarkovShi2008}.

% ----------------------------------------------------------------
\PolicyInterfaceFigure

\PolicyInterfaceBackground

For each $a\in\mathcal A$, define the action bin
$B_a:=\{x\in\{0,1\}^n:f(x)=a\}$. The diagonal projector for action $a$
is
\begin{equation}
    \Pi_a:=\sum_{x\in B_a}|x\rangle\langle x|,
    \label{eq:action-projector}
\end{equation}
so the Born policy can be written as
\begin{equation}
    \pi_\theta(a|s)=\langle\psi_\theta(s)|\Pi_a |\psi_\theta(s)\rangle.
    \label{eq:born-policy}
\end{equation}

For each bit string $x$, its computational-basis projector has the
diagonal Pauli decomposition
\begin{equation}
    \begin{aligned}
    \Pi_a&=\sum_{x\in B_a}|x\rangle\langle x|
      =\sum_{u\in\{0,1\}^n}c_{a,u}Z^u,\\[-0.2em]
    c_{a,u}&=2^{-n}\sum_{x\in B_a}(-1)^{u\cdot x}.
    \end{aligned}
    \label{eq:projector-pauli-expansion}
\end{equation}
where $u$ is a binary mask and $Z^u$ is the corresponding diagonal
Pauli string. The identity coefficient is $|B_a|/2^n$; the remaining
coefficients specify which Pauli observables the decoder exposes to the
optimizer.

% Queue Fig. 3 on page 7 for the top of page 8.
\begin{figure*}[!t]
    \centering
    \resizebox{0.96\textwidth}{!}{%
    \begin{tikzpicture}[
        wire/.style={line width=0.85pt, draw=black!75},
        control/.style={circle, fill=violet!75!black, inner sep=1.6pt},
        unitary/.style={draw=green!45!black, rounded corners=4pt,
                       fill=green!11, line width=0.9pt,
                       minimum width=1.65cm, minimum height=2.6cm,
                       align=center, font=\small},
        decoder/.style={draw=violet!65!black, rounded corners=4pt,
                       fill=violet!13, line width=0.9pt,
                       minimum width=2.25cm, minimum height=3.5cm,
                       align=center, font=\small},
        projector/.style={draw=orange!65!black, rounded corners=4pt,
                       fill=orange!12, line width=0.9pt,
                       minimum width=3.2cm, minimum height=1.0cm,
                       align=center, font=\scriptsize},
        measure/.style={draw=black!70, rounded corners=3pt, line width=0.8pt,
                        fill=orange!12, minimum width=8mm, minimum height=7mm,
                        align=center, font=\small},
        note/.style={font=\scriptsize, align=center}
    ]
        \begin{scope}[xshift=-5mm]
        \node[font=\bfseries\small] at (3.75,2.75)
            {(a) One parity ancilla: $A=2$};
        \foreach \y/\q in {2.05/1,1.35/2,0.35/n-1,-0.35/n}{
            \node[font=\scriptsize, anchor=east] at (0.50,\y)
                {$\lvert\psi_\theta\rangle_{\q}$};
            \draw[wire] (0.58,\y)--(7.35,\y);
        }
        \node[font=\small] at (0.25,0.82) {$\vdots$};
        \node[font=\scriptsize, anchor=east] at (0.50,-1.28)
            {$\lvert0\rangle_{p_0}$};
        \draw[wire] (0.58,-1.28)--(7.35,-1.28);
        \foreach \x/\y in {2.65/2.05,3.15/1.35,3.65/0.35,4.15/-0.35}{
            \draw[violet!75!black, line width=0.9pt] (\x,\y)--(\x,-1.28);
            \node[control] at (\x,\y) {};
            \node[font=\normalsize, text=violet!75!black] at (\x,-1.28) {$\oplus$};
        }
        \node[note] at (3.40,-1.90) {$p_0=x_1\oplus\cdots\oplus x_n$};
        \node[measure] at (5.30,-1.28) {$\mathcal M_Z$};
        \node[projector] at (6.45,-1.28)
            {$\displaystyle \Pi_a^{(2)}=\tfrac12\!\left[I+(-1)^aZ_{p_0}\right]$};
        \end{scope}

        \node[font=\bfseries\small] at (13.05,2.75)
            {(b) Nested parity ancillas: $A=2^k$};
        \foreach \y/\q in {2.05/1,1.35/2,0.35/n-1,-0.35/n}{
            \node[font=\scriptsize, anchor=east] at (8.90,\y)
                {$\lvert\psi_\theta\rangle_{\q}$};
            \draw[wire] (8.98,\y)--(17.75,\y);
        }
        \node[font=\small] at (8.65,0.82) {$\vdots$};
        \foreach \y/\p in {-1.00/0,-1.62/1,-2.86/k-1}{
            \node[font=\scriptsize, anchor=east] at (8.90,\y)
                {$\lvert0\rangle_{p_{\p}}$};
            \draw[wire] (8.98,\y)--(17.75,\y);
        }
        \node[font=\small] at (8.67,-2.23) {$\vdots$};
        \node[decoder, minimum height=5.45cm] at (11.20,-0.42)
            {$W_k$\\[-0.10em]\scriptsize CNOT decoder\\[-0.10em]
             \scriptsize $p_j=x_1\oplus\cdots\oplus x_{n-j}$};
        \node[projector, minimum width=2.8cm, minimum height=2.55cm,
              anchor=west] at (13.85,-1.93)
            {$\displaystyle
              \Pi_a^{(2^k)}=\prod_{j=0}^{k-1}
              \tfrac12\!\left[I+(-1)^{a_j}Z_{p_j}\right]$};
    \end{tikzpicture}}
    \caption{Ancilla realization of the recursive parity-like decoder.
    (a) For two actions, one ancilla stores the global parity and the
    action projector has one nonidentity ancilla Pauli term. (b) For
    $A=2^k$, the CNOT decoder $W_k$ writes the $k$ nested parity bits to
    $k$ ancillas. The output projectors are displayed directly in the
    ancilla Pauli basis. This is a visualization of the classical decoder
    of Eqs.~\eqref{eq:recursive-parity-bits-definition}--
    \eqref{eq:recursive-parity-policy-definition}, not an
    additional circuit in the policy.}
    \label{fig:parity-ancilla-decoder}
\end{figure*}

We suppress $s$ when analyzing a fixed environment state.  Two
established decoders provide the principal baselines below.

For $x=x_1\cdots x_n$, write
$J(x)=\sum_{i=1}^{n}2^{n-i}x_i$ as the integer representation of the binary string. 
\begin{policydefinition}[Contiguous-like Born policy]
The construction introduced by Jerbi
\emph{et al.}~\cite{Jerbi2021Parametrized} assigns adjacent computational
basis strings to the same action through
\begin{equation}
    f_{\mathrm{cont}}(x):=\left\lfloor\frac{A J(x)}{2^n}\right\rfloor,
    \label{eq:contiguous-policy-definition}
\end{equation}
for $a\in\{0,\ldots,A-1\}$. When $A=2^k$, the action is determined by
the $k$ most significant measured bits and every bin has size
$2^{n-k}$.
\end{policydefinition}

\begin{policydefinition}[Recursive parity-like Born policy]
The construction of Meyer
\emph{et al.}~\cite{Meyer2023ActionDecoding} is defined for $A=2^k$ by
the nested parity bits
\begin{equation}
    p_j(x)
    :=
    x_1\oplus\cdots\oplus x_{n-j},
    \qquad
    j=0,\ldots,k-1,
    \label{eq:recursive-parity-bits-definition}
\end{equation}
and the action map
\begin{equation}
    f_{\mathrm{RP}}(x):=
    \sum_{j=0}^{k-1}2^{k-1-j}p_j(x).
    \label{eq:recursive-parity-policy-definition}
\end{equation}
For $A=2$, this reduces to the global parity of all measured bits.
We use $\mathrm{RP}$ to denote recursive parity in policy superscripts
and subscripts throughout.
\end{policydefinition}

For non-power-of-two action counts in the finite-size numerical sweep,
we use explicit deterministic extensions of these definitions.  Set
$k=\lceil\log_2A\rceil$.  The QASP contiguous extension partitions the
$2^k$ ordered prefix labels into $A$ adjacent bins whose sizes differ by
at most one and ignores the remaining $n-k$ bits.  The recursive-parity
extension computes the $k$ bits in
Eq.~\eqref{eq:recursive-parity-bits-definition}, interprets them as a
binary integer, and reduces that integer modulo $A$.  Both extensions
coincide with the definitions above when $A=2^k$; all analytical claims
for the original recursive-parity decoder are stated explicitly for
power-of-two $A$ unless noted otherwise.

Figure~\ref{fig:parity-ancilla-decoder} gives a coherent circuit
realization of this classical decoder. It is used only to expose the
ancilla readout projectors: the trained policy still consists of the data
circuit followed by the classical map $f_{\mathrm{RP}}$.

The coefficients $\{c_{a,u}\}$ are the common object underlying the
trainability and classical-simulability results below.

\subsection{Barren plateaus and trainability}
\label{subsec:background-barren-plateaus}

Let $C_{\boldsymbol\theta}$ be a cost for an $n$-qubit model. Throughout,
$\Var_{\theta}$ denotes variance over the full parameter vector
$\boldsymbol\theta$, whose entries are sampled independently and
uniformly over a full period; $\E_{\theta}$ denotes expectation over the
same initialization unless another parameter domain is stated explicitly.
Barren plateaus were introduced through gradient concentration
~\cite{McClean2018Barren,Cerezo2021Cost}. Here we study the related
diagnostic of exponential cost concentration at random initialization:
\begin{equation}
    \Var_{\theta}[C_{\boldsymbol\theta}]=O(\alpha^{-n}),
    \qquad \alpha>1.
    \label{eq:background-bp-definition}
\end{equation}
Relations between cost and gradient concentration have been established
under specified assumptions~\cite{Arrasmith2022Equivalence}. We do not
use them to infer a lower bound for each RL parameter gradient.
Here an action probability is a decoder-selected cost observable:
its variance measures initialization nonconcentration, not the
magnitude of a return-weighted policy-gradient update.

Chebyshev's inequality makes the concentration explicit,
\begin{equation}
    \Pr_{\theta}\!\left(|C_{\boldsymbol\theta}-\E_{\theta}[C_{\boldsymbol\theta}]|\geq\epsilon\right)
    \leq\frac{\Var_{\theta}[C_{\boldsymbol\theta}]}{\epsilon^2}
    =O\!\left(\frac{\alpha^{-n}}{\epsilon^2}\right).
    \label{eq:background-chebyshev}
\end{equation}
For a sample-mean estimator with order-one per-shot variance, resolving
cost differences of order $\alpha^{-n/2}$ at constant signal-to-noise
ratio requires $N=\Omega(\alpha^n)$ independent shots. Exponential
concentration can therefore make variations in action probabilities
prohibitively expensive to resolve during optimization, even when their
absolute values are not small.

Our figure of merit for avoiding exponential policy concentration is
inverse-polynomial rather than constant initialization variance:
\begin{equation}
    \Var_{\theta}[\pi_{\boldsymbol\theta}]=\Omega(n^{-p})
    \quad\text{for some constant }p>0.
    \label{eq:background-polynomial-trainability}
\end{equation}
This criterion rules out exponentially small policy variance; it does
not guarantee successful optimization or a lower bound on every
parameter derivative. The results below concern policy probabilities,
with the decoder-induced Pauli spectrum determining their variance.
They are fixed-state, random-initialization statements.  An end-to-end RL
gradient additionally contains returns, score factors
$\partial_\theta\log\pi_\theta(a|s)$, and a state-visitation distribution
that itself depends on the policy.  We therefore use \emph{trainability
certificate} only to mean absence of exponential initialization
concentration in the stated policy quantity, not guaranteed RL
convergence.

% ================================================================
% Queue Fig. 4 on page 7 for the top of page 8.
\RankLightconeFigure

\section{Main Results}
\label{sec:main-results}
% ================================================================

The results can be read through the Pauli spectrum of an action
projector.  For a decoder $f:\{0,1\}^n\to\mathcal A$, let
$B_a=\{x:f(x)=a\}$ and
\begin{equation}
    \Pi_a=\sum_{x\in B_a}|x\rangle\langle x|
    =\sum_{u\in\{0,1\}^n}c_{a,u}Z^u,
    \qquad
    Z^u:=\prod_{i=1}^{n}Z_i^{u_i},
    \label{eq:main-pauli-decomposition}
\end{equation}
where
\begin{equation}
    c_{a,u}=2^{-n}\sum_{x\in B_a}(-1)^{u\cdot x}.
    \label{eq:main-walsh-coefficient}
\end{equation}
Here $u=(u_1,\ldots,u_n)$ is a binary mask: $u_i=1$ selects $Z_i$
and $u_i=0$ selects the identity.  Thus $S_u:=\{i:u_i=1\}$ is the
Pauli support and $c_{a,u}$ is its Walsh--Pauli coefficient.  

\begin{mainproposition}[Walsh--Pauli correspondence for decoded policies]
\label{prop:walsh-pauli-policy}
Let
\begin{equation}
    h_a(x):=\mathbf{1}[f(x)=a],
    \qquad
    \chi_u(x):=(-1)^{u\cdot x},
\end{equation}
and use the normalized Boolean Fourier transform
\begin{equation}
    \widehat h_a(u)
    :=2^{-n}\sum_{x\in\{0,1\}^n}h_a(x)\chi_u(x).
    \label{eq:policy-boolean-fourier-coefficient}
\end{equation}
Then the Boolean Fourier coefficients of the action indicator and the
Pauli-$Z$ coefficients of its action projector coincide:
\begin{equation}
    c_{a,u}=\widehat h_a(u),
    \qquad
    \Pi_a=\sum_u\widehat h_a(u)Z^u.
    \label{eq:walsh-pauli-correspondence}
\end{equation}
Equivalently, the inverse transform and the policy probability in an
arbitrary $n$-qubit pure state $|\psi_{\theta}(s)\rangle$ are
\begin{align}
    h_a(x)
    &=\sum_u c_{a,u}(-1)^{u\cdot x},
    \label{eq:policy-fourier-inverse}\\
    \pi_{\theta}(a|s)
    :=\langle\psi_{\theta}(s)|\Pi_a|\psi_{\theta}(s)\rangle
    &=\sum_u c_{a,u}\langle\psi_{\theta}(s)|Z^u|\psi_{\theta}(s)\rangle.
    \label{eq:policy-fourier-expectation}
\end{align}
Moreover,
\begin{equation}
    c_{a,0}=\frac{|B_a|}{2^n},
    \qquad
    \sum_u c_{a,u}^2=\frac{|B_a|}{2^n},
    \label{eq:policy-fourier-parseval}
\end{equation}
where the second equality is Parseval's identity for the indicator
$h_a$.
\end{mainproposition}

The proof is given in Appendix~\ref{app:proofs-single-block}.

We use the backward-light-cone notation
$\Delta_G^{\theta}(u)$ introduced in
Sec.~\ref{subsec:background-simulability}. Throughout this section,
initialization is independent and uniform over a full parameter period.
Figure~\ref{fig:rank-lightcone-generator-examples} compares the three
generator sets used below to distinguish active-generator rank from a
geometric interaction footprint and to identify the graph-IQP case in
which rank reduces exactly to backward-light-cone size.

\begin{rankdefinition}
For an IQP $X$ program, write each generator as
$P_\ell=X^{g_\ell}$ with binary support vector
$g_\ell\in\mathbb F_2^n$.  A generator is active on $Z^u$ precisely
when $g_\ell\cdot u=1$ over $\mathbb F_2$.  Let $G_u:=\begin{pmatrix}g_\ell\end{pmatrix}_{\ell\in \{ \ell:g_\ell\cdot u=1\}}$ be the binary matrix of active supports,
where $G_u$ contains one row for every active generator and
$r^u:=\rank_{\mathbb F_2}(G_u)$ is the rank associated with the
observable $Z^u$.

Thus $r^u$ counts linearly independent active support directions, rather
than the number of active gates or the number of qubits they touch.
\end{rankdefinition}

For example, consider the non-graph commuting program depicted in
Fig.~\ref{fig:rank-lightcone-generator-examples}(a).  For the observable $Z_1$, whose
mask is $u=e_1=(1,0,0)$, the active generators have supports
$g_1=(1,1,0)$ and $g_2=(1,0,1)$.  Consequently,
\begin{equation}
    G_{e_1}
    =\begin{pmatrix}
        1&1&0\\
        1&0&1
      \end{pmatrix},
    \qquad
    r^{e_1}=\rank_{\mathbb F_2}(G_{e_1})=2.
    \label{eq:main-active-generator-rank-example}
\end{equation}
The two rows are independent even though their combined support touches
three qubits.  This illustrates that active-generator rank and geometric
interaction footprint are distinct quantities.

By comparison, consider the three-qubit path $1$--$2$--$3$ in
Fig.~\ref{fig:rank-lightcone-generator-examples}(b), with local
generators and edge generators.  For $Z_2$, the active supports are
$e_2$, $e_1+e_2$, and $e_2+e_3$, corresponding respectively to
$X_2$, $X_1X_2$, and $X_2X_3$.  They span $e_1,e_2,e_3$, and hence

\begin{equation}
    r^{e_2}=3=\Delta_G^{\theta}(e_2).
    \label{eq:main-rank-lightcone-example}
\end{equation}
Here the rank equals the three-qubit backward interaction light cone.
This reduction is guaranteed by the combination of local $X_i$
generators and two-qubit edge generators in a graph IQP.  The presence
of local generators alone is not sufficient for an arbitrary
higher-body $X$ program: the external components of its active supports
need not be independently resolvable.  For example, if all local
generators and $X_1X_2X_3$ are present, then $Z_1$ activates $X_1$ and
$X_1X_2X_3$ but not $X_2$ or $X_3$; its active rank is two although the
active generators touch three qubits, as shown in
Fig.~\ref{fig:rank-lightcone-generator-examples}(c).

\subsection{Decoder-dependent initialization-variance bounds}

The single-component rank identity is also Theorem~3.8 of
Shen \emph{et al.}~\cite{Shen2026IQPTrainability}. To pass from that
identity to a policy variance, cross covariances must vanish.
Lemma~\ref{lem:pauli-orthogonality} proves this for the IQP ensemble
below. The repeated-circuit part specializes the observable
decomposition and bounds of Letcher \emph{et al.}
~\cite{Letcher2024Tight}; both parts are collected here to expose
their common dependence on the decoder.

To distinguish the condition for the policy sum from the rank of an
individual component, let $\mathsf G$ contain the supports of
\emph{all} independently parameterized generators as rows. The binary
vector $\mathsf Gu$ records which generators anticommute with $Z^u$;
we call it the active-generator signature. The condition
$\rank_{\mathbb F_2}\mathsf G=n$ means that distinct Pauli strings
have distinct signatures. It is different from the component rank
$r^u$, which uses only the active rows.

\begin{theorem}[Unified initialization-variance law for IQP Born policies]
\label{thm:main-trainability}
For an IQP $X$ program acting on $|0^n\rangle$, initialize all rotation
angles independently and uniformly over a full period. Suppose
$\rank_{\mathbb F_2}\mathsf G=n$. This holds, in particular, for an
independently parameterized local $X_i$ rotation on every qubit,
together with any additional commuting $X$-type Pauli-string
rotations. Let $r^u$ be the active-generator rank associated with
$Z^u$. For every fixed action decoder, the exact policy variance is
\begin{equation}
    \Var_{\theta}[\pi(a)]
    =\sum_{u\ne0}c_{a,u}^{2}2^{-r^u}.
    \label{eq:main-policy-rank-variance}
\end{equation}
For the local-and-edge graph-IQP model,
$r^u=\Delta_G^{\theta}(u)$ for generic parameter values,
where $\Delta_G^{\theta}(u)$ is the backward interaction
light-cone size.  Hence
\begin{equation}
    \Var_{\theta}[\pi(a)]
    =\sum_{u\ne0}c_{a,u}^{2}
      2^{-\Delta_G^{\theta}(u)}.
    \label{eq:exact-policy-variance}
\end{equation}

For the augmented $R$-block architecture of
Eq.~\eqref{eq:repeated-iqp-background},
\begin{equation}
    \Var_{\theta}[\pi_R(a)]
    \geq
    \sum_{u\ne0}c_{a,u}^{2}
    \E_{\theta}
    \left[
        4^{-\Delta_R^{\theta}(u)}
    \right],
    \label{eq:main-repeated-policy-bound}
\end{equation}
where $\boldsymbol\theta\in\{0,\pi/2\}^{M}$ is the discrete Clifford parameter
vector used in the Pauli-light-cone bounds of Letcher \emph{et al.}
~\cite{Letcher2024Tight}, and $\Delta_R^{\theta}(u)$ is the number of
qubits on which the backward-propagated Pauli is nonidentity.  Here
$\Var_{\theta}$ is over continuous full-period initialization, while
$\E_{\theta}$ is the auxiliary Clifford-restricted average.
\end{theorem}

Without full column rank, distinct Pauli strings can have identical
active-generator signatures and hence identical expectations.
An exact policy-variance decomposition still holds, but their
coefficients must be grouped before squaring.

\begin{mainproposition}[Grouped rank law for arbitrary IQP generator sets]
\label{prop:grouped-policy-variance}
For an IQP $X$ program on $|0^n\rangle$ with angles initialized
independently and uniformly over a full period and a fixed decoder,
group masks by their active-generator signature $\sigma=\mathsf Gu$.
Define
\[
    C_{a,\sigma}:=\sum_{u:\,\mathsf Gu=\sigma}c_{a,u}.
\]
Let $r_\sigma$ be the rank of the rows selected by $\sigma$;
equivalently, $r_\sigma=r^u$ for any mask with
$\mathsf Gu=\sigma$. Then
\begin{equation}
\begin{aligned}
    \E_\theta[\pi(a)]&=C_{a,0},\\
    \Var_\theta[\pi(a)]
      &=\sum_{\substack{\sigma\in\operatorname{im}\mathsf G\\
                        \sigma\neq0}}
          C_{a,\sigma}^{\,2}\,2^{-r_\sigma}.
\end{aligned}
\label{eq:grouped-policy-rank-variance}
\end{equation}
No local rotations or full-rank assumption are required.
\end{mainproposition}

The proof is given in Appendix~\ref{app:grouped-policy-variance}.
For example, on a pure-$R_{XX}$ ring
(Fig.~\ref{fig:variance-circuit-schematics}(a)), complementary masks
$u$ and $\bar u=u\oplus\mathbf1$ have identical expectations.
Each nonconstant pair therefore contributes
$(c_{a,u}+c_{a,\bar u})^2 2^{-r^u}$, while the pair
$\{0,\mathbf1\}$ contributes only to the mean. This is the
grouped policy-variance law tested in
Fig.~\ref{fig:pure-rxx-policy-variance}, with the explicit ring
formula in Eq.~\eqref{eq:pure-rxx-grouped-policy-variance}.

The local-rotation sufficient condition covers
\begin{equation}
    U_X(\boldsymbol\alpha,\boldsymbol\theta)
    =\exp\!\left[-\frac{i}{2}
      \left(\sum_{i=1}^n\alpha_iX_i
          +\sum_k\theta_kX^{g_k}\right)\right],
    \label{eq:local-plus-general-iqp}
\end{equation}
with independent uniform $\alpha_i$ and $\theta_k$, regardless of the
weights or connectivity of the additional $X^{g_k}$. Equivalently,
one may use diagonal $Z$-string rotations between two Hadamard
layers. Here ``arbitrary'' refers to the commuting Pauli strings,
not arbitrary noncommuting rotations. Fixed local angles, or local
angles tied to other parameters, do not supply the independent
averages required by this sufficient condition.

Local rotations are sufficient but not necessary. Without them,
Eq.~\eqref{eq:main-policy-rank-variance} still holds whenever the
remaining generator supports span $\mathbb F_2^n$. If they do not,
Proposition~\ref{prop:grouped-policy-variance} supplies the grouped
identity. Appendix~\ref{app:grouped-policy-variance} also gives
decoder-specific conditions and examples for recovering the
individual-term sum without local rotations. The
single-component rank law itself does not require local rotations.

Under the full-rank assumption, the IQP identities are exact nonnegative decompositions of the
policy variance into its Pauli-observable components. More generally, if
a fixed diagonal policy objective has the form
$C=\operatorname{Tr}[\rho\,O]$ with $O=\sum_u o_uZ^u$, the same
calculation gives $\Var_{\theta}[C]=\sum_{u\ne0}o_u^2 2^{-r^u}$. On a
local-and-edge graph IQP, this becomes
$\sum_{u\ne0}o_u^2 2^{-\Delta_G^{\theta}(u)}$. Thus components
cannot cancel in the variance. If even one coefficient has
inverse-polynomial magnitude and its active-generator rank is
$O(\log n)$, that term alone supplies an inverse-polynomial lower bound;
for graph IQPs, this is precisely the logarithmic-light-cone condition as established in Corollary~\ref{cor:main-protected-component}.
The remaining, possibly global, components remain part of the policy without affecting the lower bound.

The exact IQP identities follow from
Lemmas~\ref{lem:rank-law} and~\ref{lem:graph-lightcone}; the repeated
bound follows from Corollary~\ref{cor:repeated-letch}. Their proofs are collected in
Appendices~\ref{app:proofs-single-block} and
\ref{app:proofs-repeated-trainability}.  

\begin{corollary}[Logarithmic rank or propagated weight avoids concentration]
\label{cor:main-protected-component}
Under the circuit and initialization assumptions of
Theorem~\ref{thm:main-trainability}, suppose that an IQP action projector
contains a nonidentity component $Z^u$ with
$|c_{a,u}|\geq1/\poly(n)$ and associated active-generator rank
$r^u=O(\log n)$.  Then
\begin{equation}
    \Var_{\theta}[\pi(a)]\geq
    c_{a,u}^2 2^{-r^u}
    =\Omega(1/\poly(n)).
    \label{eq:main-protected-policy-component}
\end{equation}
For a local-and-edge graph-IQP policy,
$r^u=\Delta_G^{\theta}(u)$ generically,
so the condition specializes to a backward interaction light cone of
size $O(\log n)$.

More generally, under the assumptions of
Proposition~\ref{prop:grouped-policy-variance}, a nonzero signature
$\sigma$ with $|C_{a,\sigma}|\geq1/\poly(n)$ and
$r_\sigma=O(\log n)$ gives
\[
    \Var_\theta[\pi(a)]\geq
    C_{a,\sigma}^{\,2}2^{-r_\sigma}
    =\Omega(1/\poly(n)).
\]
The coefficient condition applies to the group sum: an individual
$c_{a,u}$ need not suffice because coefficients within a group can
cancel.

For the augmented repeated IQP-like architecture, the corresponding
policy-variance conclusion holds if
$|c_{a,u}|\geq1/\poly(n)$ and
$\max_{\theta\in\{0,\pi/2\}^{M}}\Delta_R^{\theta}(u)
=O(\log n)$.
\end{corollary}

To see the repeated-circuit implication, retain the $Z^u$ contribution
in Theorem~\ref{thm:main-trainability}:
\begin{equation}
    \begin{aligned}
        \Var_{\theta}[\pi_R(a)]
        &\geq c_{a,u}^{2}
        \E_{\theta}\!\left[4^{-\Delta_R^{\theta}(u)}\right]\\
        &\geq c_{a,u}^{2}
        4^{-\E_{\theta}[\Delta_R^{\theta}(u)]}\\
        &\geq c_{a,u}^{2}
        4^{-\max_{\theta}\Delta_R^{\theta}(u)}\\
        &=\Omega(1/\poly(n)).
    \end{aligned}
    \label{eq:main-protected-repeated-component}
\end{equation}
The second inequality follows from Jensen's inequality because
$x\mapsto4^{-x}$ is convex. After applying Jensen's inequality,
the third inequality uses
$\E_{\theta}[\Delta_R^{\theta}(u)]
\leq\max_{\theta}\Delta_R^{\theta}(u)$
and the fact that $4^{-x}$ decreases with $x$.
Here the expectation and maximum are over the auxiliary Clifford
ensemble $\theta\in\{0,\pi/2\}^{M}$, whereas the variance is over
continuous uniform initialization. If the maximum propagated weight
is $O(\log n)$, the variance is bounded below by an inverse polynomial.
The automatic bound $\Delta_R^\theta(u)\leq n$ alone would give only
an exponentially small lower bound.

A less restrictive, ensemble-level certificate follows by expanding
the expectation itself. In the same auxiliary Clifford ensemble, let
$p_u(\delta):=\Pr_{\theta}[\Delta_R^{\theta}(u)=\delta]$.  Since
Clifford conjugation preserves a nonidentity Pauli, its weight satisfies
$1\leq\delta\leq n$, and
\begin{equation}
    \E_{\theta}\!\left[4^{-\Delta_R^{\theta}(u)}\right]
    =\sum_{\delta=1}^{n}p_u(\delta)4^{-\delta}.
    \label{eq:mean-repeated-pauli-weight}
\end{equation}
Each possible propagated weight is weighted by its probability of
occurrence. If an $O(\log n)$ light cone occurs with at least
inverse-polynomial probability in the auxiliary Clifford ensemble,
then $\Var_\theta[\pi_R(a)]=\Omega(1/\poly(n))$, provided
$|c_{a,u}|\geq1/\poly(n)$. Unlike the worst-case condition, this
ensemble-level certificate permits extensive light cones in other
realizations. It prevents exponential policy concentration at
initialization, without guaranteeing successful optimization for
every realization.

Crucially, the graph-IQP specialization constrains only the backward
light cone of the selected Pauli component.  It does not require
$d_{\max}(G)=O(\log n)$: other vertices and other Pauli components may
have $\Theta(n)$-sized light cones.  For a 1-local component $Z_i$, the
condition is simply $\deg_G(i)=O(\log n)$.  A single protected component
lower-bounds the policy variance.  A global maximum-degree bound is a
convenient sufficient condition for controlling all 1-local light cones,
but is not necessary for policy nonconcentration.

Several consequences follow
immediately for IQP-based policies.  Contiguous and basis-modulo decoders retain
inverse-polynomial variance whenever at least one of their 1-local
Pauli components lies on a vertex of degree $O(\log n)$.  Binary
recursive parity instead contains only a global
nonidentity sector and concentrates as $2^{-n}$, whereas four or more
parity-defined actions recover a protective 1-local term, whose
certificate depends only on its own light cone.  By contrast, on the
complete graph every nontrivial support has an extensive interaction
light cone, so every decoder exhibits exponential concentration.

% Queue the recursive-parity ancilla figure while page 9 is being
% assembled so that it opens page 10.
\begin{figure*}[!t]
    \centering
    \resizebox{0.98\textwidth}{!}{%
    \begin{tikzpicture}[
        wire/.style={publication wire},
        parityone/.style={publication wire},
        paritytwo/.style={publication wire},
        controlone/.style={circle, fill=black, inner sep=1.6pt},
        controltwo/.style={circle, fill=black, inner sep=1.6pt},
        unitary/.style={publication block,
                       minimum width=2.0cm, minimum height=3.2cm,
                       align=center, font=\small},
        measure/.style={publication meter, minimum width=8mm,
                        minimum height=6mm, align=center, font=\small},
        eqnote/.style={font=\scriptsize, align=center}
    ]
        \node[font=\bfseries\small] at (4.15,3.05)
            {(a) Binary recursive parity: $A=2$};
        \foreach \y/\q in {2.25/1,1.50/2,0.40/n-1,-0.35/n}{
            \node[font=\scriptsize, anchor=east] at (0.55,\y)
                {$\lvert0\rangle_{\q}$};
            \draw[wire] (0.62,\y)--(7.70,\y);
        }
        \node[font=\small] at (0.23,0.95) {$\vdots$};
        \node[unitary] at (2.05,0.95)
            {$U_G(\boldsymbol\theta)$};
        \node[font=\scriptsize, anchor=east] at (0.55,-1.20)
            {$\lvert0\rangle_{p_0}$};
        \draw[wire] (0.62,-1.20)--(7.70,-1.20);
        \foreach \x/\y in {4.00/2.25,4.60/1.50,5.20/0.40,5.80/-0.35}{
            \draw[parityone] (\x,\y)--(\x,-1.20);
            \node[controlone] at (\x,\y) {};
            \node[font=\normalsize, text=black] at (\x,-1.20)
                {$\oplus$};
        }
        \node[measure] at (6.85,-1.20) {};
        \node[eqnote, text=violet!75!black] at (4.15,-1.85)
            {$Z_{p_0}\ \longmapsto\
              P_0Z_{p_0},\qquad P_0=Z_1\cdots Z_n$};
        \node[eqnote] at (4.15,-2.43)
            {$\displaystyle
              \Pi_{a_0}^{(2)}=\frac12\bigl[I+s_0P_0\bigr],
              \qquad
              \Var_{\theta}[\pi(a_0)]=2^{-n-2}$};

        \node[font=\bfseries\small] at (13.20,3.05)
            {(b) Two nested parity bits: $A=4$};
        \foreach \y/\q in {2.25/1,1.50/2,0.40/n-1,-0.35/n}{
            \node[font=\scriptsize, anchor=east] at (9.15,\y)
                {$\lvert0\rangle_{\q}$};
            \draw[wire] (9.22,\y)--(17.65,\y);
        }
        \node[font=\small] at (8.83,0.95) {$\vdots$};
        \node[unitary] at (10.75,0.95)
            {$U_G(\boldsymbol\theta)$};
        \node[font=\scriptsize, anchor=east] at (9.15,-1.05)
            {$\lvert0\rangle_{p_0}$};
        \node[font=\scriptsize, anchor=east] at (9.15,-1.72)
            {$\lvert0\rangle_{p_1}$};
        \draw[wire] (9.22,-1.05)--(17.65,-1.05);
        \draw[wire] (9.22,-1.72)--(17.65,-1.72);
        \foreach \x/\y in {12.35/2.25,12.80/1.50,13.25/0.40,13.70/-0.35}{
            \draw[parityone] (\x,\y)--(\x,-1.05);
            \node[controlone] at (\x,\y) {};
            \node[font=\normalsize, text=black] at (\x,-1.05)
                {$\oplus$};
        }
        \foreach \x/\y in {14.35/2.25,14.80/1.50,15.25/0.40}{
            \draw[paritytwo] (\x,\y)--(\x,-1.72);
            \node[controltwo] at (\x,\y) {};
            \node[font=\normalsize, text=black] at (\x,-1.72)
                {$\oplus$};
        }
        \node[measure] at (16.55,-1.05) {};
        \node[measure] at (16.55,-1.72) {};
        \node[eqnote] at (13.20,-2.32)
            {$\displaystyle
              \Pi_{a_0a_1}^{(4)}
              =\frac14\bigl[I+s_0P_0+s_1P_1
              +s_0s_1\,\textcolor{teal!70!black}{Z_n}\bigr]$};
        \node[eqnote, text=teal!70!black] at (13.20,-2.82)
            {$P_0P_1=Z_n:\quad$ a one-local trainable sector survives};
    \end{tikzpicture}}
    \caption{Ancilla explanation of the action-count dependence of
    recursive-parity trainability. The data circuit $U_G$ is unchanged;
    the CNOT networks coherently compute the nested parity bits before
    computational-basis readout. (a) For $A=2$, a single ancilla collects
    all data bits, so propagation of its $Z$ projector produces only the
    global Pauli string $P_0=Z_1\cdots Z_n$. (b) For $A=4$, $p_0$ collects
    all $n$ bits and $p_1$ the first $n-1$. The joint ancilla term
    $Z_{p_0}Z_{p_1}$ propagates to $P_0P_1=Z_n$, exposing the local
    component responsible for the bounded-degree trainability guarantee.
    Here $s_j=(-1)^{a_j}$ and the green meters denote measurement in the
    computational basis. The ancilla construction is an equivalent
    visualization of the classical recursive-parity decoder, not an
    extra requirement for policy execution.}
    \label{fig:parity-ancilla-trainability}
\end{figure*}
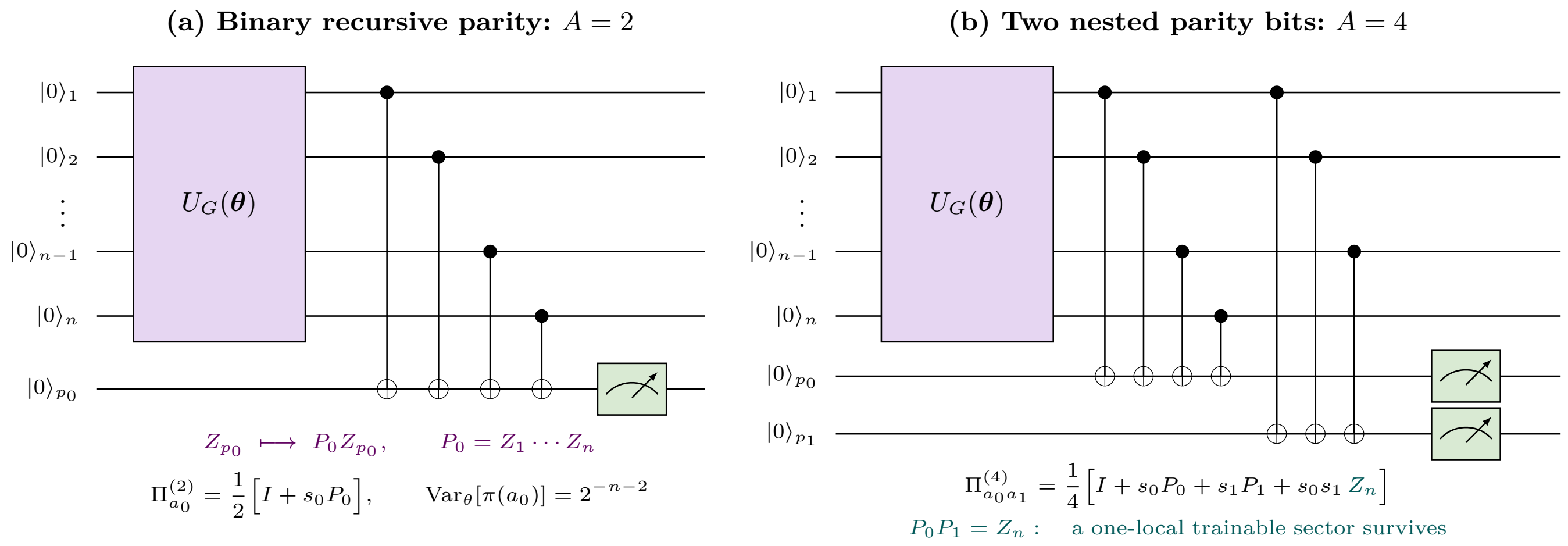

\subsection{Classical neural state embeddings}
\label{subsec:neural-state-embeddings}

An RL policy must depend on the agent's observed state.  Encoding that
classical data into circuit angles can change the parameter ensemble or
the Pauli generators on which the preceding variance laws depend.  We
therefore introduce a diagonal neural state encoder inspired by the
data-reuploading construction of P\'erez-Salinas \emph{et al.}
~\cite{PerezSalinas2020DataReuploading} and identify when,
at each fixed observation, its random initialization preserves the IQP
variance identities and the assumptions of the repeated-circuit bound.

At time step $t$, let the agent's classical observation be described by
$k$ real features,
$s_t=(s_{t,1},\ldots,s_{t,k})\in\mathbb R^k$.  A neural angle associated
with encoding gate $\ell$ is
\begin{equation}
    \phi_\ell(s_t)
    :=
    \sum_{j=1}^{k}w_j^{(\ell)}s_{t,j}+b^{(\ell)},
    \qquad \ell=1,\ldots,L_{\rm enc},
    \label{eq:neural-angle-general}
\end{equation}
where the weights $w_j^{(\ell)}$ and biases $b^{(\ell)}$ are initialized
independently and uniformly on the angle circle.  We distinguish the
number $L_{\rm enc}=\poly(n)$ of encoding gates from the repetition
depth used later.  The general diagonal Pauli encoder is
\begin{equation}
    \begin{aligned}
        E_{s_t}
        &:=\prod_{\ell=1}^{L_{\rm enc}}
           R_{P_\ell}\!\left(\phi_\ell(s_t)\right)
         =\prod_{\ell=1}^{L_{\rm enc}}
           e^{-i\phi_\ell(s_t)P_\ell/2},\\[-0.2ex]
        &\hspace{7.8em}P_\ell\in\{I,Z\}^{\otimes n}.
    \end{aligned}
    \label{eq:neural-state-embedding}
\end{equation}
The identity generator contributes only a global phase and may be
discarded.  Every remaining $P_\ell$ is a diagonal Pauli word, so the
encoder commutes with the diagonal IQP program and can be inserted as
\begin{equation}
    U_X(\boldsymbol\theta;s_t)
    :=H^{\otimes n}D_Z(\boldsymbol\theta)
       E_{s_t}H^{\otimes n}.
    \label{eq:state-conditioned-iqp}
\end{equation}
Equivalently, $E_{s_t}$ appends its Pauli supports to the commuting
generator list; when $P_\ell$ already occurs in $D_Z$, the two rotations
combine by adding their angles.  Equation~\eqref{eq:neural-state-embedding}
therefore includes local angle encoding, correlated multi-qubit feature
maps, and their data-reuploaded compositions as special cases.

\begin{embeddingcorollary}
\label{cor:neural-embedding-trainability}
Fix an observation $s_t$ and impose the circuit and initialization
hypotheses of Theorem~\ref{thm:main-trainability} on the remaining
program. If the biases $b^{(\ell)}$ are initialized
independently as $\Unif[-\pi,\pi]$ and independently of the remaining
circuit parameters, then the neural angles
$\phi_\ell(s_t)\bmod 2\pi$ are independent uniform angles, conditional
even on the encoding weights.  Hence the IQP policy-variance identities
in Theorem~\ref{thm:main-trainability} apply to the combined circuit and
encoding generator list.
\end{embeddingcorollary}

Thus, for each fixed observation, the policy variance is evaluated in
the same independent uniform-angle ensemble, with the rank recomputed
if the encoder introduces new generators.  For repeated IQP-like
policies, encoders of the form
Eq.~\eqref{eq:neural-state-embedding} may be reuploaded inside the
diagonal $D_Z^{(r)}$ programs.  The bound of
Corollary~\ref{cor:repeated-letch} then retains the same form, with the
propagated light cone evaluated for the resulting circuit.
Each repeated encoding occurrence must have its own independent
uniform offset (or an independent uniform circuit angle with which
it combines). Sharing feature weights is allowed conditional on those
weights; reusing a single bias across occurrences does not establish
the required independence. The initial orthogonal local layers must
also be retained. The proof is given in
Appendix~\ref{app:proofs-single-block}.

\subsection{Classically simulable IQP-based policies}

\begin{theorem}[Classically simulable IQP decoders]
\label{thm:main-simulability}
For a polynomial-size IQP circuit, each diagonal Pauli expectation
admits a randomized additive estimate in time
$\poly(n,\epsilon^{-1},\log\delta^{-1})$, with error at most $\epsilon$
and failure probability at most $\delta$.  Consequently, an action
probability $\pi(a)$ has such an estimator whenever either (i) its
projector has polynomially many efficiently computable Pauli terms and
polynomial coefficient $\ell_1$ norm, or (ii) its action bin is a
contiguous interval with efficiently computable boundaries.
\end{theorem}

The second route uses the interval structure directly and does not
require a sparse literal Walsh expansion. These are additive-probability
guarantees; the proof is given in
Appendix~\ref{app:proofs-classical}. Applied to the decoders studied
here, the theorem gives efficient IQP-policy estimators
for contiguous intervals, basis modulo, and recursive parity.
The theorem is deliberately restricted to a single commuting IQP block.
The repeated architectures of Sec.~\ref{sec:repeated-trainability}
alternate noncommuting $X$- and $Z$-type programs and therefore leave
the IQP class; no classical-simulability claim for that broader family
is made here.

% ================================================================
\section{Exact variance bounds for IQP policies}
\label{sec:single-iqp-trainability}
% ================================================================

Theorem~\ref{thm:main-trainability} reduces graph-IQP trainability to
the Walsh--Pauli spectrum of the decoder.  We now specialize its exact
variance decomposition to the three policy constructions introduced in
Sec.~\ref{subsec:background-interface}.

\subsection{Contiguous policies}

For the contiguous-like decoder with $A=2^k$, the $k$ most significant
qubits $Q_{\mathrm{cont}}=\{q_1,\ldots,q_k\}$ determine the action.
For action bits $a=(a_1,\ldots,a_k)$,
\begin{equation}
    \Pi_a^{\mathrm{cont}}
    =
    \frac1A
    \sum_{T\subseteq Q_{\mathrm{cont}}}
    \chi_a(T)Z_T,
    \label{eq:contiguous-projector}
\end{equation}
where $\chi_a(T)\in\{-1,+1\}$, and
\begin{equation}
    \Var_{\theta}[
        \pi^{\mathrm{cont}}(a)
    ]
    =
    \frac1{A^2}
    \sum_{\emptyset\neq T\subseteq Q_{\mathrm{cont}}}
    2^{-\Delta_G^{\theta}(T)}.
    \label{eq:contiguous-var}
\end{equation}

% ----------------------------------------------------------------
\subsection{Basis-modulo policies}

For $A=2^k$, the basis-modulo rule assigns
$a=\operatorname{int}(x)\bmod A$ and therefore depends on the $k$
least significant qubits $Q_{\mathrm{mod}}$. It has exactly the same
projector and variance expressions as Eqs.~\eqref{eq:contiguous-projector}
and~\eqref{eq:contiguous-var}, with
$Q_{\mathrm{cont}}$ replaced by $Q_{\mathrm{mod}}$.

% ----------------------------------------------------------------
\subsection{Recursive-parity policies}

For the recursive-parity-like decoder in
Eqs.~\eqref{eq:recursive-parity-bits-definition}--
\eqref{eq:recursive-parity-policy-definition}, associate the nested
parity bits with the Pauli strings

Figure~\ref{fig:parity-ancilla-trainability} gives the equivalent
coherent-ancilla picture used below to explain the action-count
dependence.

% Figure source: Figures/parity_ancilla_trainability.tex.

\begin{equation}
    P_j
    :=
    \prod_{i=1}^{n-j}Z_i.
\end{equation}

Then
\begin{equation}
    \Pi_a^{\mathrm{par}}
    =
    \prod_{j=0}^{k-1}
    \frac{
        I+(-1)^{a_j}P_j
    }2
\end{equation}
and
\begin{equation}
    \Pi_a^{\mathrm{par}}
    =
    \frac1A
    \sum_{T\subseteq[k]}
    \chi_a(T)P_T,
    \qquad
    P_T:=\prod_{j\in T}P_j.
    \label{eq:recursive-parity-projector-paper}
\end{equation}

Hence
\begin{equation}
    \Var_{\theta}[
        \pi^{\mathrm{par}}(a)
    ]
    =
    \frac1{A^2}
    \sum_{\emptyset\neq T\subseteq[k]}
    2^{-\Delta_G^{\theta}(P_T)}.
    \label{eq:recursive-parity-var-paper}
\end{equation}

For $A=2$,
\begin{equation}
    \Pi_b^{\mathrm{par}}
    =
    \frac12
    \left(
        I+
        (-1)^bZ_1\cdots Z_n
    \right),
\end{equation}
and
\begin{equation}
    \Var_{\theta}[
        \pi^{\mathrm{par}}(b)
    ]
    =
    2^{-n-2}.
    \label{eq:binary-parity-exponential}
\end{equation}

Every two-qubit generator $X_iX_j$ is inactive for $Z_1\cdots Z_n$, so
the binary parity probability is independent of the edge parameters.

For $A=4$ ($k=2$), write $a=(a_0,a_1)$ and
$s_j:=(-1)^{a_j}$.  In this case
$P_0=Z_1\cdots Z_n$, $P_1=Z_1\cdots Z_{n-1}$, and the complete Pauli
decomposition of the action projector is
\begin{equation}
\begin{aligned}
    \Pi_{a_0a_1}^{\mathrm{par}}
    & =
    \frac14
    \left(I+s_0P_0\right)
    \left(I+s_1P_1\right) \\
    & =
    \frac14\Bigl[
        I
        +s_0 Z_1\cdots Z_n
        +s_1 Z_1\cdots Z_{n-1}
        +s_0s_1 Z_n
    \Bigr].
    \label{eq:recursive-parity-four-action-projector}
\end{aligned}
\end{equation}
Here the last equality uses
$(Z_1\cdots Z_n)(Z_1\cdots Z_{n-1})=Z_n$.  Thus the two global
strings and the one-qubit term occur simultaneously, rather than the
$Z_n$ term replacing either global component.  Explicitly, the four
projectors are
\begin{equation}
\begin{aligned}
    \Pi_{00}^{\mathrm{par}}
    &=\tfrac14\left(I+P_0+P_1+Z_n\right), \\
    \Pi_{01}^{\mathrm{par}}
    &=\tfrac14\left(I+P_0-P_1-Z_n\right), \\
    \Pi_{10}^{\mathrm{par}}
    &=\tfrac14\left(I-P_0+P_1-Z_n\right), \\
    \Pi_{11}^{\mathrm{par}}
    &=\tfrac14\left(I-P_0-P_1+Z_n\right).
\end{aligned}
\label{eq:recursive-parity-four-projectors}
\end{equation}
The signs of the two global strings are selected independently by
$a_0$ and $a_1$, while the sign of the local term is their product.
Consequently, the nonidentity terms cancel when the four projectors
are summed, as required by $\sum_a\Pi_a^{\mathrm{par}}=I$.

For every $A>4$, the subset $T=\{0,1\}$ remains present in
Eq.~\eqref{eq:recursive-parity-projector-paper}, so the corresponding
projector still contains the local contribution
$(-1)^{a_0+a_1}Z_n/A$ in addition to its other Pauli terms.

The sharp change between two and four actions can also be read directly
from an equivalent coherent decoder. Introduce one ancilla $p_j$ for
each nested parity bit and compute that bit using controlled-NOT (CNOT)
gates from the data qubits.
If $W_k$ denotes this CNOT network, then Heisenberg propagation gives
\begin{equation}
    W_k^\dagger Z_{p_j}W_k=P_jZ_{p_j}.
    \label{eq:parity-ancilla-heisenberg}
\end{equation}
Because every ancilla starts in $|0\rangle$, its $Z$ factor has unit
expectation. With one ancilla, the sole nonidentity readout observable is
therefore the global string $P_0=Z_1\cdots Z_n$. With two ancillas, the
joint readout sector additionally contains
\begin{equation}
    P_0P_1
    =
    (Z_1\cdots Z_n)(Z_1\cdots Z_{n-1})
    =Z_n,
    \label{eq:parity-prefix-cancellation}
\end{equation}
because the common prefix cancels. The ancillas are only an explanatory coherent
realization of the same classical decoder; they do not alter the policy
definition or the circuit family being trained.

Thus $A=2$ concentrates exponentially, whereas every $A\geq4$ contains
this decoder-induced local sector and admits an inverse-polynomial
trainability guarantee whenever the vertex supporting that term has
degree $O(\log n)$, without changing the quantum ansatz.

% ================================================================
\section{Variance bounds for repeated IQP-like policies}
\label{sec:repeated-trainability}
% ================================================================

An IQP circuit is globally commuting. Repetition destroys this
property by alternating $X$- and $Z$-generated commuting algebras.
The results in this section are not restricted to graph-IQP generators:
each layer may contain any polynomial-size collection of commuting
Pauli strings of the appropriate axis, including arbitrary one-qubit
rotations and higher-body interactions.  Equation
~\eqref{eq:one-two-three-body-repeated-example}, containing all
one-, two-, and three-qubit $Z$ strings and their $X$-type counterparts,
is one concrete non-graph example.  Graph-IQP layers are recovered as
the 1- and 2-local edge-generated specialization.

For fixed classical input, we use the augmented architecture introduced
in Eq.~\eqref{eq:repeated-iqp-background}, written compactly as
\begin{equation}
    C_R
    =
    D_X^{(R)}D_Z^{(R)}
    \cdots
    D_X^{(1)}D_Z^{(1)}
    D_X^{(0)},
    \label{eq:repeated-circuit-paper}
\end{equation}
All layer definitions and the expressivity-preserving zero-angle slice
$D_X^{(0)}=I$ are given in Sec.~\ref{sec:model}.  For the variance
statements below, every trainable angle is independently uniform over a
full period.  Commuting the first $Z$ layer next to $D_X^{(0)}$ supplies
the two orthogonal local layers required by
Ref.~\cite{Letcher2024Tight}; the product input makes its state-dependent
prefactor equal to one.  The resulting bounds therefore certify the
\emph{augmented} random-initialization ensemble.  Although
$D_X^{(0)}=I$ recovers the standard repeated circuit as a model class,
the randomized-layer bound does not automatically apply on that
zero-angle slice.

% ----------------------------------------------------------------
\subsection{Pauli-light-cone bounds}

The added $D_X^{(0)}$ layer supplies the first of the two local
orthogonal layers required by Theorem~1 of Letcher, Woerner, and
Zoufal~\cite{Letcher2024Tight}.  That theorem applies to circuits made
of independently and uniformly initialized Pauli rotations, preceded by
two orthogonal one-qubit rotation layers, and to an arbitrary Pauli
decomposition of the measured observable.  These are precisely the
hypotheses verified here for the augmented ensemble.

For the Clifford-restricted draw
$\boldsymbol\theta\sim\operatorname{Unif}(\{0,\pi/2\}^{M})$ used by that result, backward
propagation maps $Z^u$ to one Pauli word.  We write
$\Delta_R^{\theta}(u)$ for the number of qubits on which that propagated
word is nonidentity.  Thus $\Var_{\theta}$ below is taken over the
continuous full-period initialization, whereas $\E_{\theta}$ in the
light-cone bounds is the auxiliary average over the explicitly stated
Clifford-restricted domain.  Using the same symbol avoids introducing a
second parameter vector solely for this bound.

\begin{corollary}[Repeated IQP-like policy variance, from Letcher et al.~\cite{Letcher2024Tight}]
\label{cor:repeated-letch}
For the augmented ensemble above, let $\Pi_a=\sum_u c_{a,u}Z^u$ be an
action projector. The repeated IQP-like policy variance satisfies
\begin{equation}
    \Var_{\theta}[\pi_R(a)]
    =\sum_{u\neq0}c_{a,u}^2\Var_{\theta}[L_u^{(R)}].
    \label{eq:repeated-policy-var-decomposition}
\end{equation}
For every nonidentity component $u\neq0$, the termwise bounds are
\begin{equation}
    \E_{\theta}
    \left[
        4^{-\Delta_R^{\theta}(u)}
    \right]
    \leq
    \Var_{\theta}[
        L_u^{(R)}
    ]
    \leq
    \E_{\theta}
    \left[
        2^{-\Delta_R^{\theta}(u)}
    \right].
    \label{eq:repeated-letch-bound-paper}
\end{equation}
\end{corollary}

Applying the componentwise lower bound to the nonnegative variance sum
gives the policy-level consequence
\begin{equation}
    \Var_{\theta}[\pi_R(a)]
    \geq
    \sum_{u\neq0}c_{a,u}^2
    \E_{\theta}\!\left[4^{-\Delta_R^{\theta}(u)}\right].
    \label{eq:repeated-policy-lightcone-bound-paper}
\end{equation}

The proof is given in Appendix~\ref{app:proofs-repeated-trainability}.

This result depends just on the support weight of
the Pauli word obtained by backward propagation.  It therefore applies
unchanged to higher-body generator families such as
Eq.~\eqref{eq:one-two-three-body-repeated-example}.

% ----------------------------------------------------------------
% ================================================================
\section{Classical simulability bounds for IQP policies}
\label{sec:single-iqp-simulability}
% ================================================================

The background results in
Sec.~\ref{subsec:background-simulability} can now be applied directly to
the decoder projectors.  For action $a$, let
\begin{equation}
    h_a(x):=\mathbf1[f(x)=a].
    \label{eq:action-indicator-SCP}
\end{equation}
Then $\pi(a)=\E[h_a(x)]$, and the Fourier coefficients of $h_a$ are
exactly the Pauli coefficients $c_{a,u}$ of $\Pi_a$. For the explicit
polynomial-size expansions used below, estimating and aggregating the
listed Pauli terms with
Proposition~\ref{prop:van-den-nest-iqp-estimator} suffices; the error and
runtime bounds are proved in
Proposition~\ref{prop:pauli-sparse-policy} in the appendix. More
generally, the sparse-postprocessing result of Takahashi \emph{et al.}
also covers efficiently evaluable action indicators with a known
polynomial Fourier-sparsity bound when their nonzero support is not
supplied~\cite{TakahashiMiyamotoKunihiro2026}.

\begin{corollary}[Power-of-two Fourier-sparse policies]
\label{cor:power-two-classical}

If
\begin{equation}
    A=2^k=\poly(n),
\end{equation}
then the action probabilities of the contiguous, basis-modulo, and
recursive-parity IQP policies can be estimated in classical
polynomial time.

\end{corollary}

The proof is given in Appendix~\ref{app:proofs-classical}.

Each of these projectors contains $A$ Pauli terms with coefficient
magnitude $1/A$.
\\
\\
The $A=2$ full-parity case is already covered by
Corollary~\ref{cor:power-two-classical}, but graph-IQP models admit an
equivalent and more intuitive direct argument that uses neither the Van
den Nest expectation estimator nor the sparse-postprocessing
characterization.  Let
$U_G(\boldsymbol\alpha,\boldsymbol\beta)$ be the graph-IQP circuit in
Eq.~\eqref{eq:graph-iqp-x}, acting on $|0^n\rangle$, with the action
determined by the parity of all $n$ measured qubits.  Then
\begin{equation}
    \pi^{\rm P}(a)
    =\frac12\left[
        1+(-1)^a\prod_{i=1}^{n}\cos\alpha_i
      \right],
    \qquad a\in\{0,1\}.
    \label{eq:exact-global-parity-policy}
\end{equation}
The action distribution is independent of the graph and of every
two-qubit angle $\beta_{ij}$.  The closed form is exact; for efficiently
specified angles it can be evaluated to additive error $\epsilon$ in
time $\poly(n,\log\epsilon^{-1})$ and then sampled classically.

To see this directly, write $P=Z_1\cdots Z_n$.  Every edge generator
$X_iX_j$ overlaps $P$ on two qubits and therefore commutes with it.
Because the entire $X$ program is commuting, the edge rotations cancel
from $\langle P\rangle$.  The remaining one-qubit rotations factorize on
$|0^n\rangle$, giving
$\langle P\rangle=\prod_i\cos\alpha_i$ and hence
Eq.~\eqref{eq:exact-global-parity-policy}.  Thus the parity policy is
equivalent to measuring a product-state circuit even though the physical
IQP state and its full $2^n$-outcome Born distribution can remain
entangled and classically hard to sample.  The decoder discards all
graph-dependent information carried only by the edge rotations.  This
argument is specific to full-register parity in a single local-and-edge
graph-IQP block: a parity string supported on a proper subset
anticommutes with boundary edges, and intervening noncommuting layers in
a repeated IQP-like circuit invalidate the cancellation.

\begin{corollary}[Arbitrary contiguous partitions]
\label{cor:arbitrary-contiguous-partitions}
Let an IQP policy use a contiguous decoder with efficiently computable
bin boundaries.  For any number of actions $A$, including
$A\neq2^k$, every action probability can be estimated to additive error
$\epsilon$ and failure probability $\delta$ in randomized time
$\poly(n,\epsilon^{-1},\log\delta^{-1})$.  In particular, the result
does not require a Fourier-sparse action indicator.
\end{corollary}

Indeed, each action corresponds to one integer interval
$B_a=[L_a,U_a)$, although the interval sizes need not be equal.  Its
probability is the difference of two threshold probabilities,
\begin{equation}
    \pi(a)=\Pr[x<U_a]-\Pr[x<L_a].
\end{equation}
The binary expansion of a threshold partitions the event $x<t$ into at
most $n$ disjoint prefix events.  A length-$j$ prefix projector has
$2^j$ Pauli terms, but these terms have equal magnitude and known
signs.  Sampling a uniformly random subset of the prefix therefore
turns the full Pauli sum into a bounded unbiased estimator of one IQP
Pauli expectation.  Since an interval uses at most $2n$ signed prefix
events, its action probability is estimated with only polynomially many
calls to the Van den Nest estimator.  The construction, formal error
bound, and runtime are given in
Appendix~\ref{app:arbitrary-contiguous-sim}, which proves
Corollary~\ref{cor:arbitrary-contiguous-partitions}.  Thus
non-power-of-two contiguous policies remain classically easy even when
their literal Walsh expansions are exponentially large: their interval
structure, rather than Fourier sparsity, supplies the efficient
algorithm.
\\
For one action bin, direct quantum estimation executes the IQP circuit,
measures the output, and records the Bernoulli variable
$\mathbf 1[x\in B_a]$.  Hoeffding's inequality therefore requires
$O(\epsilon^{-2}\log\delta^{-1})$ circuit shots to estimate $\pi(a)$
to additive error $\epsilon$ and failure probability $\delta$.  The
classical construction in Appendix~\ref{app:arbitrary-contiguous-sim}
samples from at most $2n$ signed prefix events and requires
$O(n^2\epsilon^{-2}\log\delta^{-1})$ evaluations of the bounded IQP
estimator.  If the IQP block contains $m$ commuting rotations, a direct
evaluation of one classical sample costs $O(mn)$ bit operations, giving
the explicit, nonoptimized upper bound
\begin{equation}
    T_{\rm C}
    =O\!\left(mn^3\epsilon^{-2}\log\delta^{-1}\right),
    \qquad
    N_{\rm Q}
    =O\!\left(\epsilon^{-2}\log\delta^{-1}\right).
    \label{eq:contiguous-classical-quantum-cost}
\end{equation}
Thus the theorem excludes a superpolynomial advantage for this
additive action-probability task, but it does not prove equal quantum
and classical costs.  The displayed algorithms differ by a polynomial
overhead, which may leave room for a polynomial quantum speedup.
Neither upper bound is an optimality or lower-bound statement, and
hardware circuit executions and classical bit operations are not
directly interchangeable resource units.

% ================================================================
\section{Majority--modulo policies: protected components and dense exact support}
\label{sec:majority-modulo-construction}
% ================================================================

We now construct a decoder with inverse-polynomial low-weight
coefficients and exponentially large exact Pauli support.  Its
trainability certificate is componentwise: one surviving Pauli term of
rank $O(\log n)$ suffices for nonconcentration.  For graph IQPs, only
that term's backward light cone must contain $O(\log n)$ qubits; the
rest of the graph may contain vertices of much larger degree.  Its dense exact expansion fails the polynomial-sparsity certificate
above. This excludes direct polynomial-size enumeration, not
efficient classical simulation: aggregate estimators, approximations,
and circuit-specific contractions may avoid that expansion entirely.

The construction begins with the Boolean majority function. For $n$
bits, write $z_i=(-1)^{x_i}$ and define
\begin{equation}
    M_n(z):={\rm sgn}\!\left(\sum_{i=1}^{n}z_i\right)\in\{-1,+1\}.
    \label{eq:majority-function-paper}
\end{equation}
For even $n$ we use the deterministic convention ${\rm sgn}(0)=+1$,
so ties are assigned to majority bit zero.
Its associated majority bit is
$m_n(x)=[1-M_n((-1)^x)]/2$: a majority of zeros gives $m_n=0$, and a
majority of ones gives $m_n=1$.

Every function $g:\{-1,+1\}^n\rightarrow\mathbb R$ has a unique
Fourier--Walsh expansion as a sum of parity functions
~\cite{ODonnell2014BooleanFunctions}:
\begin{align}
    g(z)
    &=\sum_{S\subseteq[n]}\widehat g(S)\chi_S(z),
    &
    \chi_S(z)&:=\prod_{i\in S}z_i,
    \label{eq:boolean-walsh-expansion-majority-section}\\
    \widehat g(S)
    &=2^{-n}\sum_{z\in\{-1,+1\}^n}g(z)\chi_S(z).
    \nonumber
\end{align}
The characters $\chi_S$ are subset parities. On computational-basis
states they are exactly the eigenvalues of the Pauli strings
$Z_S:=\prod_{i\in S}Z_i$, because
$Z_S|x\rangle=\chi_S((-1)^x)|x\rangle$. Therefore the diagonal majority
observable is
\begin{equation}
    \mathcal M_n
    :=\sum_{x\in\{0,1\}^n}M_n((-1)^x)|x\rangle\langle x|
    =\sum_{S\subseteq[n]}\widehat M_n(S)Z_S.
    \label{eq:majority-diagonal-observable}
\end{equation}
Permutation symmetry implies that $\widehat M_n(S)$ depends only on the
Pauli weight $\ell=|S|$, so Eq.~\eqref{eq:majority-diagonal-observable}
can be grouped as
\begin{equation}
    \mathcal M_n
    =\sum_{\ell=0}^{n}
      \sum_{\substack{S\subseteq[n]\\|S|=\ell}}
      \widehat M_n(S)Z_S.
    \label{eq:majority-general-pauli-expansion}
\end{equation}
The exact odd-majority spectrum of Theorem~5.19 in
Ref.~\cite{ODonnell2014BooleanFunctions}, together with the restriction
$M_n(z)=M_{n+1}(z,+1)$ for even $n$, gives
\begin{equation}
\begin{aligned}
\widehat M_n(S)
&=
\begin{cases}
c_{n,\ell},&n\ \mathrm{odd},\\
d_{n,\ell},&n\ \mathrm{even},
\end{cases}
\qquad \ell=|S|,\\[1ex]
c_{n,\ell}
&=
\frac{\binom{n-1}{(n-1)/2}}{2^{n-1}}
\begin{cases}
\displaystyle
(-1)^{(\ell-1)/2}
\frac{\binom{(n-1)/2}{(\ell-1)/2}}
     {\binom{n-1}{\ell-1}},
&\ell\ \mathrm{odd},\\[2.5ex]
0,&\ell\ \mathrm{even},
\end{cases}\\[1ex]
d_{n,\ell}
&=
\frac{\binom n{n/2}}{2^n}
\begin{cases}
\displaystyle
(-1)^{(\ell-1)/2}
\frac{\binom{n/2}{(\ell-1)/2}}{\binom n{\ell-1}},
&\ell\ \mathrm{odd},\\[2.5ex]
\displaystyle
(-1)^{\ell/2}
\frac{\binom{n/2}{\ell/2}}{\binom n\ell},
&\ell\ \mathrm{even}.
\end{cases}
\end{aligned}
    \label{eq:majority-general-weight-coefficient}
\end{equation}
For odd $n$, the complement symmetry $M_n(-z)=-M_n(z)$ eliminates every
even-weight coefficient, whereas every odd-weight coefficient is
nonzero. Hence
\begin{equation}
    N_{\rm Pauli}(\mathcal M_n)
    =\sum_{\substack{0\leq\ell\leq n\\ \ell\ \mathrm{odd}}}\binom n\ell
    =2^{n-1}.
    \label{eq:majority-support-paper}
\end{equation}
For even $n$ under the stated tie convention, both branches in
Eq.~\eqref{eq:majority-general-weight-coefficient} are nonzero, so all
$2^n$ Pauli strings occur. Majority therefore retains both local and
global correlations despite being defined by a single deterministic
bit.

Its two eigenspaces give the binary majority projectors directly:
\begin{equation}
    \Pi_a^{\rm Maj}
    :=\sum_{x:m_n(x)=a}|x\rangle\langle x|
    =\frac{I+(-1)^a\mathcal M_n}{2},
    \qquad a\in\{0,1\}.
    \label{eq:binary-majority-projector}
\end{equation}
Thus the parity expansion of the classical decoder becomes the Pauli
expansion of its diagonal quantum projector. For odd $n$, each binary
action projector therefore contains the identity together with the
$2^{n-1}$ odd-weight strings in $\mathcal M_n$, for a total of
$2^{n-1}+1$ nonzero Pauli terms.

For even $n$, assigning both members of a tied complementary pair to the
same output breaks $M_n(-z)=-M_n(z)$ and restores the identity and the
other even-weight sectors. The four-qubit example below makes this
effect explicit.

To construct a policy with
$A\geq2$ actions, we generate $k=\lceil\log_2A\rceil$ majority bits on
disjoint blocks and reduce their binary value modulo $A$.

\begin{definition}[Majority--modulo policy]
\label{def:majority-modulo-policy}
Let $A\geq2$, $k=\lceil\log_2A\rceil$, and $C=2^k$.
Partition the measured register into $k$ disjoint blocks
$V_1,\ldots,V_k$. For a measured bit string $x$, define the majority
sign and corresponding bit of block $V_r$ directly as
\begin{align}
    M_r(x)&:={\rm sgn}\!\left(\sum_{i\in V_r}(-1)^{x_i}\right),\nonumber\\
    y_r(x)&:=\frac{1-M_r(x)}2\in\{0,1\}.
    \label{eq:majmod-majority-bit}
\end{align}
When an even block is used, we set ${\rm sgn}(0)=+1$, assigning ties to
majority bit $0$.
Interpreting $y(x)=(y_1,\ldots,y_k)$ as
$J(y)=\sum_{r=1}^k2^{k-r}y_r$, the majority--modulo (MM) decoder and its
Born policy are
\begin{align}
    f_{\rm MM}(x)&:=J(y(x))\bmod A,
    \label{eq:majmod-action-map}\\
    \pi_{\boldsymbol\theta}^{\rm MM}(a)
    &:=\langle\psi_{\boldsymbol\theta}|\Pi_a^{\rm MM}
       |\psi_{\boldsymbol\theta}\rangle,
    \qquad
    \Pi_a^{\rm MM}:=
    \sum_{x:f_{\rm MM}(x)=a}|x\rangle\langle x|.
    \label{eq:majmod-policy-definition}
\end{align}
\end{definition}

For the exact asymptotic results below, the blocks have a common odd
size $b$, so $n=kb$. Odd blocks avoid ties and make every majority bit
balanced. Approximately balanced blocks, padding qubits, and the preceding
tie rule give valid finite-size decoders, but their exact spectra must be
used rather than the odd equal-block formula.

\subsection{Examples}

\subsubsection{Binary five-qubit policy}

For $A=2$, $k=1$ and the full register forms one majority block. Hence
$f_{\rm MM}$ is ordinary binary majority and
\begin{equation}
    \Pi_a^{\rm MM}
    =\frac{I+(-1)^a\mathcal M_5}{2}.
    \label{eq:majmod-binary-projector-example}
\end{equation}
The five-bit majority observable contains every odd-weight Pauli string.
There are $\binom51=5$ 1-local, $\binom53=10$ 3-local, and
$\binom55=1$ five-local strings. Substituting $n=5$ into
Eq.~\eqref{eq:majority-general-weight-coefficient} gives
$c_{5,1}=3/8$, $c_{5,3}=-1/8$, and $c_{5,5}=3/8$. Hence, with
$Z_S:=\prod_{i\in S}Z_i$,
\begin{equation}
    \mathcal M_5
    =\frac38\underbrace{\sum_{\substack{S\subseteq[5]\\|S|=1}}Z_S}_{5\ \mathrm{strings}}
     -\frac18\underbrace{\sum_{\substack{S\subseteq[5]\\|S|=3}}Z_S}_{10\ \mathrm{strings}}
     +\frac38\underbrace{\sum_{\substack{S\subseteq[5]\\|S|=5}}Z_S}_{1\ \mathrm{string}}.
    \label{eq:majmod-five-compact}
\end{equation}
Thus the same binary action probability contains local and global
correlations. For example, $x=00011$ has majority sign $+1$ and maps to
action $0$, whereas $x=00111$ has sign $-1$ and maps to action $1$.
The term-by-term expansion of Eq.~\eqref{eq:majmod-five-compact} is
given in Appendix~\ref{app:supplementary-majority-examples}.

\subsubsection{Three-action modulo collision}

Now $k=2$ and $C=4$. Choose
$V_1=\{1,2,3\}$ and $V_2=\{4,5,6\}$. Each block has
\begin{align}
    \mathcal M_1&=\tfrac12(Z_1+Z_2+Z_3-Z_1Z_2Z_3),\nonumber\\
    \mathcal M_2&=\tfrac12(Z_4+Z_5+Z_6-Z_4Z_5Z_6).
    \label{eq:majmod-three-bit-blocks}
\end{align}
The majority codes satisfy
$00,11\mapsto0$, $01\mapsto1$, and $10\mapsto2$. Therefore,
the four codeword projectors are
\begin{align}
    Q_{00}
    &=\tfrac14(I+\mathcal M_1)(I+\mathcal M_2)\nonumber\\
    &=\tfrac14(I+\mathcal M_1+\mathcal M_2
                    +\mathcal M_1\mathcal M_2),\nonumber\\
    Q_{01}
    &=\tfrac14(I+\mathcal M_1)(I-\mathcal M_2)\nonumber\\
    &=\tfrac14(I+\mathcal M_1-\mathcal M_2
                    -\mathcal M_1\mathcal M_2),\nonumber\displaybreak[1]\\
    Q_{10}
    &=\tfrac14(I-\mathcal M_1)(I+\mathcal M_2)\nonumber\\
    &=\tfrac14(I-\mathcal M_1+\mathcal M_2
                    -\mathcal M_1\mathcal M_2),\nonumber\\
    Q_{11}
    &=\tfrac14(I-\mathcal M_1)(I-\mathcal M_2)\nonumber\\
    &=\tfrac14(I-\mathcal M_1-\mathcal M_2
                    +\mathcal M_1\mathcal M_2).
    \label{eq:majmod-three-action-codewords}
\end{align}
Grouping these projectors according to the modulo map gives
\begin{align}
    \Pi_0^{\rm MM}&=Q_{00}+Q_{11}
       =\tfrac12(I+\mathcal M_1\mathcal M_2),\nonumber\\
    \Pi_1^{\rm MM}&=Q_{01}\nonumber\\
       &=\tfrac14(I+\mathcal M_1-\mathcal M_2
                         -\mathcal M_1\mathcal M_2),\nonumber\\
    \Pi_2^{\rm MM}&=Q_{10}\nonumber\\
       &=\tfrac14(I-\mathcal M_1+\mathcal M_2
                         -\mathcal M_1\mathcal M_2).
    \label{eq:majmod-three-action-example}
\end{align}
The collision cancels the individual block observables from action $0$,
but leaves their product. Its lowest-weight contribution is
\begin{equation}
    \Pi_0^{\rm MM}
    =\frac12I+\frac18
      \sum_{\substack{i\in V_1\\j\in V_2}}Z_iZ_j
      +\text{higher-weight terms}.
    \label{eq:majmod-three-action-lowweight}
\end{equation}
Thus collided actions retain 2-local information, while the
one-preimage actions also retain 1-local terms through
$\mathcal M_1$ or $\mathcal M_2$.

By contrast, when $A=C=2^k$, every action has one code preimage and no
standalone block observable cancels.  The resulting projectors therefore
retain their 1-local sectors as well as cross-block and higher-weight
correlations.  The explicit $A=4$ construction is given in
Appendix~\ref{app:supplementary-majority-examples}.

\subsubsection{Even four-qubit block with deterministic ties}

For an even block, majority requires a tie convention. Mapping ties to
action $0$ means assigning majority sign $+1$ when
$\sum_i(-1)^{x_i}=0$. This function equals five-bit majority with one
sign fixed to $+1$, and its Pauli expansion is
\begin{align}
    \mathcal M_4^{(+)}
    ={}&\frac38 I+\frac38\sum_iZ_i
       -\frac18\sum_{i<j}Z_iZ_j\nonumber\\
      &-\frac18\sum_{i<j<k}Z_iZ_jZ_k
       +\frac38Z_1Z_2Z_3Z_4.
    \label{eq:majmod-even-four-example}
\end{align}
Unlike odd-block majority, deterministic tie-breaking produces both
even- and odd-weight terms. With
$\Pi_a^{\rm MM}=[I+(-1)^a\mathcal M_4^{(+)}]/2$, the identity
coefficients are $11/16$ and $5/16$, reflecting the six tied strings
assigned to action $0$. This finite extension preserves local terms but
is not substituted into the odd-block theorem below.

Modulo collisions and within-block ties are distinct effects.  When
even blocks are combined with $A<2^k$, the identity component created by
the tie convention can restore 1-local terms inside a product of block
observables even when the standalone observables cancel.  The explicit
$A=3$, $n=8$ construction and its action masses are given in
Appendix~\ref{app:supplementary-majority-examples}.

% ----------------------------------------------------------------
\subsection{Fourier structure and local coefficients}

\paragraph{Odd blocks.}
For an odd block of size $b$, the general spectrum in
Eq.~\eqref{eq:majority-general-pauli-expansion} contains exactly
$2^{b-1}$ nonzero strings, all at odd Pauli weight.  Its 1-local
coefficient is
\begin{equation}
    \begin{aligned}
    \eta_b:=\widehat M_b(\{i\})
    &=\frac{\binom{b-1}{(b-1)/2}}{2^{b-1}} \\
    &=\sqrt{\frac{2}{\pi(b-1)}}
      \left[1+O(b^{-1})\right]
      =\Theta(b^{-1/2}),
    \end{aligned}
    \label{eq:majority-eta-paper}
\end{equation}
where the exact expression is the $\ell=1$ specialization of
Eq.~\eqref{eq:majority-general-weight-coefficient} and the asymptotic
form follows by writing $b=2m+1$ and applying Stirling's formula to
$\binom{2m}{m}/4^m$.

More generally, for every fixed odd weight $\ell$,
Eq.~\eqref{eq:majority-general-weight-coefficient} gives
\begin{equation}
    |c_{b,\ell}|=\Theta(b^{-\ell/2}).
    \label{eq:majority-fixed-low-weight-scaling}
\end{equation}
Thus there are no intrinsic 2-local terms in an odd block, while its
3-local coefficient satisfies
\begin{equation}
    |c_{b,3}|=\frac{\eta_b}{b-2}=\Theta(b^{-3/2}).
    \label{eq:majority-three-local-scaling}
\end{equation}
The block observable can therefore be organized as
\begin{equation}
    \mathcal M_r
    =\eta_b\sum_{i\in V_r}Z_i
     +\sum_{\substack{S\subseteq V_r\\|S|\geq3,\ |S|\ {\rm odd}}}
      \widehat M_b(S)Z_S.
    \label{eq:block-majority-pauli}
\end{equation}
The 1-local and 3-local sectors are the components needed for the
trainability bounds, while the same observable retains exponentially
many higher-weight strings and hence global correlations.

\paragraph{Even blocks.}
Let $b$ be even and assign every tie to majority bit $0$, or equivalently
majority sign $+1$.  The two even-$b$ branches in
Eq.~\eqref{eq:majority-general-weight-coefficient} are nonzero at every
weight $\ell=0,1,\ldots,b$, and therefore
\begin{equation}
    N_{\rm Pauli}(\mathcal M_b^{(+)})
    =\sum_{\ell=0}^{b}\binom b\ell=2^b.
    \label{eq:even-majority-full-support}
\end{equation}
Its identity and 1-local coefficients coincide:
\begin{equation}
    \mu_b:=d_{b,0}
    =\eta_b^{(+)}:=d_{b,1}
    =\frac{\binom b{b/2}}{2^b}
    =\Theta(b^{-1/2}).
    \label{eq:even-majority-identity-one-local}
\end{equation}
The first intrinsic 2- and 3-local coefficients also coincide and obey
\begin{equation}
    d_{b,2}=d_{b,3}=c_{b+1,3}
    =-\frac{\eta_{b+1}}{b-1}
    =\Theta(b^{-3/2}).
    \label{eq:even-majority-two-three-local-scaling}
\end{equation}
Thus, on block $V_r$,
\begin{equation}
    \mathcal M_r^{(+)}
    =\mu_b I+\eta_b^{(+)}\sum_{i\in V_r}Z_i
     +\sum_{\substack{S\subseteq V_r\\|S|\geq2}}d_{b,|S|}Z_S.
    \label{eq:even-block-majority-pauli}
\end{equation}
Assigning every tie to majority bit $1$ reverses the sign of the
even-weight sector, including the identity, while leaving the odd-weight
sector unchanged.

% ----------------------------------------------------------------
\subsection{Initialization-nonconcentration guarantees}

Theorem~\ref{thm:main-trainability} decomposes the variance of an IQP
policy exactly into nonnegative Pauli contributions: if
$\Pi_a=\sum_u c_{a,u}Z^u$, then each nonidentity component contributes
$c_{a,u}^2 2^{-r^u}$, where $r^u$ is the active-generator rank
associated with $Z^u$.  Hence
the exponentially large Pauli decomposition of a majority policy need
not be controlled term by term.  A single component with a sufficiently
large coefficient and sufficiently small rank already lower-bounds the
entire policy variance; all remaining terms can only increase it.

\begin{corollary}[Protected-component trainability of majority policies]
\label{cor:majmod-trainability-paper}
Under the circuit and initialization assumptions of
Theorem~\ref{thm:main-trainability}, suppose an action projector of a
majority policy contains a nonidentity component $Z^u$ with
$|c_{a,u}|\geq1/\poly(n)$.  If its associated rank satisfies
$r^u=O(\log n)$ for an IQP $X$ program, then
$\Var_{\theta}[\pi(a)]=\Omega(1/\poly(n))$.  For a local-and-edge graph IQP this
condition is equivalently
$\Delta_G^{\theta}(u)=O(\log n)$.  For an augmented repeated
IQP-like model, the same conclusion holds if
$\max_{\theta\in\{0,\pi/2\}^{M}}\Delta_R^{\theta}(u)=O(\log n)$.
\end{corollary}

The corollary is independent of whether the majority blocks are odd or
even and of how ties are assigned.  Those choices determine which
coefficients survive, but once one surviving coefficient is inverse
polynomial, logarithmic rank or propagated light-cone size is sufficient
to avoid exponential policy concentration.
For repeated models, the weaker ensemble-probability criterion
discussed after Eq.~\eqref{eq:mean-repeated-pauli-weight} also suffices.

For the common odd equal-block construction with $A=\poly(n)$, the
coefficient condition is explicit.  A one-preimage action has 1-local
terms of magnitude $\eta_b/C=\Omega(1/(A\sqrt b))$.  A collided action
has cross-block 2-local terms of magnitude
$2\eta_b^2/C=\Omega(1/(Ab))$, obtained by selecting two blocks on
which its codewords differ.  Thus each action has an
inverse-polynomial low-weight coefficient; its variance is protected
whenever at least one such term has logarithmic rank.

For graph IQPs this statement should not be read as a maximum-degree
assumption.  If the protected component is $Z_i$, then
$\Delta_G^{\theta}(i)=1+\deg_G(i)$, so a single vertex of degree
$O(\log n)$ certifies the action-probability variance even when other
vertices have degree $\Theta(n)$.  If a modulo collision removes all
1-local terms and the selected component is $Z_iZ_j$, the relevant
condition is instead that the union of $\{i,j\}$ and its external graph
neighbours has size $O(\log n)$.

% ----------------------------------------------------------------
\subsection{Fourier density and classical-simulability status}

\begin{proposition}[Fourier support of majority--modulo actions]
\label{prop:majmod-fourier-support-paper}
Let $A=\poly(n)$, $k=\lceil\log_2A\rceil$, and partition $n=kb$ qubits
into equal odd blocks of size $b\geq3$.  Every majority--modulo action
satisfies
\begin{equation}
    N_{\mathrm{Pauli}}
    (
        \Pi_a^{\mathrm{MM}}
    )
    =
    2^{\Omega(n)}.
    \label{eq:majmod-exponential-support-paper}
\end{equation}

\end{proposition}

The proof is given in Appendix~\ref{app:proofs-majority}.
For common even blocks with deterministic ties, every one-preimage code
projector has the full $2^n$ Pauli support by
Eq.~\eqref{eq:even-majority-full-support}; the collided $A=3,n=8$
example also has full support. For arbitrary even-block collisions,
the exact weighted cancellations depend on the chosen action map and
deterministic tie convention.

Consequently, for the odd-block family of
Proposition~\ref{prop:majmod-fourier-support-paper}, the majority--modulo
action indicator is not Fourier sparse. Enumerating its full exact
Pauli expansion and estimating every term separately with the Van den
Nest estimator would require $2^{\Omega(n)}$ termwise estimates,
while the sparse-postprocessing guarantee of Takahashi \emph{et al.}
does not apply because it requires polynomial exact Fourier
support~\cite{TakahashiMiyamotoKunihiro2026}. These observations exclude
direct polynomial-size
enumeration of the exact expansion and the application of that
exact-sparsity certificate, not efficient approximate or randomized
aggregation.

These failures do not exhaust classical simulation.  Any efficient weak
sampler for the complete IQP output distribution immediately yields an
efficient estimator of every decoded action probability by applying the
decoder to its samples.  Such circuit-level routes are known when the
circuit tensor network has logarithmic treewidth
~\cite{MarkovShi2008}, when the output distribution is promised to be
approximately sparse~\cite{SchwarzVanDenNest2013}, when the non-Clifford
resource admits an efficient stabilizer decomposition
~\cite{BravyiEtAl2019Stabilizer}, or in several sufficiently noisy IQP
regimes~\cite{RajakumarWatsonLiu2024}.  Strong permutation symmetry can
also reduce the dynamics to polynomial-size representation blocks
~\cite{AnschuetzEtAl2023Symmetric}.  In particular, bounded-treewidth
path instances can be contracted while carrying a polynomial-size
counter for the majority threshold, so Fourier density does not make
those instances hard.  None of these results currently gives a generic
polynomial-time estimator for noiseless, independently parameterized
bounded-degree expander families of the type motivated by our finite
cubic instance.  The ten-qubit experiment below is a fixed connectivity
stress test, not evidence for an asymptotic expander-family separation.

Fourier density therefore obstructs these particular exact-expansion
routes, not classical simulation in general. Approximate truncation,
sampling of Fourier terms, or a direct estimator of the majority event
could still exploit structure invisible to the exact support size.
Establishing or excluding such an algorithm is left open.

\section{Numerical Experiments}
\label{sec:experiments}
% ================================================================

We next test whether the finite-size behavior of simulated IQP policies
matches the exact concentration laws and whether the choice of action
decoder has a measurable effect on policy optimization. The experiments
are designed as validation and mechanism studies: they test the derived
equalities under their stated initialization ensemble and then isolate
the readout in a matched RL comparison. They do not
attempt to demonstrate quantum computational advantage.

\subsection{Simulation and statistical protocol}
\label{subsec:experimental-protocol}

Circuit definitions use QASP. Most variance and small-system training
calculations use statevectors, including independently checked
differentiable implementations; the final connected model instead uses
an exact tensor-network contraction. The finite-shot study samples
measurement counts. These backends and their checks are specified in
the relevant experiments.

The public companion experiment repository, identified in the Data
availability statement, maps these studies to their notebooks, scripts,
and available saved results. The repository contains the studies
reported here and their experiment-specific helpers; QASP is installed
separately from its maintained software repository.

The path and ring variance tests use 2,000 independent parameter vectors
with angles uniform on $[-\pi,\pi]$. Error bars are twice the plug-in
Monte Carlo standard error of the unbiased sample variance, using the
empirical fourth central moment. The random-cubic study instead uses
100 topologies with 20 parameter draws each and its stated pooled
estimator. Repeated-circuit tests also sample the auxiliary Clifford
ensemble for the bounds. None of these error bars represent hardware
noise. Active ranks and Walsh coefficients are evaluated exactly, and
a common set of statevectors is reused across decoders where possible.

CartPole episodes are limited to 200 environment steps, with a
study-specific success threshold of 195 over a trailing 100-episode
window. The independent experimental unit is the seed, not an episode.
Moving averages are formed within runs, then uncertainty is summarized
across seeds. Individual studies differ in initialization, sample size,
and stopping rules; their detailed protocols are retained rather than
pooled into one statistical comparison. The final connected-model
study requires the full 100-episode window before freezing learning.

We first test initialization laws, then compare majority with valid
local readouts and MLPs. Secondary depth sweeps, three-body extensions,
finite-shot tests, and full gradient-sector diagnostics are collected
in Appendix~\ref{app:supplementary-numerics}. They are finite-instance
mechanism studies, not demonstrations of quantum advantage or of
asymptotic RL convergence.

\subsection{Initialization circuits}
\label{subsec:experiment-rank-law}

\begin{figure*}[!htbp]
    \centering
    \begin{subfigure}[t]{0.485\textwidth}
        \centering
        \includegraphics[width=\linewidth]{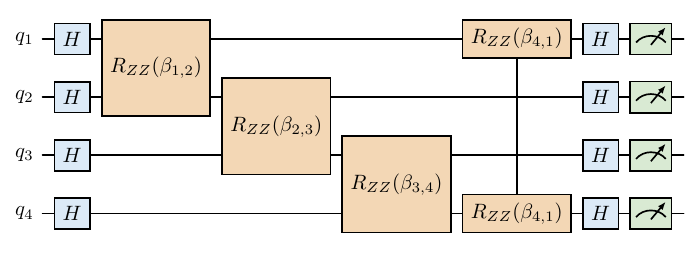}
        \caption{Pure-$R_{XX}$ periodic ring.}
    \end{subfigure}\hfill
    \begin{subfigure}[t]{0.485\textwidth}
        \centering
        \includegraphics[width=\linewidth]{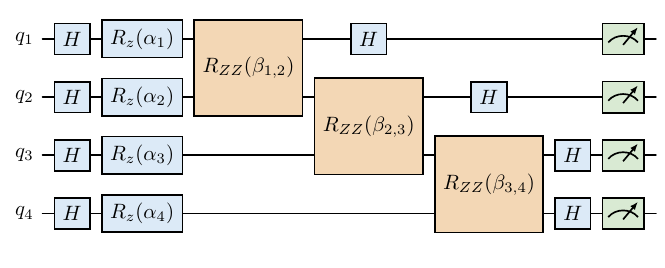}
        \caption{Local-plus-edge graph-IQP path.}
    \end{subfigure}
    \caption{Circuit families used in the initialization-variance
    experiments, exported directly from the QASP builders at the
    representative size $n=4$. (a) The pure-$R_{XX}$ control contains
    $ZZ$ rotations on the periodic ring $C_n$ and no local rotation.
    (b) The graph-IQP circuit contains every local $Z$ rotation and
    nearest-neighbor $ZZ$ rotations on the path $P_n$. Hadamard
    conjugation gives the equivalent commuting $R_{XX}$ and
    $R_X$--$R_{XX}$ programs, respectively. The terminal meter symbols
    denote measurement in the computational basis; action probabilities
    are evaluated by analytic statevector simulation.}
    \label{fig:variance-circuit-schematics}
\end{figure*}

Figure~\ref{fig:variance-circuit-schematics} shows QASP exports of the
one-layer circuits used for the policy-variance comparisons. Both are
implemented in the native $H^{\otimes n}D_ZH^{\otimes n}$ form and are
identical to the displayed commuting $X$-rotation circuits by Hadamard
conjugation. The pure-$R_{XX}$ family removes all local rotations and
uses the cycle $C_n$; the graph-IQP family retains local rotations and
uses the path $P_n$.

These one-layer path and ring circuits are deliberately classically
tractable validation models. Their nearest-neighbor interactions can
be scheduled in constant depth, yielding bounded-width tensor networks
that permit efficient classical sampling of the full output
distribution~\cite{MarkovShi2008}. Applying any of the efficiently
computable decoders studied here to those samples also gives efficient
additive estimates of its action probabilities, including majority
decoding despite its dense exact Fourier expansion. We chose these
simple geometries to verify the rank and light-cone predictions and
isolate decoder effects, not to test classical sampling hardness.
The statevector calculations used below provide a direct finite-size
reference; they are not the most efficient classical simulation method
available for these architectures.

The policy study below uses the same initialization ensemble and
statevector backend. In addition to the graph-IQP path, it includes a
pure-$R_{XX}$ periodic ring to isolate the effect of removing all
single-qubit generators. We therefore proceed directly to the policy
variances, which are the quantities used by the RL
objective.

\subsection{Action decoding controls concentration}
\label{subsec:experiment-action-decoding}

\begin{figure*}[!htbp]
    \centering
    \includegraphics[width=0.98\textwidth]{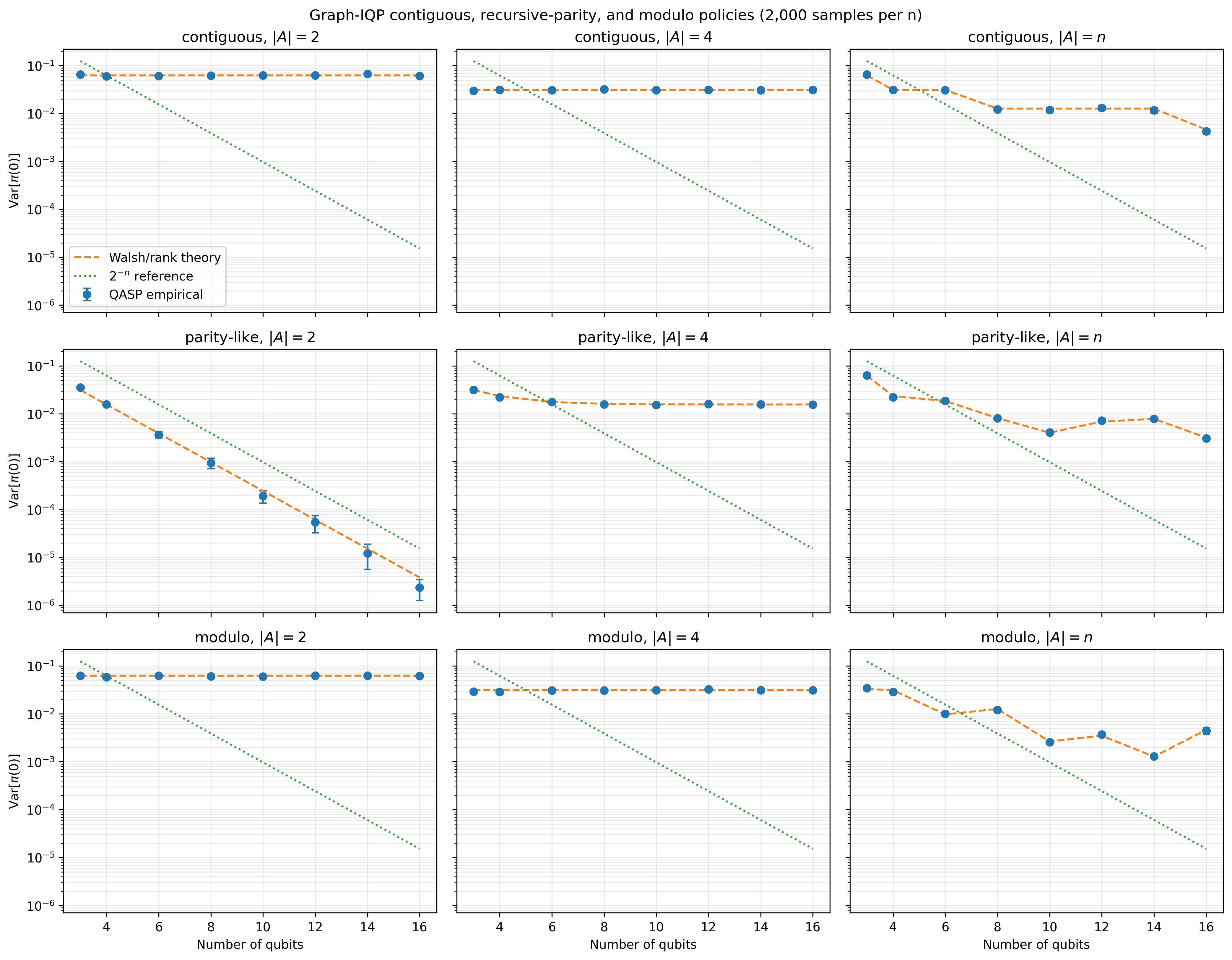}
    \caption{Variance of the action-$0$ probability for contiguous,
    recursive-parity, and modulo policies on graph-IQP paths. Columns
    correspond to $A=2$, $A=4$, and $A=n$. Each point uses 2,000 random
    initializations; orange dashed curves are the exact Walsh/rank
    prediction and green dotted curves show $2^{-n}$ for scale.}
    \label{fig:iqp-policy-variance}
\end{figure*}

We next evaluate action-$0$ projectors for contiguous, recursive-parity,
and modulo decoding on graph-IQP paths and pure-$R_{XX}$ rings at
$n\in\{3,4,6,8,10,12,14,16\}$ and
$A\in\{2,4,n\}$. Figure~\ref{fig:iqp-policy-variance} reports the
graph-IQP results. The exact curve is obtained from the full Walsh
spectrum using Eq.~\eqref{eq:exact-policy-variance}; no scaling law is
fitted to the samples.

\begin{figure*}[!htbp]
    \centering
    \includegraphics[width=0.98\textwidth]{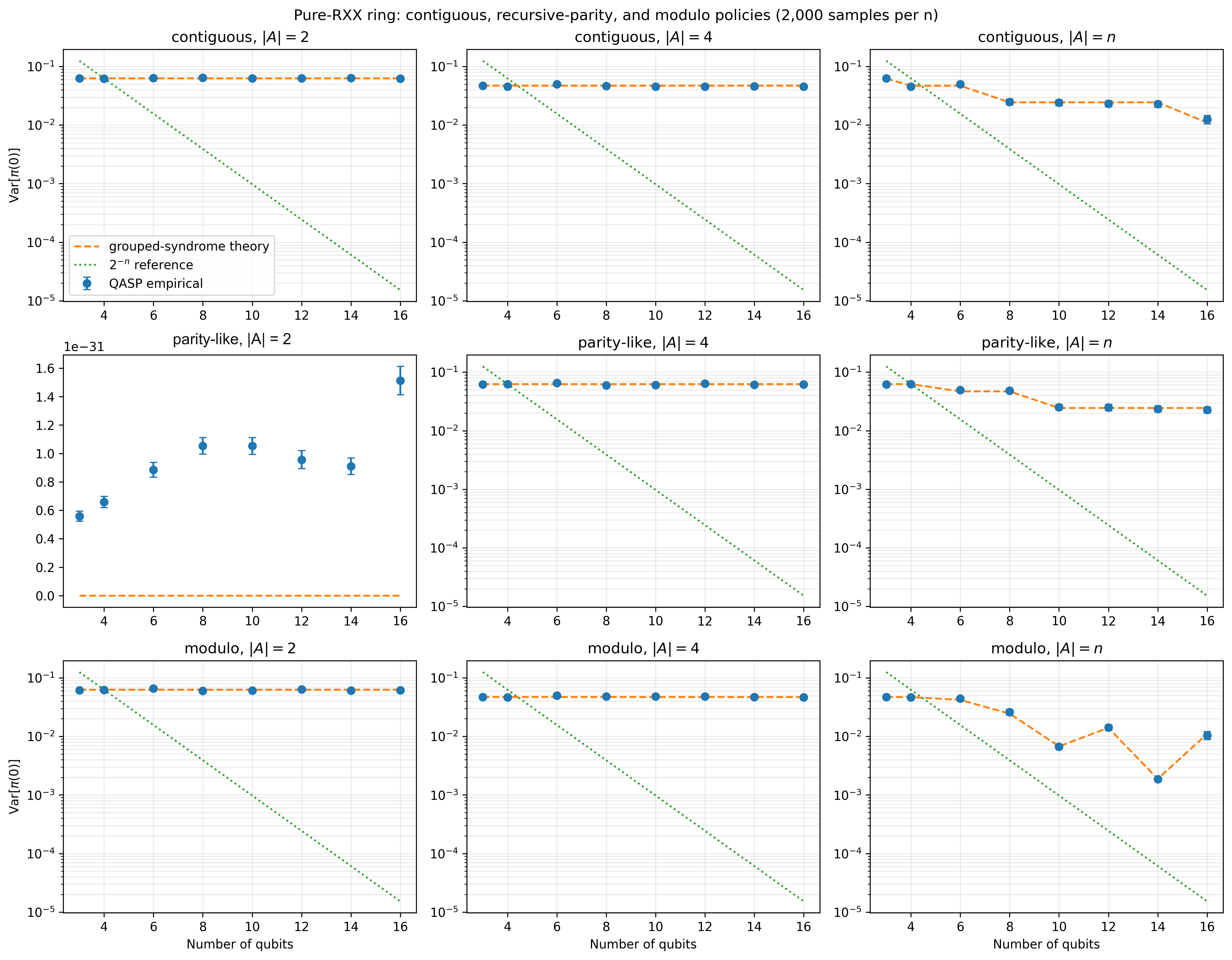}
    \caption{Variance of the action-$0$ probability for contiguous,
    recursive-parity, and modulo policies on pure-$R_{XX}$ periodic
    rings. Columns correspond to $A=2$, $A=4$, and $A=n$. Each point
    uses 2,000 random initializations; orange dashed curves are the exact
    grouped policy-variance predictions from
    Eq.~\eqref{eq:pure-rxx-grouped-policy-variance}, and green dotted curves show $2^{-n}$
    for scale. For binary recursive parity (middle-left), the theoretical
    variance is exactly zero. The displayed values of order $10^{-31}$
    are floating-point residue from statevector evaluation, not a
    nonzero physical variance.}
    \label{fig:pure-rxx-policy-variance}
\end{figure*}

The binary comparison isolates the decoder-induced locality predicted
in Sec.~\ref{sec:single-iqp-trainability}. Contiguous and modulo
readouts contain a 1-local Pauli component, and their variance remains
constant once the path contains the corresponding closed neighborhood.
Binary recursive parity contains only the global nonidentity component
and follows $2^{-n-2}$. At $A=4$, recursive parity no longer follows the
binary curve because products of nested parity constraints generate the
local term $Z_{n-1}$; its variance consequently approaches a nonzero
constant. The $A=n$ panels use the deterministic non-power-of-two QASP
extensions defined in Sec.~\ref{sec:model} at each $n$, explaining their
piecewise and nonmonotone finite-size structure.
Figure~\ref{fig:pure-rxx-policy-variance} repeats the comparison for a
periodic ring containing only $R_{XX}$ rotations. These are still
\emph{policy} variances, but the applicable exact law is the grouped
identity in Proposition~\ref{prop:grouped-policy-variance}, not the
unrestricted individual-term sum. The ring's generator matrix has
rank $n-1$: complementary masks $u$ and $u\oplus\mathbf1$ have the
same active-generator signature (also called an edge syndrome) and
identical Pauli expectations. Their coefficients must therefore be
added before squaring, as in
Eq.~\eqref{eq:pure-rxx-grouped-policy-variance}. The resulting
prediction follows the numerical policy variances across all three
decoders and action-space choices. Thus this experiment verifies
the grouped extension without local rotations; it does not imply
that all individual Pauli contributions are uncorrelated.

The exactly vanishing binary-parity variance is enforced by a circuit
symmetry. For a Pauli string $Z^s=\prod_{k\in S}Z_k$ and an edge
generator $X_iX_j$,
\begin{equation}
    Z^sX_iX_j
    =(-1)^{|S\cap\{i,j\}|}X_iX_jZ^s.
    \label{eq:pure-rxx-z-commutation}
\end{equation}
The operators therefore commute when the support $S$ contains neither
or both endpoints of the edge and anticommute when it contains exactly
one. The global string $P=Z_0Z_1\cdots Z_{n-1}$ contains both endpoints
of every edge, so $[P,X_iX_j]=0$ for every generator and hence
$[P,U_{XX}]=0$ for any pure-$R_{XX}$ interaction graph. Because
$|0^n\rangle$ is a $+1$ eigenstate of $P$, the binary recursive-parity
projector satisfies
\begin{equation}
    \pi_{\boldsymbol\theta}^{\mathrm{RP}}(0)
    =\left\langle\frac{I+P}{2}\right\rangle=1,
    \qquad
    \Var_{\theta}
    [\pi_{\boldsymbol\theta}^{\mathrm{RP}}(0)]=0.
    \label{eq:pure-rxx-parity-conservation}
\end{equation}
Thus, the zero is not an asymptotic barren-plateau scaling but exact
symmetry-induced determinism: the binary parity probability and all of
its parameter gradients are identically fixed. More generally, a
$Z$ string is conserved precisely when no interaction edge crosses the
cut between its support and its complement. Its support must therefore
be a union of connected components; on a connected interaction graph,
only $I$ and $Z^{\otimes n}$ satisfy this condition. Across all 144
graph-IQP and pure-$R_{XX}$ policy instances, every empirical variance
agrees with its exact Walsh/rank or grouped-syndrome prediction within
five estimated Monte Carlo standard errors.

\subsection{Majority--modulo: protected low-weight sectors with dense support}
\label{subsec:experiment-majority-modulo}

The majority--modulo study uses balanced contiguous blocks whose sizes
differ by at most one. When a finite-size block is even, ties are mapped
deterministically to majority bit zero. This extension is useful for a
dense size sweep but is not substituted into the odd, equal-block
asymptotic Fourier formulas derived above.
Instead, for every plotted $(n,A,a)$ instance we enumerate the actual
finite decoder, compute its complete Walsh spectrum, and evaluate the
exact light-cone sum. Odd equal-block cases coincide directly with those
analytic formulas.

\begin{figure*}[!htbp]
    \centering
    \includegraphics[width=0.85\textwidth]{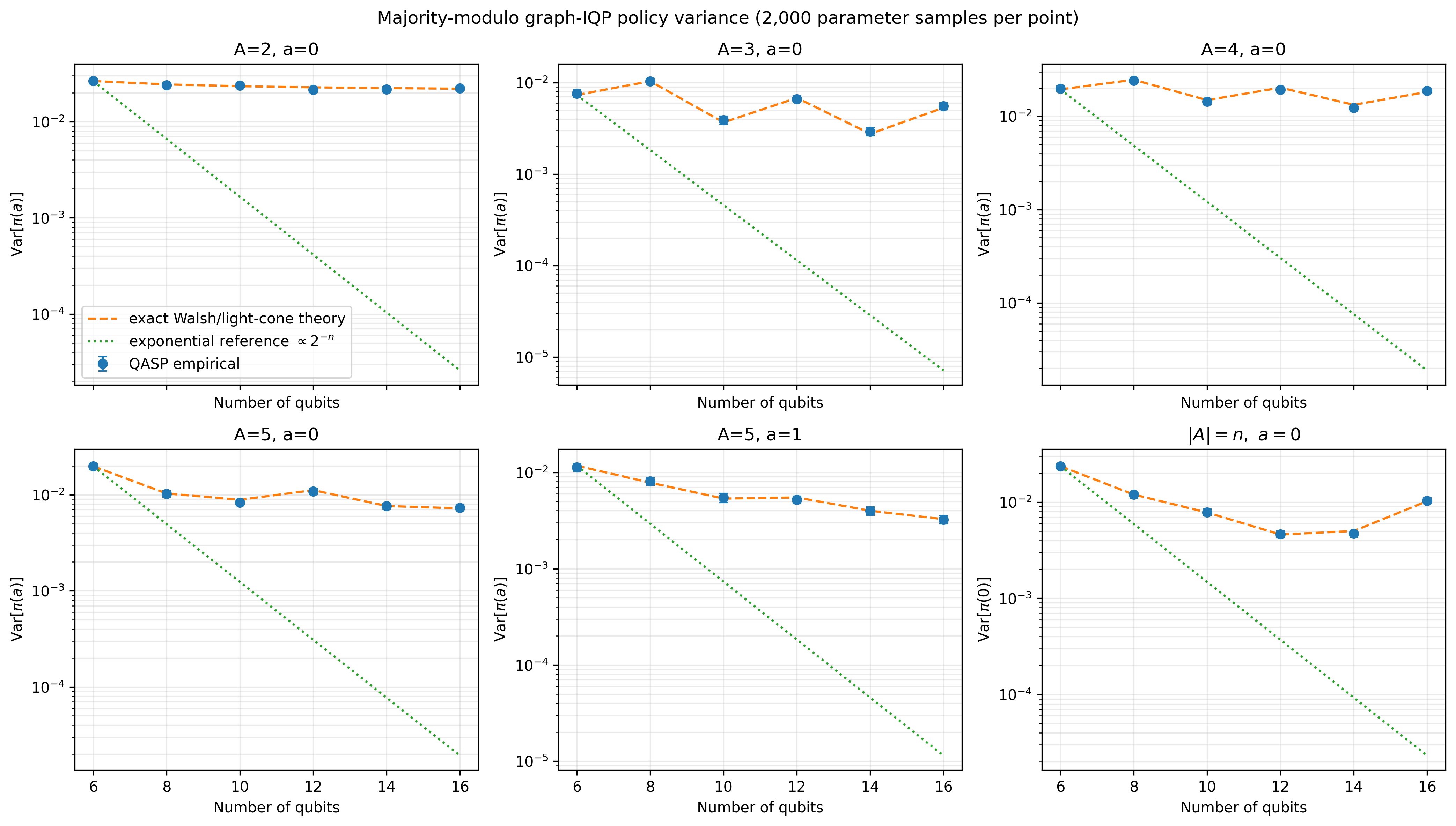}
    \caption{Finite-size majority--modulo policy variance on graph-IQP paths.
    Each panel compares 2,000-sample empirical variances with the exact
    Walsh/light-cone prediction; the exponential reference is normalized
    at the first point in each panel.}
    \label{fig:majmod-experiments}
    \label{fig:majmod-variance-panel}
\end{figure*}

Figure~\ref{fig:majmod-variance-panel} covers $n=6$ through $16$, fixed
$A\in\{2,3,4,5\}$, and the growing action space $A=n$. All 36 empirical
variances agree with the exact finite-size prediction within five
estimated Monte Carlo standard errors. In particular, at $n=16$ the
two collided $A=5$ actions shown in the lower row have variances
$7.34\times10^{-3}$ and $3.23\times10^{-3}$, respectively, well above
the normalized $2^{-n}$ references. The distinction between the two
actions is structural: action $0$ retains a 1-local sector from its
unchanged majority block, whereas the fully changed action $1$ is
protected by products of majority blocks in the odd equal-block case.
For the illustrative odd-block instance $(n,A)=(9,5)$, the exact
calculation gives zero 1-local terms and 27 2-local terms for action
$1$, with total variance $0.0040436$; action $0$ retains three 1-local
and nine 2-local terms, with total variance $0.0078716$.

The large Fourier sector is independently visible in
Fig.~\ref{fig:majmod-support-panel}. At $n=16$, the implemented projectors
contain 65,535 nonidentity terms for the displayed $A=3$ and $A=4$
actions, and 17,424 and 2,304 terms for the two $A=5$ actions. These
counts corroborate the mechanism behind
Proposition~\ref{prop:majmod-fourier-support-paper}: low-weight terms can
protect trainability even while the exact action projector is far from
Fourier sparse.

\subsection{Neurally encoded IQP policies on random cubic graphs}
\label{subsec:neural-cubic-variance}
\begingroup
\setlength{\parskip}{0pt}

The preceding one-layer path and ring experiments use efficiently
classically simulable circuits: their purpose is to verify the variance
laws and isolate the effect of the decoder. We now ask whether majority
policies avoid barren plateaus beyond these restricted geometries.
We consider random
3-regular graphs and then augment them with random three-body
$R_{ZZZ}$ rotations. This choice is motivated by conditional hardness
results for sparse IQP sampling~\cite{BremnerMontanaroShepherd2017},
but those results do not establish sampling hardness for the particular
3-regular ensemble used here. Unlike paths and rings, this connectivity
does not supply the same bounded-width contraction guarantee.
The additional three-body interactions enlarge the parameterized
circuit family and test variance under richer interactions; their
effect on classical simulation cost is not assumed.

Let $G$ be a connected simple 3-regular graph on $n$ qubits, with
edge set $E(G)$, so each qubit participates in exactly three
two-qubit interactions. At a fixed observation $s$, each qubit first
receives a neural angle
$\phi_i(s)=\sum_{j=1}^{4}w_j^{(i)}s_j$ and an independent trainable
bias $\alpha_i$. Between the initial and final Hadamard layers,
we apply these local rotations, one $R_{ZZ}$ rotation per graph
edge, and optionally one $R_{ZZZ}$ rotation per triple in a set $T$:
\begin{equation}
    \begin{aligned}
        U_{G,T}(s)
        &=H^{\otimes n}
          \left[\prod_{i=1}^{n}
          R_{Z_i}\!\left(\phi_i(s)+\alpha_i\right)\right]\\
        &\quad\times
          \left[\prod_{\{i,j\}\in E(G)}
          R_{Z_iZ_j}(\beta_{ij})\right]\\
        &\quad\times
          \left[\prod_{\{i,j,k\}\in T}
          R_{Z_iZ_jZ_k}(\gamma_{ijk})\right]H^{\otimes n}.
    \end{aligned}
    \label{eq:neural-cubic-circuits}
\end{equation}
The base family has $T=\varnothing$,
so the third product is the identity. The augmented family has
$|T|=n$ distinct unordered triples sampled uniformly without
replacement. All rotations between the Hadamard layers commute,
so both families remain IQP circuits. Representative circuit diagrams
are given in Appendix~\ref{app:neural-cubic-circuits},
Fig.~\ref{fig:neural-cubic-circuits}.
We measure all $n$ qubits and use binary majority decoding with
$|\mathcal A|=2$:
action $0$ is assigned to strings of Hamming weight at most $n/2$,
including ties, and action $1$ to the remaining strings.
For every circuit realization,
$\pi_{\theta}(1\mid s_0)=1-\pi_{\theta}(0\mid s_0)$, and therefore
\[
    \Var_{\theta}[\pi_{\theta}(1\mid s_0)]
    =\Var_{\theta}[\pi_{\theta}(0\mid s_0)].
\]
We consequently plot only action $0$; the action-$1$ variance is
identical, including for the pooled samples, despite their unequal
mean probabilities under deterministic tie-breaking.

We fix $s_0=(0.2,-0.3,0.4,-0.5)$ and consider
$n\in\{4,6,8,10,12,14,16\}$. For each size, we sample 100 graphs
and 20 independent paired initializations per graph, giving 2,000
policy realizations per family. Each triple set is held fixed across
its graph's 20 initializations. Paired circuits share the graph,
neural weights, and local and edge angles; the augmented circuit
additionally receives independent three-body angles. All weights
and angles are sampled independently from $\Unif[-\pi,\pi]$, subject
to this pairing. Consequently,
$\phi_i(s_0)+\alpha_i\bmod2\pi$ remains independently uniform
conditional on the neural weights, preserving the initialization
assumptions of Theorem~\ref{thm:main-trainability}.

Figure~\ref{fig:neural-cubic-variance} reports the pooled sample
variance of the exact action-$0$ probabilities over topology and
initialization, with denominator $2{,}000-1$. This is an ensemble
diagnostic, rather than a variance estimate for one selected graph.
Error bars are pointwise 95\% percentile intervals from 2,000
topology-block bootstrap replicates: each replicate resamples 100
topology blocks with replacement, retaining all 20 initializations
in each selected block. The same blocks are selected in both
conditions when estimating their variance difference.

\begin{figure*}[!t]
    \centering
    \begin{subfigure}[t]{0.48\textwidth}
        \centering
        \includegraphics[width=\linewidth]{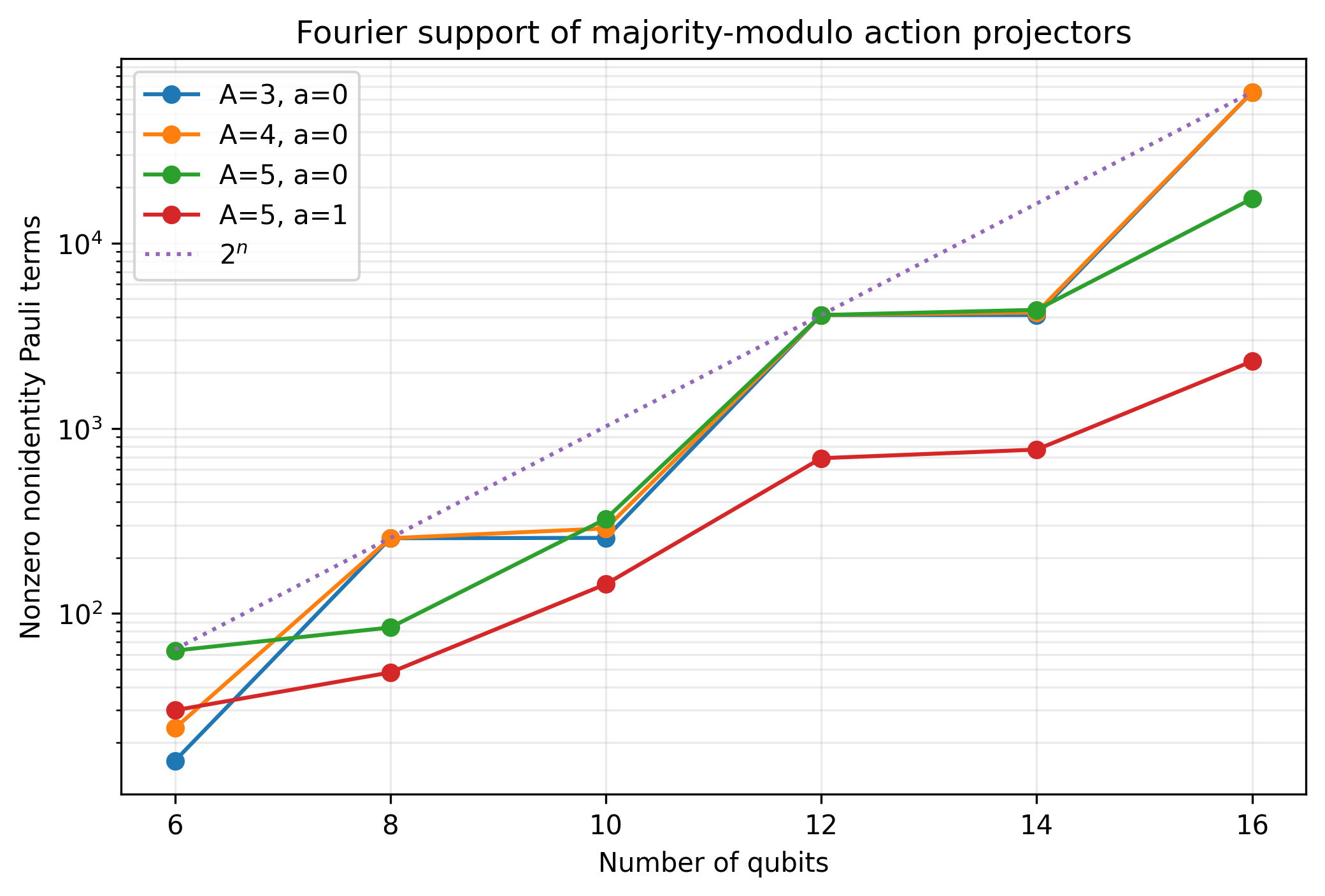}
        \caption{Exact nonidentity Pauli support on graph-IQP paths.}
        \label{fig:majmod-support-panel}
    \end{subfigure}\hfill
    \begin{subfigure}[t]{0.50\textwidth}
        \centering
        \includegraphics[width=\linewidth]{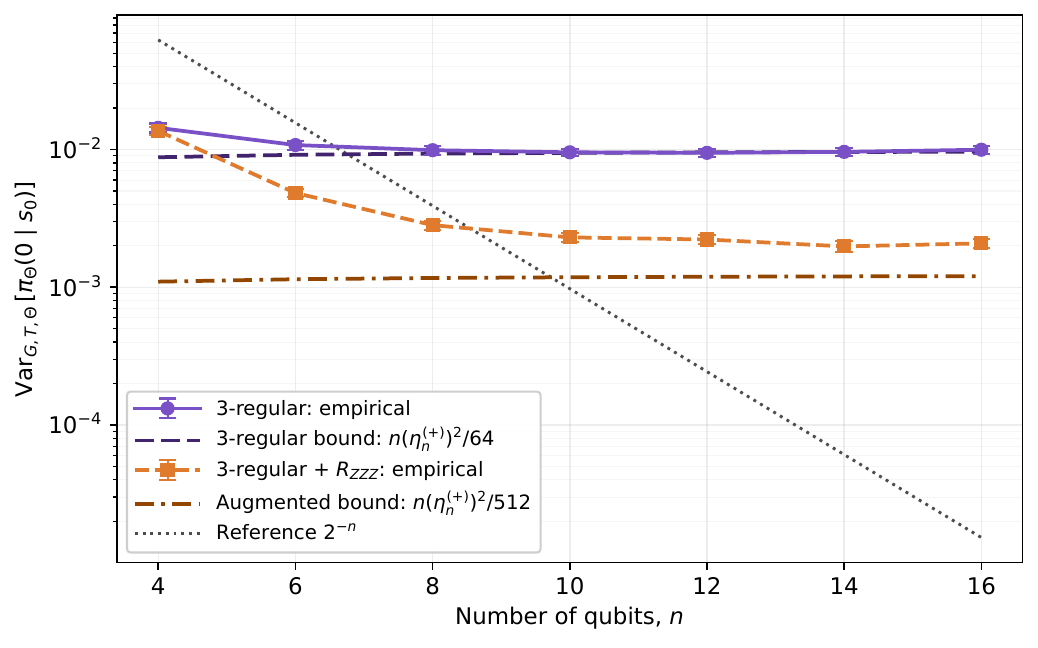}
        \caption{Neurally encoded majority-policy variance on cubic graphs.}
        \label{fig:neural-cubic-variance}
    \end{subfigure}
    \caption{Fourier support and initialization variance in two distinct
    majority-policy experiments. (a) Exact numbers of nonzero
    nonidentity Walsh--Pauli terms in the majority--modulo action
    projectors used in Fig.~\ref{fig:majmod-experiments}.
    (b) Initialization variance of the action-$0$ probability for
    neurally encoded binary-majority policies ($|\mathcal A|=2$)
    on random cubic graphs, with and without $n$ additional
    three-body rotations. Action $1$ has exactly the same variance
    because its probability is complementary. Each point pools 100 topologies and 20
    initializations per topology; error bars are 95\% topology-block
    bootstrap intervals. Here $G$ is the sampled graph, $T$ the sampled
    triple set (empty for the base family), and $\Theta$ collects the
    circuit angles and neural weights at fixed input $s_0$.
    The dark-purple dashed and dark-orange dash-dotted curves are
    the finite-$n$ theoretical lower bounds
    $n(\eta_n^{(+)})^2/64$ and $n(\eta_n^{(+)})^2/512$ for the base
    and augmented families, respectively, where
    $\eta_n^{(+)}=\binom n{n/2}/2^n$. These bounds apply to the
    ensemble variances, rather than individual finite-sample estimates.
    The dotted $2^{-n}$ curve is a scale reference, not a prediction
    for these families.}
    \label{fig:majority-support-neural-variance}
\end{figure*}

The base-family variance remains near $10^{-2}$ for $n\geq6$,
whereas the augmented-family variance decreases initially and
levels near $2\times10^{-3}$ at the largest sizes. At $n=16$, the
estimates are $0.00996$ and $0.00208$, with respective 95\%
intervals $[0.00932,0.01060]$ and $[0.00192,0.00225]$.
The averages of the 100 within-topology unbiased variance estimates,
$0.00991$ and $0.00210$, show that the pooled contrast is also present
within individual topologies. Paired intervals for the
augmented-minus-base variance exclude zero for every tested
$n\geq6$, but not for $n=4$. Both curves exceed the $2^{-n}$
reference for $n\geq10$ and show no sustained exponential
policy concentration over the tested range.

The exact rank law explains the persistence of policy variance.
For the even sizes and tie convention used here, the majority
coefficient in Eq.~\eqref{eq:even-majority-identity-one-local} gives
$|c_{0,e_i}|=\eta_n^{(+)}/2$, where
$\eta_n^{(+)}=\binom n{n/2}/2^n\sim\sqrt{2/(\pi n)}$.
Every $Z_i$ component of the base family has rank
$r^{e_i}=1+\deg_G(i)=4$. Keeping just these $n$ nonnegative
contributions in Theorem~\ref{thm:main-trainability} yields
\[
    \Var_{\theta}[\pi_{\mathrm{base}}(0\mid s_0)]
    \geq \frac{n(\eta_n^{(+)})^2}{64}
    \longrightarrow \frac{1}{32\pi}.
\]
Here the limit refers to the lower bound, not necessarily to the
full variance. This protected 1-local sector explains why increasing
the number of qubits need not suppress the base-policy variance.

The three-body rotations can reduce the variance without removing
this protection. Adding generators cannot decrease any active rank,
while the decoder coefficients remain unchanged; hence the exact
parameter-averaged variance of an augmented circuit is no larger
than that of its base circuit. To bound it from below, let $t_i$
count the added triples containing vertex $i$. Since $|T|=n$,
$\sum_i t_i=3n$, and each additional active generator increases
the rank by at most one, giving $r^{e_i}\leq4+t_i$.
The arithmetic--geometric mean inequality implies
$n^{-1}\sum_i2^{-t_i}\geq2^{-3}$. Consequently,
\[
    \begin{aligned}
    \Var_{\theta}[\pi_{\mathrm{aug}}(0\mid s_0)]
    &\geq\frac{(\eta_n^{(+)})^2}{64}\sum_i2^{-t_i}\\
    &\geq\frac{n(\eta_n^{(+)})^2}{512}
    \longrightarrow\frac{1}{256\pi}.
    \end{aligned}
\]
The two finite-size lower bounds are overlaid in
Fig.~\ref{fig:neural-cubic-variance} for direct comparison with the
empirical estimates and their confidence intervals.
Both families therefore have $\Omega(1)$ policy variance for every
allowed topology. Their conditional mean action probability is the
same decoder-bin fraction for every topology, so the law of total
variance transfers these bounds to the topology-averaged ensemble
shown in the plot. At $n=16$, the lower bounds are $0.009641$ and
$0.001205$, respectively, consistent with the empirical estimates.
The lower augmented curve thus reflects reduced Pauli contributions
as ranks increase, while the surviving low-rank sector prevents
exponential policy concentration. These bounds explain the
nonconcentration of both actions without fixing the precise height
of either variance curve.
\endgroup

\subsection{Repeated-IQP policy variance across depth}
\label{subsec:experiment-repeated-iqp}

We next test the repeated-circuit bounds of
Sec.~\ref{sec:repeated-trainability}. The simulated architecture is the
augmented circuit in Eq.~\eqref{eq:repeated-circuit-paper}: a leading
$D_X^{(0)}$ layer is followed by $L$ repetitions of local $Z$ rotations,
periodic nearest-neighbor $ZZ$ rotations on the cycle $C_n$, local $X$
rotations, and open-chain nearest-neighbor $XX$ rotations. Every angle,
including those in $D_X^{(0)}$, is sampled independently from
$\Unif[-\pi,\pi]$. We use $n\in\{4,6,8,10,12,14,16\}$ and the five depth
families $L\in\{1,2,3,\lceil\log_2 n\rceil,n\}$.

For this nearest-neighbor architecture, a spatial tensor-network
contraction has width $O(L)$ and cost polynomial in $n$ and $L$
times $\exp[O(L)]$~\cite{MarkovShi2008}. Thus the fixed-depth and
logarithmic-depth families also admit efficient classical sampling
and additive estimation of the decoded action probabilities. The
same argument does not give a polynomial-time guarantee for $L=n$:
one-dimensional connectivity alone does not ensure efficient
simulation as depth grows. That sweep is a finite-size statevector
test of the variance bounds and concentration mechanism, not a
hardness demonstration.

\begin{figure*}[!htbp]
    \centering
    \includegraphics[width=0.98\textwidth]{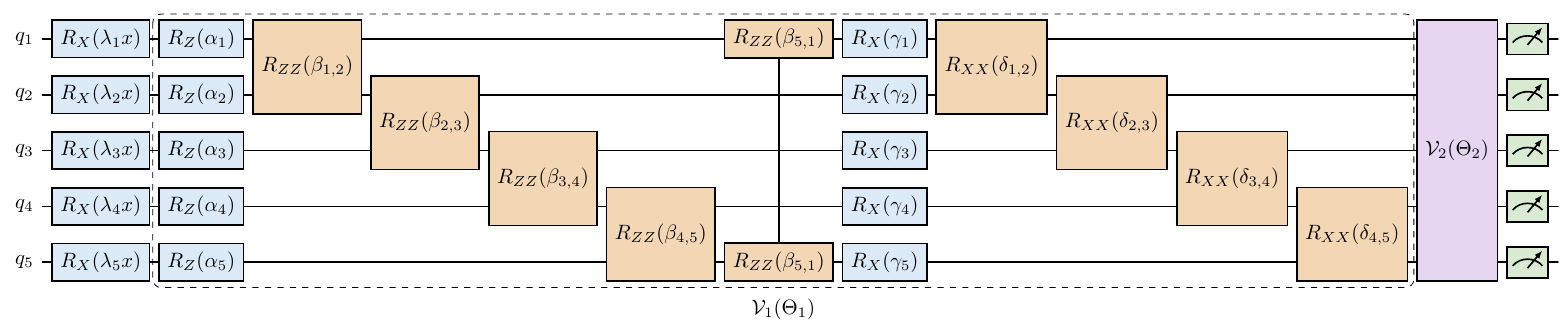}
    \caption{Representative repeated IQP-like circuit used in the
    variance experiment, exported directly from QASP for $n=5$ and
    $L=2$. The leading $R_X(\lambda_i x)$ gates form the initial
    trainability layer. The first repeated block is shown gate by gate:
    local $R_Z$, periodic nearest-neighbor $R_{ZZ}$, local $R_X$, and
    open-chain nearest-neighbor $R_{XX}$ rotations. The second block is
    compressed as $\mathcal V_2(\Theta_2)$ to preserve legibility. The
    terminal meter symbols denote computational-basis measurement on
    every qubit.}
    \label{fig:repeated-iqp-experimental-circuit}
\end{figure*}

Figure~\ref{fig:repeated-iqp-experimental-circuit} is a direct QASP
export of a representative $n=5$, $L=2$ instance. It makes explicit
the initial trainability layer, the periodic $ZZ$ program, and the
open-chain $XX$ program used in every repetition.

\begin{figure*}[!htbp]
    \centering
    \includegraphics[width=0.99\textwidth]{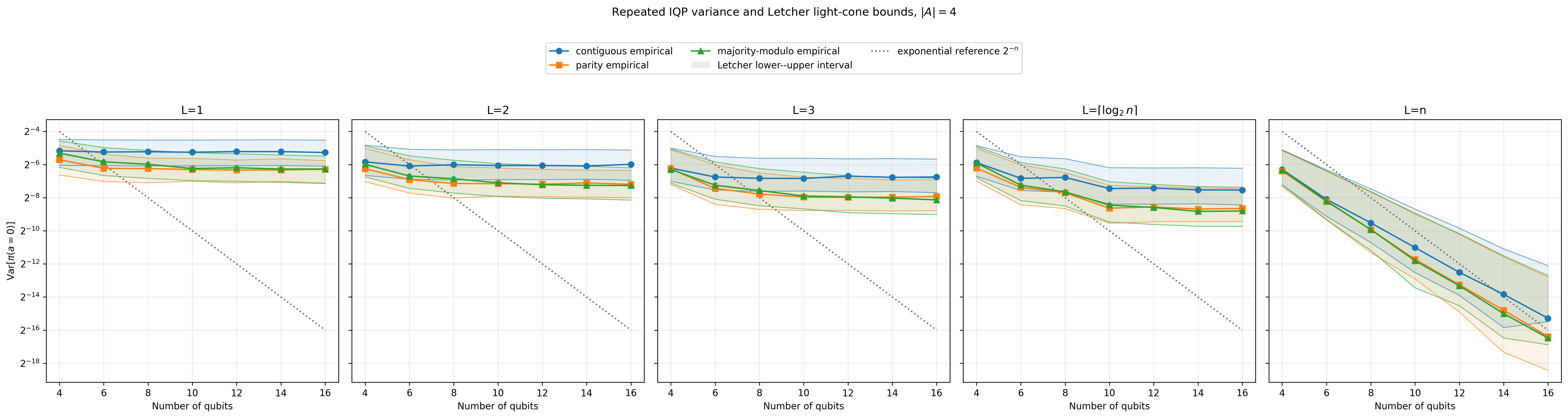}
    \caption{Variance of the action-$0$ probability for repeated-IQP
    policies with $A=4$. From left to right, the repetition depth is
    $L=1$, $2$, $3$, $\lceil\log_2 n\rceil$, and $n$. Markers show the
    empirical variance over 2,000 independent continuous-angle
    initializations; error bars are twice its Monte Carlo standard error.
    Light colored bands show the policy-level Letcher lower--upper
    interval estimated with 10,000 discrete parameter samples. The gray
    dotted curve is the reference $2^{-n}$. All three policies are
    evaluated on the same circuit samples at each $(n,L)$.}
    \label{fig:repeated-iqp-policy-variance}
\end{figure*}

For each circuit realization, one analytic statevector probability
vector is decoded into the action-$0$ probabilities of the contiguous,
recursive-parity, and majority--modulo policies with $A=4$. Thus the
three curves at each $(n,L)$ share exactly the same 2,000 circuit
initializations. The shaded bands in
Fig.~\ref{fig:repeated-iqp-policy-variance} are obtained independently
from 10,000 discrete parameter samples: for every
Walsh--Pauli term, the propagated support gives the Letcher lower and
upper factors $4^{-\Delta_R^{\theta}(u)}$ and
$2^{-\Delta_R^{\theta}(u)}$, which are combined with
the exact squared Walsh coefficients of the corresponding action
projector.

The Letcher inequalities are satisfied at all 105 tested
$(L,n,\text{policy})$ points within five combined Monte Carlo standard
errors.

The depth dependence is the dominant effect. Over the tested range, the
fitted slopes of $\log_2\Var_{\theta}[\pi(0)]$ against $n$, across the three fixed
depths $L=1,2,3$, range from $-0.0045$ to $-0.0269$ for contiguous,
from $-0.0360$ to $-0.1119$ for parity, and from $-0.0706$ to $-0.1332$
for majority--modulo. At $L=\lceil\log_2 n\rceil$, the corresponding
slopes are $-0.125$, $-0.193$, and $-0.229$, still substantially
shallower than the $2^{-n}$ reference over these sizes. At linear depth
they steepen to $-0.740$, $-0.830$, and $-0.847$. These finite-size data support the
light-cone mechanism: shallow repetition preserves appreciable policy
variance, whereas extensive repetition spreads the decoder observables
through most of the register and drives all three policies toward
exponential concentration. The fitted slopes are descriptive and are
not used as asymptotic estimates.

\subsection{CartPole decoder comparisons}
\label{subsec:cartpole-overview}

The matched depth sweeps compare majority with recursive parity on a
five-qubit ring and a ten-qubit cubic graph. Majority has higher
late-training mean returns at each tested depth, but these comparisons
alone do not establish that higher-weight terms improve learning:
parity is an intentionally concentration-prone comparator. The full
protocols and outcomes are in Appendix~\ref{app:supplementary-numerics},
together with the three-body augmentation and finite-shot tests.
The latter preserve the majority--parity ordering with seven SPSA
directions, without reaching the strict solve criterion. The controlled
local-readout comparison below addresses the more informative question
of whether majority improves over valid policies that already retain
a local initialization signal.

\subsection{Controlled local-sector comparison and held-out evaluation}
\label{subsec:decoder-sector-comparison}

The parity comparison establishes that a global readout can suppress
learning, but it does not determine whether majority's higher-weight
Fourier terms improve control beyond the 1-local sector.  We therefore
repeat the five-qubit $L=3$ and ten-qubit $L=1$ experiments with three
valid 1-local controls.  The first two are deterministic projectors on
individual measured bits,
$\Pi^{(q_2)}_0=(I+Z_2)/2$ and
$\Pi^{(q_1)}_0=(I+Z_1)/2$; the latter is precisely the binary contiguous
decoder.  The third is a permutation-symmetric positive-operator-valued
measure (POVM)
\begin{equation}
    \Pi^{\mathrm{lin}}_0
    =\frac{1}{2}\left(I+\frac{1}{m}\sum_{i=1}^{m}Z_i\right),
    \qquad
    \Pi^{\mathrm{lin}}_1=I-\Pi^{\mathrm{lin}}_0,
    \label{eq:local-linear-povm}
\end{equation}
whose computational-basis eigenvalue is
$1-|x|/m\in[0,1]$.  Operationally, it chooses one of the $m$ measured
bits uniformly and applies the corresponding binary readout.  It is
therefore a realizable local-only decoder rather than a formal cost
observable.

A coefficient-matched truncation of the majority projector is not, in
general, positive semidefinite: deleting the higher-weight terms can
move its eigenvalues outside $[0,1]$.  We consequently use
Eq.~\eqref{eq:local-linear-povm} and the two local projectors as valid
1-local controls, while the full-policy gradient attribution below
diagnoses how different Pauli localities participate during training.
Global parity is retained as a negative control.  We also train a size-matched classical
multilayer perceptron (MLP) with 65 parameters on the ten-qubit instance and
93 parameters on the five-qubit instance, compared with 65 and 90
quantum parameters, respectively.  The classical baseline calibrates
the task; it is not a surrogate for a classical simulation algorithm.

Every condition uses the ten matched seeds
$\{3,7,11,23,37,41,43,47,53,59\}$, 500 training episodes, and 100
additional episodes with the final policy frozen.  All quantum decoders
within an architecture and seed share the same initial trainable tensor.
The held-out episodes use disjoint environment and action-sampling seeds.
The production runs use an exact differentiable statevector
implementation generated from the live QASP circuit specification.  On
both architectures and all five quantum decoders, its action
probabilities and analytic gradients agreed with the native QASP models
to maximum absolute errors below $3\times10^{-6}$ and
$3\times10^{-5}$, respectively, before the full run.

\begin{figure*}[t]
    \centering
    \includegraphics[width=0.98\textwidth]{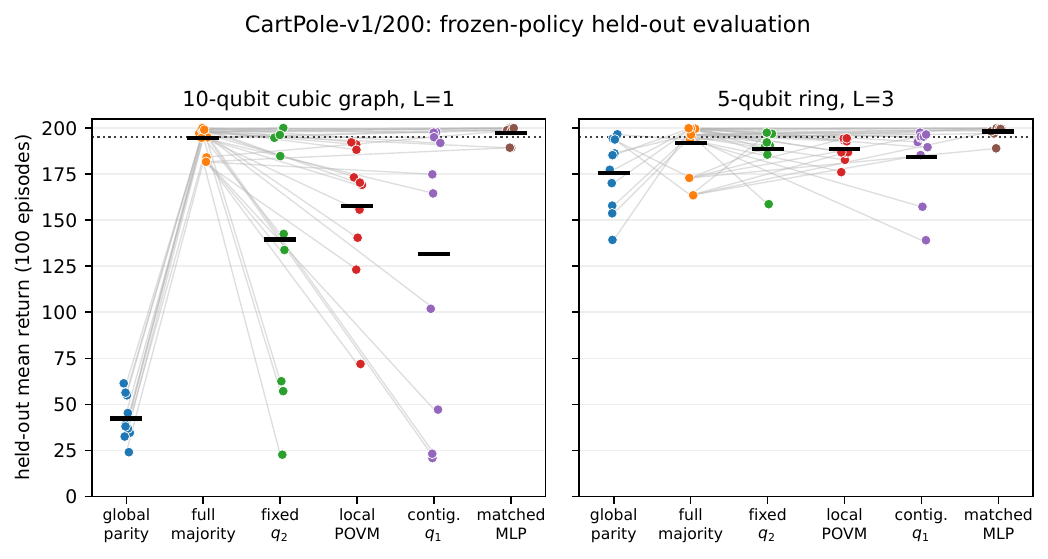}
    \caption{Frozen-policy held-out evaluation for the controlled decoder comparison.
    Each point is the mean return over 100 evaluation episodes for one
    trained seed; black bars are means over ten seeds.  Gray segments
    connect full majority to each same-seed comparator, preserving the
    matched design.  The dotted line marks the study-specific return-195
    threshold.  The local POVM is Eq.~\eqref{eq:local-linear-povm};
    contiguous $q_1$ and fixed $q_2$ are deterministic one-bit readouts.
    The MLP is size matched and is included as a task-calibration
    baseline, not as a circuit simulator.}
    \label{fig:decoder-comparison-heldout}
\end{figure*}

\begin{table*}[t]
\caption{Held-out CartPole performance in the controlled decoder comparison}
\label{tab:decoder-comparison-heldout}
\centering
\small
\setlength{\tabcolsep}{11pt}
\begin{tabular}{lcc}
\toprule
Decoder or model & Five-qubit ring, $L=3$ & Ten-qubit cubic graph, $L=1$ \\
\midrule
Majority--modulo        & $191.68\pm4.03$ & $194.69\pm2.06$ \\
Recursive parity        & $175.44\pm6.21$ & $42.35\pm3.77$ \\
Fixed $q_2$             & $188.42\pm3.49$ & $139.46\pm21.60$ \\
Local-linear POVM       & $188.76\pm1.93$ & $157.55\pm11.84$ \\
Contiguous $q_1$        & $184.32\pm6.27$ & $131.49\pm23.88$ \\
Matched classical MLP   & $198.12\pm1.05$ & $197.35\pm1.36$ \\
\bottomrule
\end{tabular}
\vspace{0.4ex}

\footnotesize Entries are mean held-out returns $\pm$ standard errors over
ten matched training seeds; each seed is evaluated for 100 additional
episodes with its final policy frozen.  The MLP calibrates the task and is
not a classical simulation of the quantum circuit.
\end{table*}

Figure~\ref{fig:decoder-comparison-heldout} shows every held-out seed mean,
while Table~\ref{tab:decoder-comparison-heldout} collects the corresponding
mean $\pm$ standard error.  On the five-qubit ring, majority has the
largest mean among all quantum policies, exceeding parity, fixed $q_2$,
the local-linear POVM, and contiguous $q_1$.  The paired intervals are
nevertheless too wide to resolve those differences with ten seeds; for
example, the majority advantages over the local-linear and fixed-bit
controls are $2.92$ ($[-6.05,10.33]$) and $3.25$
($[-7.92,14.27]$), respectively.  This should be read as an unresolved
but directionally consistent five-qubit result, not as evidence that the
decoders perform equally.  The different five- and ten-qubit outcomes do not isolate size alone:
the interaction graph, measured register, and circuit depth also differ.

A resolved decoder separation is observed on the ten-qubit graph.
Majority reaches $194.69\pm2.06$ and the study threshold in six of ten
seeds.  Its paired advantages are $152.34$ over parity (95\% interval
$[146.79,157.53]$), $55.22$ over fixed $q_2$ ($[19.67,95.01]$),
$37.13$ over the local-linear POVM ($[20.40,57.66]$), and $63.20$ over
contiguous $q_1$ ($[21.39,109.03]$).  Thus every majority-versus-quantum
control interval excludes zero.  The matched MLP reaches
$197.35\pm1.36$, and its difference from majority remains unresolved
($\text{majority}-\text{MLP}=-2.66$, $[-6.16,0.75]$); neither circuit
comparison therefore establishes an end-to-end quantum advantage.

The contrast with the valid local-linear POVM is particularly
informative.  That decoder is trainable through 1-local observables,
yet its ten-qubit held-out return is lower than full majority by a
resolved $37.13$ returns.  Thus access to trainable 1-local observables
alone does not determine final performance.  The gap does not, however,
identify a higher-weight mechanism: full majority also differs from the
local POVM in coefficient scale, permutation-symmetric aggregation, and
its complete action partition.  This motivates the direct gradient-sector
tracking below.

Finally, the decoder structures used by all quantum controls in this
controlled comparison other than majority fall inside the classical-estimation regimes
proved in Sec.~\ref{sec:single-iqp-simulability} when they are paired with
a genuine IQP circuit.  Recursive parity and the two single-bit readouts
are Fourier sparse, contiguous $q_1$ is an interval decoder, and the
local-linear POVM contains only $m+1$ Pauli terms.  For polynomial-size
IQP circuits, their action probabilities are therefore classically
additively estimable by the results of this paper.  The ten-qubit $L=1$ comparison is a genuine IQP instance and is
covered by these estimator results. The five-qubit $L=3$ comparison
uses a repeated IQP-like architecture; the IQP estimator theorem
does not automatically extend to that circuit.  Majority is the only tested
decoder that already lies outside these sufficient certificates at the
genuine-IQP level, but this does not constitute a classical-hardness proof.

\subsection{What gradient-sector diagnostics show}
\label{subsec:gradient-summary}

Appendix~\ref{subsec:majority-gradient-locality-training} resolves the
full-policy gradient into Pauli-weight sectors during training.
Its diagnostic is each sector-vector norm divided by the sum of
sector-vector norms, not its fraction of the net gradient after
cancellation. Optimization continues after the usual solve threshold
in this diagnostic, so the trajectories are not an additional matched
performance comparison.

On five qubits, the combined higher-weight share changes from
$39.9\%$ to $41.4\%$, with uncertainty that includes no change.
On ten qubits it increases from $29.2\%$ to $43.6\%$
(paired 95\% interval for the increase, $[11.6,17.2]$ percentage points),
mainly through weights three and five. The maximal-weight sector
does not grow: its share decreases from $5.2\%$ to $4.0\%$.
Thus higher-weight terms participate, but the observations do not
show that global terms take over, that their absolute gradients
increase, or that they cause the return improvement. Those claims
would require different measurements or interventions.

\subsection{Connected majority-IQP policies and MLP size scaling}
\label{subsec:connected-iqp-mlp}

As a final numerical experiment, we test whether a neurally encoded
majority policy retains competitive learning performance when its
entangled three-qubit blocks are connected, and whether increasing
the capacity of a tuned MLP closes the performance gap. This is a
separate CartPole comparison with a 200-step episode limit; it measures
episode efficiency, not a quantum computational advantage. Because the
circuit and majority readout admit an exact tensor-network simulator
of the type introduced in Sec.~\ref{subsec:background-tensor-networks},
this experiment also tests a quantum-inspired classical policy model.

The model has $n=21$ qubits partitioned into seven triples
$B_j=\{3j-2,3j-1,3j\}$. With the neural encoding of
Sec.~\ref{subsec:neural-state-embeddings}, the state at observation $s$
is $U(s)\ket{0}^{\otimes21}$, where
\begin{equation}
\begin{aligned}
 U(s)&=H^{\otimes21}D(s)H^{\otimes21},\\
 D(s)&=\left[\prod_{i=1}^{21}R_{Z_i}(\phi_i(s))\right]
       \left[\prod_{j=1}^{7}D_j\right]
       \left[\prod_{j=1}^{6}R_{Z_{3j}Z_{3j+1}}(\beta_j)\right],\\
 D_j&=R_{Z_{3j-2}Z_{3j-1}}(\alpha_{j,1})
       R_{Z_{3j-2}Z_{3j}}(\alpha_{j,2})\\
    &\quad\times R_{Z_{3j-1}Z_{3j}}(\alpha_{j,3})
       R_{Z_{3j-2}Z_{3j-1}Z_{3j}}(\gamma_j),
\end{aligned}
\label{eq:connected-majority-circuit}
\end{equation}
where $\phi_i(s)=w_i^\top\widetilde s+b_i$ and $\widetilde s$ is the
four-component normalized observation. The first Hadamard layer
prepares $\ket{+}^{\otimes21}$. The diagonal block $D(s)$ then applies
one neural $R_Z$ encoding per qubit, all three pair rotations and one
three-body rotation within each triple $B_j$, and six bridge rotations
on $(3,4),(6,7),\ldots,(18,19)$. Each bridge joins adjacent triples,
so the interaction graph is connected. All gates in $D(s)$ commute;
their order is immaterial. A final Hadamard layer precedes
computational-basis measurement of all 21 qubits.

\begin{figure}[H]
    \centering
    \resizebox{0.90\textwidth}{!}{\scriptsize
        \input{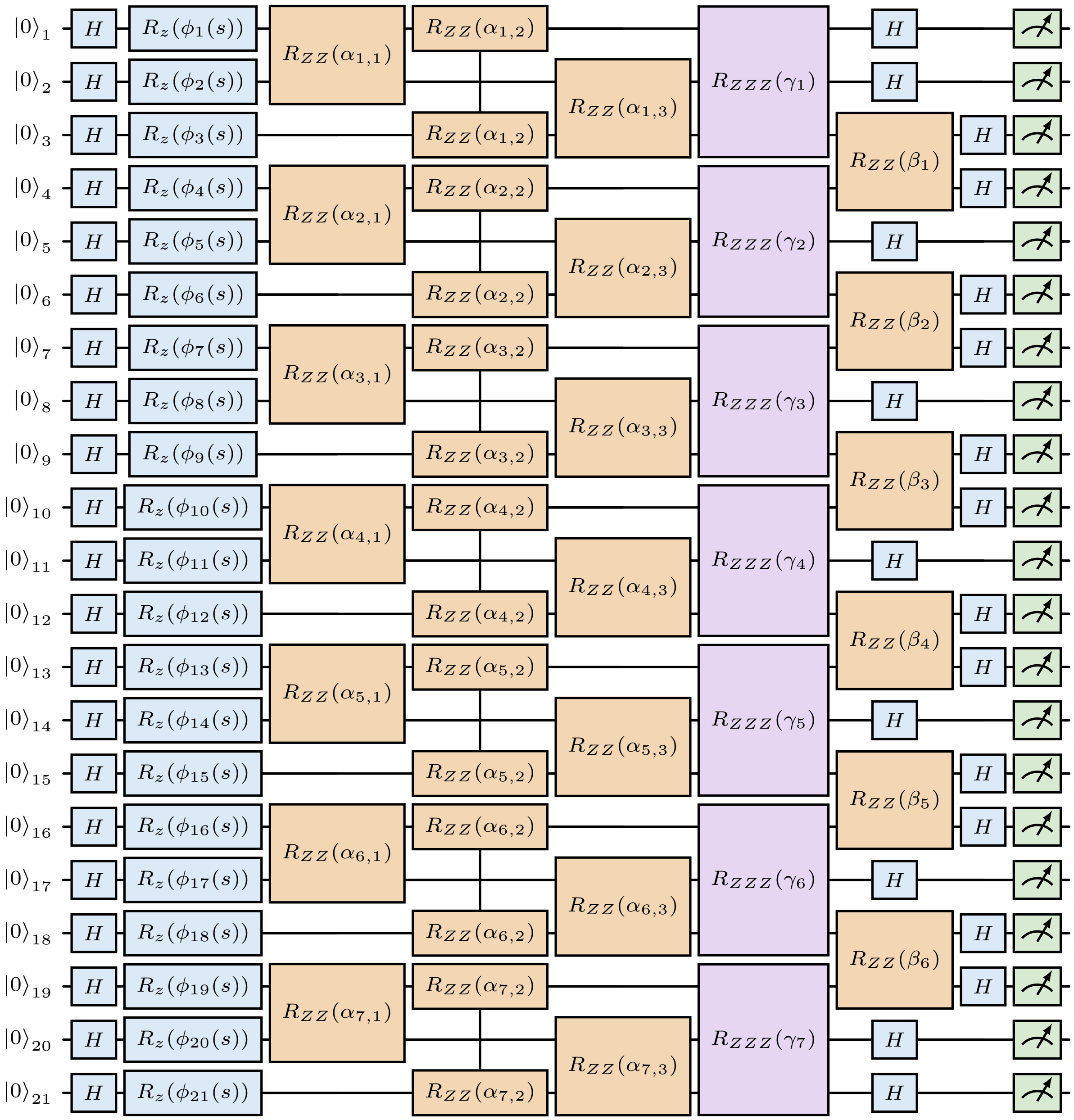}}
    \caption{Complete 21-qubit connected majority-IQP policy used in
    Sec.~\ref{subsec:connected-iqp-mlp}, implementing
    Eq.~\eqref{eq:connected-majority-circuit}, exported using QASP's
    TikZ functionality. Blue gates encode
    $\phi_i(s)=w_i^\top\widetilde s+b_i$; orange gates are two-qubit
    rotations, including the six inter-block bridges $\beta_j$;
    purple gates are three-qubit rotations. Every wire starts in
    $\ket{0}$ and is measured in the computational basis. One global
    majority decoder acts on all 21 output bits, as in
    Eq.~\eqref{eq:connected-majority-policy}. The two boxes joined
    across a skipped wire denote a single rotation on the marked
    endpoints only. QASP schedules disjoint gates in parallel; final
    Hadamards can therefore appear in different columns.}
    \label{fig:connected-majority-circuit}
\end{figure}

A single global binary majority decoder, rather than separate
decisions within the triples, defines the action probabilities:
\begin{equation}
\begin{aligned}
 \pi_\theta(0\mid s)
   &=\sum_{\substack{x\in\{0,1\}^{21}\\ |x|\leq10}}
       |\langle x|U(s)|0^{21}\rangle|^2,\\
 \pi_\theta(1\mid s)&=1-\pi_\theta(0\mid s),
\end{aligned}
\label{eq:connected-majority-policy}
\end{equation}
where $|x|$ is the Hamming weight and $\theta$ collects all trainable
parameters. The model has $84$ neural weights, $21$ biases, $21$ internal
pair angles, seven triple angles, and six bridge angles: $139$
trainable parameters in total. The complete QASP TikZ circuit is
shown in Fig.~\ref{fig:connected-majority-circuit},
Appendix~\ref{app:connected-majority-circuit}.

For the connected IQP, neural weights are initialized independently in
$[0,2\pi)$, biases at $\pi/2$, and all interaction angles independently
from $\mathcal N(0,0.5^2)$. Biases remain trainable. The bridges use an
independent random stream. These task-oriented initializations differ
from the uniform-angle ensemble used in the variance theorems.
Observations are divided componentwise by $(2.4,3,0.2095,3.5)$ and
clipped to $[-3,3]$ for every model.

The classical comparison uses $4\!\to\!h\!\to\!2$ MLPs with a
hidden $\tanh$ activation and a softmax output, giving $7h+2$
parameters. We test $h\in\{1,2,4,6,9,12,16,19\}$ and learning rates
$\{0.003,0.01,0.03,0.1\}$ on eight development seeds. The largest rate
was added during development, before any test runs. We then lock the
highest-mean-AUC rate for each width, as well as the fastest-converging
configuration across the full grid. The latter is an additional
$86$-parameter MLP at rate $0.1$; the best development AUC is attained
by the $30$-parameter MLP. The connected IQP uses rate $0.03$, with no
further architecture or bridge-strength search in this comparison.
Twenty new paired test seeds evaluate these nine MLP configurations
and the connected IQP. The reported comparison comprises $264$
development and $200$ test training runs, with no selection based on
the test outcomes.

All models receive $500$ episodes of REINFORCE training with discount
$0.99$, Adam AMSGrad, five-episode batches, an exponential-moving-average
baseline with momentum $0.05$, normalized batch advantages, entropy
coefficient $0.005$, and gradient-norm clipping at $1$. Updates freeze
only after a complete trailing 100-episode return mean reaches $195$,
checked at update boundaries; rollouts continue to episode $500$.
Each final policy is additionally evaluated on $100$ fresh episodes.
For episodic returns $R_e$, we report
\begin{equation}
    \mathrm{AUC}=\frac{1}{200\cdot500}\sum_{e=1}^{500}R_e.
    \label{eq:connected-normalized-auc}
\end{equation}
The second primary metric is the first episode at which the full
100-episode moving mean reaches $195$. Runs not reaching this
criterion are assigned $501$, rather than excluded. We call this
the restricted number of episodes to criterion.

\begin{figure*}[tbp]
    \centering
    \includegraphics[width=\textwidth]{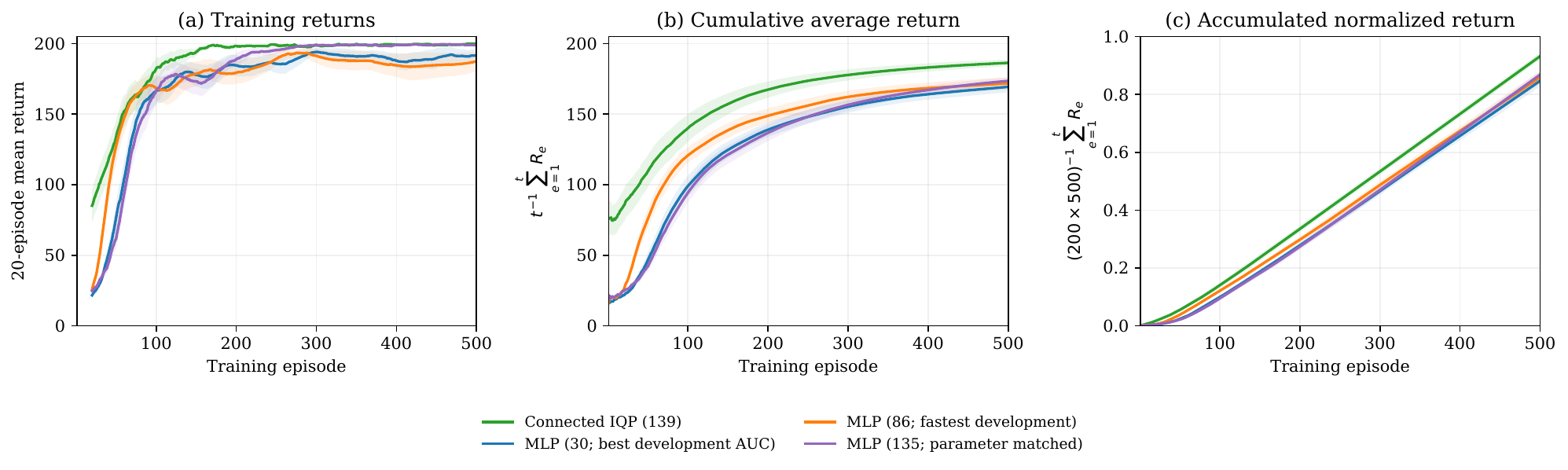}
    \caption{Learning curves for the connected IQP and three MLP
    comparisons selected before testing:
    best development AUC, fastest development convergence, and the
    approximately parameter-matched model. (a) Trailing 20-episode
    mean training return. (b) Cumulative average return.
    (c) Accumulated return normalized by the complete training budget;
    its endpoint is exactly the AUC in
    Eq.~\eqref{eq:connected-normalized-auc}. Lines average 20 new test
    seeds and bands show one standard error across seeds. Numbers in
    parentheses are trainable parameter counts.}
    \label{fig:connected-iqp-learning}
\end{figure*}

\begin{table*}[tbp]
    \centering
    \caption{Connected-IQP comparison on 20 new test seeds. MLP rates
    are chosen by development AUC within each width; the final MLP row
    is the additional fastest-development configuration. All entries
    are test means; ``reached'' counts runs meeting the criterion by
    episode 500. Evaluation returns average 100 fresh episodes per run.}
    \label{tab:connected-iqp-mlp}
    \small
    \begin{tabular}{lrrrrrr}
    \toprule
    Model & Parameters & Learning rate & AUC & Episodes to criterion & Reached & Evaluation return\\
    \midrule
    MLP ($h=1$)  & 9   & 0.10 & 0.7229 & 374.45 & 15/20 & 191.25\\
    MLP ($h=2$)  & 16  & 0.10 & 0.7985 & 277.70 & 18/20 & 197.10\\
    MLP ($h=4$)  & 30  & 0.10 & 0.8469 & 263.65 & 17/20 & 191.58\\
    MLP ($h=6$)  & 44  & 0.10 & 0.8631 & 261.90 & 17/20 & 193.86\\
    MLP ($h=9$)  & 65  & 0.10 & 0.8428 & 309.75 & 16/20 & 187.60\\
    MLP ($h=12$) & 86  & 0.03 & 0.8413 & 274.70 & 18/20 & 198.24\\
    MLP ($h=16$) & 114 & 0.10 & 0.7941 & 323.15 & 16/20 & 185.79\\
    MLP ($h=19$) & 135 & 0.03 & 0.8680 & 236.50 & 20/20 & 198.96\\
    MLP ($h=12$, fastest) & 86 & 0.10 & 0.8591 & 261.60 & 17/20 & 188.72\\
    \midrule
    Connected IQP & 139 & 0.03 & 0.9319 & 171.45 & 20/20 & 199.11\\
    \bottomrule
    \end{tabular}
\end{table*}

\begin{figure*}[tbp]
    \centering
    \includegraphics[width=0.95\textwidth]{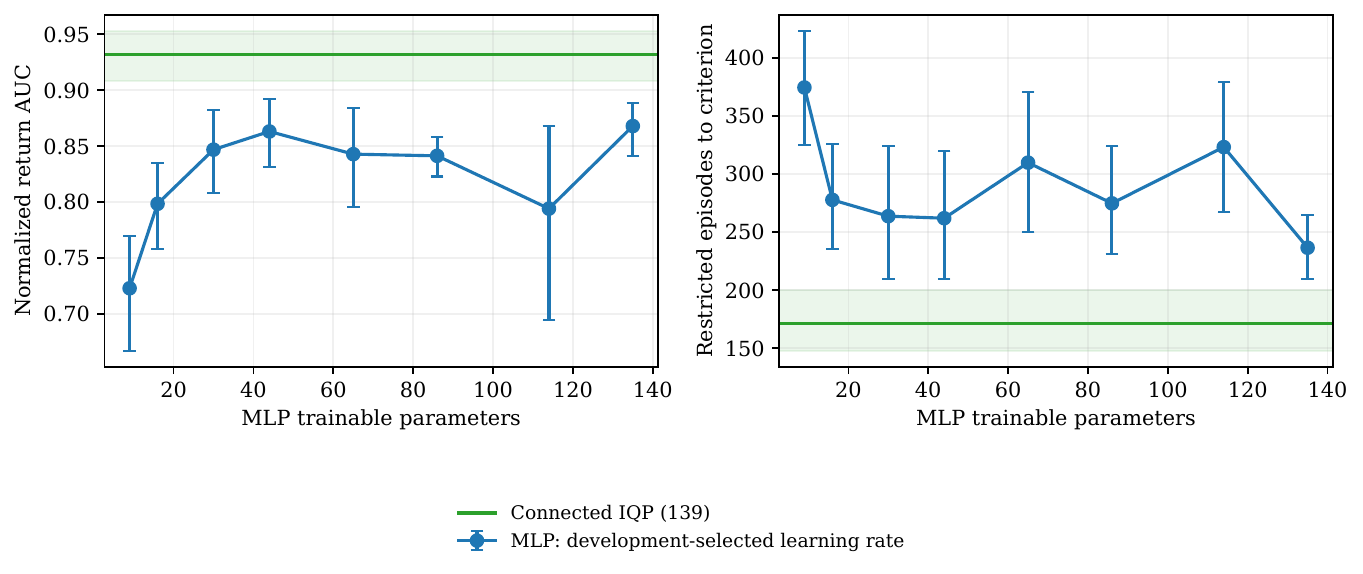}
    \caption{MLP capacity comparison using the development-selected
    learning rate at each width. Points are means over 20 test seeds;
    error bars are bootstrap 95\% intervals. Each horizontal line shows
    the connected IQP mean, with its 95\% interval shaded.
    The additional fastest-development $86$-parameter MLP is reported
    separately in Table~\ref{tab:connected-iqp-mlp}. Higher AUC and fewer
    restricted episodes indicate better performance. The nonmonotonic
    size curve does not establish saturation.}
    \label{fig:connected-iqp-size}
\end{figure*}

Figures~\ref{fig:connected-iqp-learning} and
\ref{fig:connected-iqp-size}, together with
Table~\ref{tab:connected-iqp-mlp}, show that the connected IQP retains
an advantage over all tested MLP configurations. Relative to the
$135$-parameter MLP, its mean normalized AUC increases by $0.06395$
(paired-bootstrap 95\% interval $[0.03435,0.09782]$), and it reaches
the criterion $65.05$ episodes earlier on average
($[31.65,98.40]$). Exact two-sided paired sign-randomization tests,
with one Holm correction over all 18 IQP-versus-MLP primary comparisons
(nine configurations and two metrics), give adjusted
$p=0.00248$ and $p=0.01254$, respectively. All 18 comparisons favor
the connected IQP with adjusted $p<0.05$. The confidence intervals
are individual paired intervals, not simultaneous intervals.

Increasing MLP size does not give monotonic improvements. Although the
$135$-parameter MLP has the highest mean test AUC among the tested
MLPs, it is not the development-selected best-AUC model. Nor does the
size sweep establish saturation: comparisons with the $65$-, $86$-,
and $114$-parameter models fail to establish equivalence within the
prospectively specified margins of $0.02$ AUC and $25$ episodes, using
98.33\% paired-bootstrap intervals for the three size contrasts.

The bridges genuinely entangle the blocks, but only in a restricted
way. Across each cut between successive triples, exactly one $R_{ZZ}$
gate crosses the cut. Its two-term decomposition gives at most two
Schmidt components, regardless of the total number of blocks.
The reduced-state purity is $1-\tfrac12\sin^2\beta_j$, independently
of the observation; all 120 inter-block cuts in the 20 trained models
remain entangled. Nevertheless, their bounded Schmidt rank permits a
matrix-product-state tensor network, with each triple treated as one
site and inter-block bond dimension at most two.
Pairing amplitudes with their conjugates produces a probability
network with bond dimension at most four.

Majority readout is tractable because it depends only on the total
number of measured ones. Let $q_w(s)$ denote the probability of
Hamming weight $w$, and collect these $n+1$ probabilities in
\begin{equation}
    F_s(z)=\sum_{w=0}^{n}q_w(s)z^w
          =\sum_{x\in\{0,1\}^n}p_\theta(x\mid s)z^{|x|}.
    \label{eq:connected-majority-generating-function}
\end{equation}
Since $|x|$ is the sum of the block weights, the factor $z^{|x|}$
separates across blocks. Summing the eight outputs of each triple
therefore gives a $4\times4$ transfer matrix. Multiplying these
matrices evaluates $F_s(z)$ in $O(n)$ operations while retaining
all inter-block correlations. Evaluations at $n+1$ roots of unity
and an inverse Fourier transform recover every $q_w(s)$ in
$O(n^2)$ arithmetic operations overall. Summing over
$w\leq\lfloor n/2\rfloor$ gives the majority action probability;
a Bernoulli draw then samples the policy.
Appendix~\ref{app:connected-iqp-simulation} derives each step.

Analytical training uses this custom differentiable PyTorch
tensor-network simulator coupled to the QASP-based REINFORCE trainer, rather
than a full-statevector evaluation at each observation. The
implementation was checked against full-state simulation, with
maximum Hamming-weight-probability error $7.1\times10^{-8}$ at
21 qubits. Thus this experiment is deliberately classically
simulable, despite its nonfactorizing state and global majority
readout. No bond truncation or Pauli-expansion enumeration is used.
The generating-polynomial identity holds for any majority policy,
but its efficient contraction here relies on the chain of
constant-size blocks and single bridges. It is not a polynomial-time
simulation guarantee for arbitrary majority-based IQP policies.
The practical interpretation is therefore a quantum-inspired model
with competitive episode efficiency, not a quantum computational
advantage.

\subsubsection{Uniform-initialization follow-up}

To test sensitivity to initialization, we retain the circuit, majority
decoder, neural weights, preprocessing, optimizer, learning rate,
batching, and stopping rule, but initialize all 21 biases, 28
within-triple interaction angles, and six bridge angles independently
from $\Unif[-\pi,\pi]$. The neural weights remain bitwise identical
to their original per-seed values. Thus the effective encoding angles
are uniform conditional on any fixed observation, as required by the
variance analysis.

This follow-up uses the same 20 seed IDs and stored MLP baselines.
It was requested after the original outcomes were examined, so it is
an exploratory sensitivity analysis, not an untouched confirmatory
cohort. No hyperparameters or seeds were reselected.

\begin{table}[tbp]
\centering
\caption{Initialization sensitivity on the same 20 seeds.
All three conditions reach the criterion in 20/20 runs.
Entries are means; AUC uses Eq.~\eqref{eq:connected-normalized-auc}.}
\label{tab:connected-initialization}
\small
\setlength{\tabcolsep}{3pt}
\begin{tabular}{lrrr}
\toprule
Condition & Initial return & AUC & Episodes\\
\midrule
IQP, original & 73.63 & 0.9319 & 171.45\\
IQP, uniform & 22.13 & 0.7615 & 315.90\\
MLP, 135 parameters & 20.22 & 0.8680 & 236.50\\
\bottomrule
\end{tabular}
\end{table}

Uniform initialization lowers the IQP's AUC by $0.1704$ relative to
the original initialization (paired-bootstrap 95\% interval
$[-0.1990,-0.1407]$ for uniform minus original). Relative to the
parameter-matched MLP, its AUC is lower by $0.1065$
($[-0.1428,-0.0703]$), and it takes $79.4$ more episodes to criterion
($[40.4,118.0]$). All uniform-IQP runs nevertheless reach the criterion;
their final held-out mean return is $196.99$.

The original episode-efficiency advantage therefore depends strongly
on initialization and includes a substantially stronger starting policy.
Uniform initialization preserves the theorem's ensemble assumptions
but does not preserve that performance ordering. This distinguishes
policy nonconcentration from fast RL learning. Episode efficiency
also does not imply wall-clock efficiency: the matched three-seed
analytical-runtime audit in Appendix~\ref{app:analytical-runtime}
finds that the optimized IQP remains slower than the MLP. No
finite-shot evaluation is required for this classically tractable
model.

\section{Discussion}
\label{sec:discussion}
% ================================================================

The central message is that action decoding can be analyzed through a
nonnegative initialization-variance decomposition. For IQP circuits
with independent uniform angles and generator supports spanning
$\mathbb F_2^n$, each squared decoder coefficient is weighted by the
variance $2^{-r^u}$ of its Pauli expectation. Independent local
rotations on every qubit are sufficient, but not necessary. The component identity overlaps
with Shen \emph{et al.}~\cite{Shen2026IQPTrainability}; the policy sum
also requires the covariance cancellation proved here. Consequently,
a single inverse-polynomial coefficient with logarithmic rank
protects the whole policy against exponential concentration. For
local-and-edge graph IQPs, this criterion constrains only the
selected component's backward light cone, not the maximum degree
of the full graph.

If the generator support list is rank deficient, equal active-generator
signatures identify identical Pauli expectations. Their decoder
coefficients must be added before squaring, as in
Proposition~\ref{prop:grouped-policy-variance}. The resulting grouped
sum remains nonnegative, but coefficients within one group may cancel.
This distinction explains the pure-$R_{XX}$ ring experiment and makes
clear why an individual protected coefficient is not sufficient for
an arbitrary generator set.

For the augmented repeated IQP-like architecture, applying the
observable-decomposition theorem of Letcher \emph{et al.}
~\cite{Letcher2024Tight} gives the corresponding sum of propagated-Pauli
bounds. An inverse-polynomial probability of logarithmic propagated
weight in the auxiliary Clifford ensemble is already sufficient;
every realization need not have a small light cone. This is useful
beyond globally commuting circuits and beyond pairwise interactions.
It remains conditional on the initial local layers and independent
parameter ensemble. Neither the decomposition nor its lower bounds
are universal claims for arbitrary PQCs.

These statements are about initialization nonconcentration at a
fixed observation. They do not prove efficient RL optimization:
the learning update also depends on score factors, rewards,
state visitation, and correlations between gradient components.
Neural encoding preserves the ensemble statement only with the
independent offsets specified in Sec.~\ref{subsec:neural-state-embeddings}.
In particular, favorable learning with a task-oriented initialization
cannot be presented as a direct test of a uniform-angle theorem.

Classical simulation is a second, logically separate question.
A difficult full-output sampling problem may become easy after
decoding. Sparse Pauli expansions and contiguous intervals supply
explicit sufficient conditions for additive estimation of IQP action
probabilities. Conversely, exponential \emph{exact} Fourier support
only excludes direct polynomial-size enumeration of that exact
representation. It does not exclude approximate truncation,
importance sampling, symmetry-based aggregation, or tensor-network
contraction. The majority--modulo construction therefore establishes
compatibility of a protected component with dense exact support,
not classical sampling hardness.

The CartPole evidence illustrates both the promise and the limits of
this design principle. On the ten-qubit instance, majority outperforms
the tested valid local readouts, whereas the five-qubit contrasts are
unresolved. MLPs are competitive on both. The gradient diagnostics
show increased \emph{relative} participation of higher-weight sectors
on ten qubits, driven mainly by weights three and five, not an
increasing maximal-weight sector. They do not establish that these
sectors cause improved control or that exponentially many terms are
needed. Letcher \emph{et al.} already discuss the participation of
global terms beyond initialization~\cite{Letcher2024Tight}; the present
diagnostics examine that mechanism for action-decoded RL policies.

The connected 21-qubit experiment provides a concrete counterexample
to equating dense support or entanglement with simulation hardness.
Its state is nonfactorizing, but a chain of constant-size blocks and
single bridges has bounded inter-block bond dimension. Majority
depends only on total Hamming weight, so a generating-polynomial
tensor-network contraction evaluates the policy in $O(n^2)$ arithmetic
operations.
This is an exact classical method up to floating-point error,
without Pauli enumeration or bond truncation. It is specific to the
circuit/readout structure, not a simulator guaranteed to be efficient
for every majority-based IQP policy.

Under its original initialization, this quantum-inspired model has
higher episode-based AUC than the tested MLP configurations. The
uniform-initialization follow-up, however, reverses the ordering:
AUC falls from $0.9319$ to $0.7615$, below the parameter-matched
MLP's $0.8680$, while all 20 runs still meet the solve criterion.
The original IQP also starts from a substantially stronger policy.
Thus the learning comparison reflects the combined architecture and
initialization bias, not an initialization-independent advantage.
Because the follow-up reuses the previously examined seed cohort,
it is exploratory rather than a fresh confirmatory test. The
three-seed runtime check likewise shows no classical speed advantage:
the optimized analytical IQP takes about $34.0$ seconds versus
$7.9$ seconds for the comparably optimized MLP
(Appendix~\ref{app:analytical-runtime}). Polynomial scaling does not
make two implementations equally fast at a fixed size.

Classically tractable quantum-inspired policies nevertheless deserve
study as useful structured models. The experiments establish that
this architecture can learn CartPole using exact classical probability
evaluation; quantum hardware and finite-shot estimation are not
needed for that result. Further tests should separate initialization
quality from subsequent learning, compare equal environment-step
and wall-clock budgets, and include feature-matched classical
models, additional tasks, and prospectively held-out seeds.
The current MLP grid is a controlled finite comparison, not a
claim against all classical models or of asymptotic sample advantage.

Sparse and low-degree decoder surrogates are another priority.
Retraining a valid local or low-degree policy to similar returns
would demonstrate an alternative solution of the control task,
not simulation of the original majority policy. A simulation claim
instead requires accurate action probabilities at the same inputs
and parameters, with a polynomial resource bound. A coefficient
truncation must also preserve positivity and normalization if it
is to define a policy. Fourier-tail bounds under uniform bit strings
alone do not control errors under the circuit's Born distribution.

For repeated IQP-like and other noncommuting circuits, the decoder
certificates provide a starting point, but simulation needs a
separate study of observable propagation, circuit connectivity,
depth, approximation error, and contraction width. Dense exact
decoder support may obstruct one representation while leaving
others efficient. Establishing hardness or an efficient algorithm
for specific protected-component families is left to future work
by the author. Noise must also be included in that assessment:
known layerwise Pauli-noise results can make sufficiently deep IQP
circuits classically sampleable~\cite{RajakumarWatsonLiu2024}.
The present noiseless concentration laws do not establish hardware
robustness or end-to-end quantum advantage.

\section*{Acknowledgements}
The single-Pauli component identity in
Lemma~\ref{lem:rank-law} was derived independently in this work.
The author became aware of the overlapping result of Shen
\emph{et al.}~\cite{Shen2026IQPTrainability} only during the final
stages of preparing the manuscript and subsequently incorporated
the citation and comparison. The self-contained proof is retained
for completeness; no priority claim is made for the overlapping
identity.

\section*{Data availability}
No external dataset was used. Numerical data were generated by the
simulation protocols described in Sec.~\ref{sec:experiments} and
Appendix~\ref{app:supplementary-numerics}. Available data are publicly
accessible in the companion experiment repository at
\url{https://github.com/andre-sequeira10/iqp-policy-experiments},
including seed-level records for the
controlled decoder studies, gradient diagnostics, random-cubic
variance study, connected-IQP/MLP comparison, and uniform-initialization
follow-up, together with the analytical-runtime records.

\section*{Competing interests}
The author declares no competing interests.

% ================================================================
\appendices
% Distinguish appendix subsections from the appendix letters.
\renewcommand{\thesubsection}{\thesection-\Roman{subsection}}
% ================================================================

\section{Proofs of the Walsh--Pauli Correspondence and IQP Trainability Results}
\label{app:proofs-single-block}

\subsection{Proof of Proposition~\ref{prop:walsh-pauli-policy}}
\begin{proof}
For one computational-basis bit,
$|x_i\rangle\langle x_i|=[I+(-1)^{x_i}Z_i]/2$.  Taking tensor products
gives the character expansion
\begin{equation}
    |x\rangle\langle x|
    =2^{-n}\sum_{u\in\{0,1\}^n}(-1)^{u\cdot x}Z^u.
    \label{eq:basis-projector-character-expansion}
\end{equation}
Multiplying by $h_a(x)$ and summing over $x$ yields
\begin{equation}
    \Pi_a
    =\sum_u\left(2^{-n}\sum_x h_a(x)(-1)^{u\cdot x}\right)Z^u,
\end{equation}
so the coefficient of $Z^u$ is exactly $\widehat h_a(u)$.  Evaluating
the diagonal matrix element in $|x\rangle$ gives
$\langle x|Z^u|x\rangle=(-1)^{u\cdot x}$ and hence the inverse formula;
taking the expectation value in $|\psi\rangle$ gives
Eq.~\eqref{eq:policy-fourier-expectation}.  Finally, the zero-frequency
coefficient is the mean of $h_a$, namely $|B_a|/2^n$, while Walsh
orthogonality gives
$\sum_u\widehat h_a(u)^2=2^{-n}\sum_x h_a(x)^2$.  Because $h_a$ is an
indicator, $h_a(x)^2=h_a(x)$, proving
Eq.~\eqref{eq:policy-fourier-parseval}.
\end{proof}

\subsection{Exact active-generator rank law}

For a diagonal Pauli $Z^u$, define
\begin{equation}
    L_u(\boldsymbol\theta)
    :=\langle0^n|U_X^\dagger Z^uU_X|0^n\rangle.
    \label{eq:Lu}
\end{equation}
Write $P_\ell=X^{g_\ell}$, where $g_\ell$ records the qubits on which
the generator acts. A generator is active when it anticommutes with
$Z^u$, equivalently when $g_\ell\cdot u=1$ modulo two. Let
$\mathcal I_u$ index these generators and let $G_u$ contain their
binary supports as rows. Its binary rank $r^u$ counts the independent
active supports; all binary arithmetic below is modulo two.

The following component identity was obtained independently in this
work and also appears in Theorem~3.8 of
Shen \emph{et al.}~\cite{Shen2026IQPTrainability}. We retain the proof
for completeness and to make the subsequent covariance argument
self-contained. This single-observable result does not require local
rotations or a full-rank generator support list. The individual-term
policy sum additionally requires covariance cancellation; the grouped
identity in Proposition~\ref{prop:grouped-policy-variance} covers
arbitrary generator sets.

\begin{lemma}[Exact rank law]
\label{lem:rank-law}
For the input $|0^n\rangle$, initialize each gate angle independently
as $\theta_\ell\sim\Unif[-\pi,\pi]$. Let
$r^u:=\rank_{\mathbb F_2}(G_u)$. If
$\mathcal I_u\neq\emptyset$, then
\begin{equation}
    \E_{\theta}[L_u]=0,
    \qquad
    \Var_{\theta}[L_u]=2^{-r^u}.
    \label{eq:rank-law}
\end{equation}
If $\mathcal I_u=\emptyset$, then $L_u\equiv1$ and
$\Var_{\theta}[L_u]=0$.
\end{lemma}

\begin{proof}
\emph{Step 1: express the observable as a phase average.}
Use the $X$ basis $|b\rangle_X:=H^{\otimes n}|b\rangle$, for
$b\in\{0,1\}^n$. In this basis, each generator supplies a sign and
$Z^u$ flips the bits selected by $u$:
\begin{equation}
    X^{g_\ell}|b\rangle_X=(-1)^{g_\ell\cdot b}|b\rangle_X,
    \qquad
    Z^u|b\rangle_X=|b\oplus u\rangle_X.
    \label{eq:rank-proof-x-actions}
\end{equation}
Since $|0^n\rangle=2^{-n/2}\sum_b|b\rangle_X$, all basis states have
equal weight. In $U_X^\dagger Z^uU_X$, an inactive generator has the
same eigenvalue on $b$ and $b\oplus u$, so its two phases cancel.
For an active generator the eigenvalue changes sign, giving the phase
$e^{-i\theta_\ell(-1)^{g_\ell\cdot b}}$. Consequently,
\begin{equation}
    L_u=2^{-n}\sum_{b\in\{0,1\}^n}
       \exp\!\left[-i\sum_{\ell\in\mathcal I_u}
       \theta_\ell(-1)^{g_\ell\cdot b}\right].
    \label{eq:Lu-xbasis}
\end{equation}
If there are no active generators, every summand equals one. Thus
$L_u\equiv1$ and its variance is zero. Assume henceforth that at least
one generator is active.

\emph{Step 2: the mean vanishes.}
Fix a bit string $b$. For each active generator, the sign
$(-1)^{g_\ell\cdot b}$ is either $+1$ or $-1$, never zero.
The uniform average of the corresponding phase is zero for either
sign:
\[
    \E_{\theta_\ell}\!\left[e^{\pm i\theta_\ell}\right]
    =\frac{1}{2\pi}\int_{-\pi}^{\pi}e^{\pm it}\,dt=0.
\]
Indeed, both $\cos t$ and $\sin t$ integrate to zero on this interval.
Writing the exponential in Eq.~\eqref{eq:Lu-xbasis} as a product,
independence of the angles gives
\[
    \begin{aligned}
        &\E_{\theta}\!\left[
          \prod_{\ell\in\mathcal I_u}
          e^{-i\theta_\ell(-1)^{g_\ell\cdot b}}
        \right]\\
        &\quad=
          \prod_{\ell\in\mathcal I_u}
          \E_{\theta_\ell}\!\left[
          e^{-i\theta_\ell(-1)^{g_\ell\cdot b}}
          \right]=0.
    \end{aligned}
\]
The product contains at least one zero factor because
$\mathcal I_u\neq\emptyset$. Linearity of expectation then yields
\[
    \E_{\theta}[L_u]
    =2^{-n}\sum_{b\in\{0,1\}^n}0=0.
\]
Equivalently, averaging over any one independent active angle while
holding the others fixed already makes each summand vanish. No linear
independence of the binary supports $g_\ell$ is needed for this step;
the independence used here is that of the random angles.

\emph{Step 3: count the terms surviving in the second moment.}
Unlike the first moment, the second moment contains two factors with
the same random angles. Their phases can therefore cancel before
averaging. For example,
\[
    \E_{\theta_\ell}\!\left[
       e^{-i\theta_\ell}e^{+i\theta_\ell}\right]
    =\E_{\theta_\ell}[1]=1,
\]
although each phase separately has zero mean. These two factors are
not independent, so their expectation cannot be factorized.

\emph{Multiplying the two expansions.}
The observable expectation $L_u$ is real because $Z^u$ is Hermitian,
so $L_u^2=|L_u|^2$. Multiplying Eq.~\eqref{eq:Lu-xbasis} by its
complex conjugate introduces two bit strings $b,c\in\{0,1\}^n$:
\[
    |L_u|^2
    =4^{-n}\sum_{b,c}\prod_{\ell\in\mathcal I_u}
      e^{-i\theta_\ell
      \left[(-1)^{g_\ell\cdot b}-(-1)^{g_\ell\cdot c}\right]}.
\]
The factor $4^{-n}$ is the product of the two normalizations
$2^{-n}$. Independence between different angles now gives
\begin{equation}
    \begin{aligned}
        &\E_{\theta}[|L_u|^2]\\
        &\quad=4^{-n}\sum_{b,c}
          \prod_{\ell\in\mathcal I_u}
          \E_{\theta_\ell}\!\left[
          e^{-i\theta_\ell
          \left[(-1)^{g_\ell\cdot b}-(-1)^{g_\ell\cdot c}\right]}
          \right].
    \end{aligned}
    \label{eq:rank-proof-second-moment}
\end{equation}

\emph{Identifying the surviving pairs.}
For each active generator, equal signs give a zero exponent, so the
phase and its average are both one. Unequal signs differ by $+2$ or
$-2$, giving instead
\[
    \E_{\theta_\ell}\!\left[e^{\pm2i\theta_\ell}\right]
    =\frac{1}{2\pi}\int_{-\pi}^{\pi}e^{\pm2it}\,dt=0.
\]
Hence the product of averages in
Eq.~\eqref{eq:rank-proof-second-moment} is one precisely when
$g_\ell\cdot c=g_\ell\cdot b$ modulo two for every active
generator; otherwise at least one factor is zero. Since these
supports are the rows of $G_u$, the simultaneous conditions are
\[
    G_uc=G_ub.
\]
The second moment is therefore $4^{-n}$ times the number of such
pairs. In particular, every pair with $c=b$ survives, so the second
moment cannot vanish.

\emph{Counting with the binary rank.}
Fix any $b$. The system $G_uc=G_ub$ is consistent because $c=b$ is
a solution. Binary row reduction leaves $r^u$ independent equations
in the $n$ bits of $c$: $r^u$ pivot bits are determined by the
remaining $n-r^u$ free bits. Each free bit can be zero or one,
giving exactly $2^{n-r^u}$ solutions for $c$. This count is the same
for every $b$. Since there are $2^n$ choices of $b$, the number of
surviving pairs is
\[
    \#\{(b,c):G_uc=G_ub\}=2^n\,2^{n-r^u}.
\]
Using the zero mean established in Step~2, we conclude that
\begin{equation}
    \begin{aligned}
        \Var_{\theta}[L_u]
        &=\E_{\theta}[|L_u|^2]
          -\bigl(\E_{\theta}[L_u]\bigr)^2\\
        &=4^{-n}\,2^n\,2^{n-r^u}=2^{-r^u}.
    \end{aligned}
\end{equation}
Thus each independent binary matching condition halves the fraction
of surviving pairs.

\emph{Example: three qubits and rank two.}
Take $Z^u=Z_1$ and the two independently parameterized generators
$X_1X_2$ and $X_1X_3$. Both are active, with
\[
    G_u=\begin{pmatrix}1&1&0\\1&0&1\end{pmatrix},
    \qquad r^u=2.
\]
For $b=000$, the equations $G_uc=G_ub$ read
\[
    c_1+c_2=0,\qquad c_1+c_3=0\pmod2.
\]
Hence $c_2=c_1$ and $c_3=c_1$. The one free bit $c_1$ gives
exactly two solutions, $c=000$ and $c=111$. For any other $b$,
the two solutions are $c=b$ and $c=b\oplus111$. There are therefore
$8\times2=16$ surviving pairs out of $8\times8=64$, so
\[
    \Var_{\theta}[L_u]=\frac{16}{64}=\frac14=2^{-2}.
\]
\end{proof}

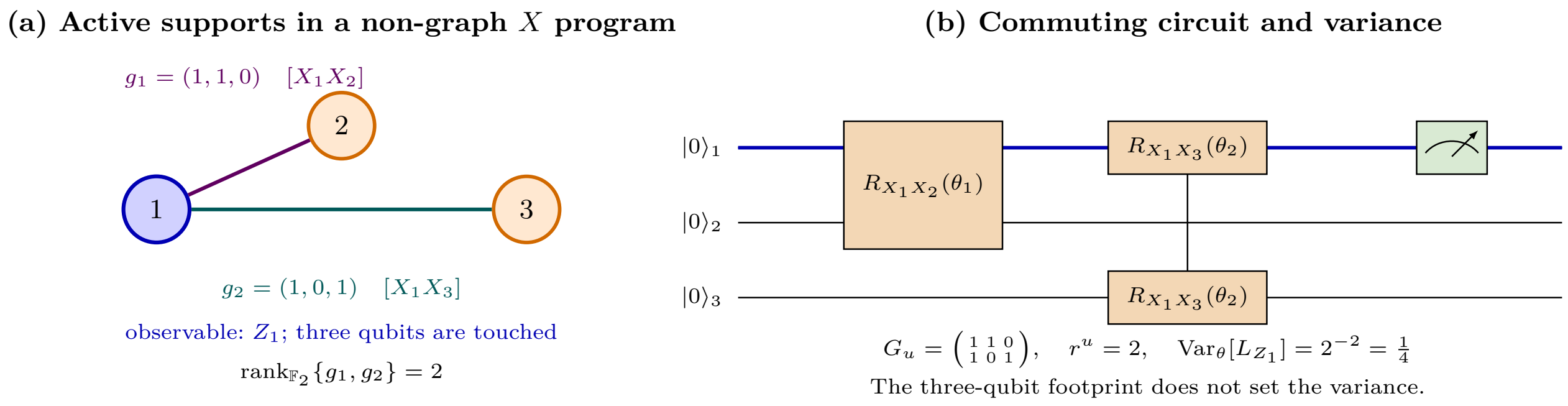
\begin{figure*}[!htbp]
    \centering
    \resizebox{0.92\textwidth}{!}{%
    \begin{tikzpicture}[
        wire/.style={publication wire},
        activewire/.style={line width=1.15pt, draw=blue!70!black},
        vertex/.style={circle, draw=black!55, fill=black!8,
                       line width=0.9pt, minimum size=7.5mm,
                       inner sep=0pt, font=\small},
        observed/.style={vertex, draw=blue!70!black, fill=blue!18,
                          line width=1.25pt},
        reached/.style={vertex, draw=orange!82!black, fill=orange!18,
                         line width=1.15pt},
        hyperone/.style={draw=violet!75!black, line width=1.2pt,
                         rounded corners=5pt, fill=violet!8},
        hypertwo/.style={draw=teal!70!black, line width=1.2pt,
                         rounded corners=5pt, fill=teal!8},
        gate/.style={publication gate, fill=paperTwoGate,
                     minimum width=18mm, font=\scriptsize},
        twogate/.style={gate, minimum height=14.5mm},
        measure/.style={publication meter, font=\scriptsize},
        label/.style={font=\scriptsize, align=center}
    ]
        \node[font=\bfseries\small] at (3.10,2.35)
            {(a) Active supports in a non-graph $X$ program};
        \node[observed] (v1) at (1.00,0.25) {$1$};
        \node[reached] (v2) at (3.10,1.20) {$2$};
        \node[reached] (v3) at (5.20,0.25) {$3$};
        \draw[hyperone] (v1) -- (v2);
        \draw[hypertwo] (v1) -- (v3);
        \node[label, text=violet!75!black] at (2.00,1.74)
            {$g_1=(1,1,0)\quad [X_1X_2]$};
        \node[label, text=teal!70!black] at (3.10,-0.65)
            {$g_2=(1,0,1)\quad [X_1X_3]$};
        \node[label, text=blue!70!black] at (3.10,-1.16)
            {observable: $Z_1$; three qubits are touched};
        \node[label] at (3.10,-1.61)
            {$\operatorname{rank}_{\mathbb F_2}\{g_1,g_2\}=2$};

        \node[font=\bfseries\small] at (12.65,2.35)
            {(b) Commuting circuit and variance};
        \foreach \y/\q in {0.95/1,0.10/2,-0.75/3}{
            \node[font=\scriptsize, anchor=east] at (7.55,\y)
                {$\lvert0\rangle_{\q}$};
            \draw[wire] (7.60,\y)--(16.95,\y);
        }
        \draw[activewire] (7.60,0.95)--(16.95,0.95);
        \node[twogate] at (9.70,0.525)
            {$R_{X_1X_2}(\theta_1)$};
        \draw[wire] (12.70,0.95)--(12.70,-0.75);
        \node[gate] at (12.70,0.95)
            {$R_{X_1X_3}(\theta_2)$};
        \node[gate] at (12.70,-0.75)
            {$R_{X_1X_3}(\theta_2)$};
        \node[measure] at (15.70,0.95) {};
        \node[label] at (12.25,-1.34)
            {$G_u=\begin{psmallmatrix}1&1&0\\1&0&1\end{psmallmatrix},\quad
              r^u=2,\quad \Var_{\theta}[L_{Z_1}]=2^{-2}=\tfrac14$};
        \node[label] at (12.25,-1.78)
            {The three-qubit footprint does not set the variance.};
    \end{tikzpicture}}
    \caption{Binary rank versus geometric interaction footprint. For the
    non-graph commuting $X$ program, $Z_1$ activates the generators
    $X_1X_2$ and $X_1X_3$. Their union touches three qubits, but their
    support vectors span only two independent binary directions. The
    exact rank law therefore gives $\Var_{\theta}[L_{Z_1}]=1/4$, not $1/8$.
    The linked boxes represent one nonadjacent $X_1X_3$ rotation,
    leaving qubit 2 untouched. The green meter denotes
    computational-basis measurement of qubit 1.}
    \label{fig:rank-versus-footprint}
\end{figure*}

The rank counts independent active directions, not active gates or the
number of qubits in their union.  Figure~\ref{fig:rank-versus-footprint}
illustrates the distinction for a non-graph commuting program; the
local-and-edge graph specialization is established next.

\subsection{Graph light-cone variance law}

\begin{lemma}[Graph light-cone variance law]
\label{lem:graph-lightcone}
Consider the graph-IQP circuit in Eq.~\eqref{eq:graph-iqp-x}, with
a local $X_i$ generator on every vertex, $X_iX_j$ generators on its
edges, and all angles independently uniform on $[-\pi,\pi]$.
For every nonidentity $Z^u$,
\begin{equation}
    r^u=\Delta_G^{\theta}(u),
    \qquad
    \Var_{\theta}[L_u]=2^{-\Delta_G^{\theta}(u)}
    \label{eq:graph-lightcone-variance}
\end{equation}
where the light-cone size is evaluated at generic parameter values.
\end{lemma}

\begin{proof}
Let $S_u=\{i:u_i=1\}$ be the observable's support, and let $e_i$
denote the binary vector with a single one at position $i$.
By Lemma~\ref{lem:rank-law}, it suffices to identify the active
binary rank with the generic light-cone size. Since $Z^u\neq I$,
$S_u$ is nonempty, and any local $X_i$ with $i\in S_u$ is active.
Thus the nonempty-active-set condition of that lemma is satisfied.
This includes a single-qubit observable $Z_i$: its support is
just $\{i\}$, not the empty set.

\emph{Step 1: identify the active generators.}
The commutation signs are
\[
    X_iZ^u=(-1)^{u_i}Z^uX_i,\qquad
    X_iX_jZ^u=(-1)^{u_i+u_j}Z^uX_iX_j.
\]
A local $X_i$ is therefore active exactly when $i\in S_u$.
An edge generator is active when exactly one endpoint lies in
$S_u$; edges with both endpoints inside or both outside commute
with the observable. Hence $G_u$ has rows $e_i$ for $i\in S_u$
and $e_i+e_j$ for edges crossing from $S_u$ to an external neighbour.

\emph{Step 2: compute the binary rank.}
Every active row is zero outside $S_u$ and its external neighbours,
so the rank cannot exceed the number of these vertices.
To prove the reverse inequality, each vertex $i\in S_u$ already
supplies the row $e_i$. For every external neighbour $j$, choose
one boundary edge $(i,j)$ with $i\in S_u$. Adding its row to
the local row modulo two gives
\[
    (e_i+e_j)+e_i=e_j.
\]
Thus the row span contains the coordinate vector of every vertex
in $S_u$ and every external neighbour. These vectors are independent,
because their nonzero entries occupy distinct coordinates.
The rank is consequently exactly the number of these vertices.
Multiple boundary edges reaching the same external vertex do not
increase this count. This is binary row reduction, not the addition
of physical gates to the circuit.

\emph{Step 3: identify these vertices with the light cone.}
Write $U_G=U_{\rm inactive}U_{\rm active}$. All gates commute,
and every inactive gate commutes with $Z^u$, so
\[
    \begin{aligned}
        U_G^\dagger Z^uU_G
        &=U_{\rm active}^\dagger
          U_{\rm inactive}^\dagger Z^u
          U_{\rm inactive}U_{\rm active}\\
        &=U_{\rm active}^\dagger Z^uU_{\rm active}.
    \end{aligned}
\]
Inactive gates therefore cancel exactly; they cannot spread the
observable farther after an active edge reaches a new vertex.
For example, on the path $1$--$2$--$3$, the observable $Z_1$
can reach vertex $2$ through $X_1X_2$, but $X_2X_3$ is inactive
and cannot extend its support to vertex $3$.

Active gates act only on the vertices counted in Step~2, giving
an upper bound on the propagated support. Conversely, every
nonzero Pauli term in the expansion retains the original support:
multiplication by $X$ toggles a $Z$ factor to a $Y$ factor or back,
up to phase, never to the identity.
Every external neighbour also appears for generic angles.
Indeed, for a boundary edge $(i,j)$, set all angles except
$\beta_{ij}$ to zero. Since $X_iX_j$ anticommutes with $Z^u$,
\[
    \begin{aligned}
        &e^{i\beta_{ij}X_iX_j/2}Z^u
          e^{-i\beta_{ij}X_iX_j/2}\\
        &\qquad=\cos(\beta_{ij})Z^u
           +i\sin(\beta_{ij})X_iX_jZ^u.
    \end{aligned}
\]
The second term acts nontrivially on $j$. Its coefficient in the
full expansion is a trigonometric polynomial that is therefore
not identically zero, and hence is nonzero for generic angles.
Setting the other angles to zero here only establishes this
nonvanishing property; it does not change the initialization
ensemble. Since there are finitely many external neighbours,
they all belong to the light cone at generic parameter values.
The light cone is the union of supports across the Pauli expansion;
individual terms need not occupy every vertex in that union.

\emph{Step 4: apply the rank law.}
Steps~2 and~3 count the same vertices, so
$r^u=\Delta_G^{\theta}(u)$. The active set is nonempty, and
Lemma~\ref{lem:rank-law} gives
\[
    \E_\theta[L_u]=0,\qquad
    \Var_\theta[L_u]=2^{-r^u}=2^{-\Delta_G^{\theta}(u)}.
\]

\emph{Example: a single-qubit observable $Z_i$.}
Its active generators are $X_i$ and the incident edges $X_iX_j$.
The rows $e_i$ and $e_i+e_j$ span $e_i$ and one coordinate vector
for each neighbour. Consequently,
\[
    r^{e_i}=\Delta_G^\theta(e_i)=1+\deg_G(i).
\]
One can also check the variance directly. In the backward expansion,
choosing the sine term of any incident edge leaves an $X_j$ on its
external endpoint, whose expectation in $|0^n\rangle$ vanishes.
The sine term of the local rotation similarly leaves a $Y_i$.
Only the product of cosine terms contributes, yielding
\[
    L_{e_i}(\theta)
    =\cos\alpha_i\prod_{j:(i,j)\in E}\cos\beta_{ij}.
\]
Each independent angle has $\E[\cos\theta]=0$ and
$\E[\cos^2\theta]=1/2$, so
\[
    \Var_\theta[L_{e_i}]=2^{-(1+\deg_G(i))}.
\]
Thus a one-qubit measurement need not have a one-qubit light cone.
For $Z_2$ on the path $1$--$2$--$3$,
\[
    G_{e_2}=
    \begin{pmatrix}
        0&1&0\\
        1&1&0\\
        0&1&1
    \end{pmatrix}
    \ \longrightarrow\
    \begin{pmatrix}
        0&1&0\\
        1&0&0\\
        0&0&1
    \end{pmatrix},
\]
where the first row is added to each of the other rows.
The rank and generic light-cone size are both three, and the
variance is $1/8$.

\emph{Why the local rotations cannot generally be omitted.}
Remove the local generators from the same three-qubit circuit.
For $Z_2$, the active matrix is now
\[
    G_{e_2}=
    \begin{pmatrix}
        1&1&0\\
        0&1&1
    \end{pmatrix},
    \qquad r^{e_2}=2.
\]
Both edges still spread the observable, so its generic light cone
contains all three qubits. Nevertheless,
$L_{e_2}=\cos\beta_{12}\cos\beta_{23}$ and its variance is $1/4$,
not $1/8$. The rank law remains valid, but rank and light-cone
size no longer coincide. Local generators supply the coordinate
rows used in Step~2. For a fixed observable, local generators on
its support suffice for that argument; including them on every
vertex guarantees the equality for every nonidentity $Z^u$.
\end{proof}

The generic-angle qualification concerns the realized light cone, not
the ensemble variance: special angles can shrink the propagated
support, but they form a set of zero probability under independent
continuous initialization. The exact variance uses the structural,
generic light-cone size, not its value at an exceptional angle
configuration such as the identity circuit.

\subsection{Orthogonality of diagonal Pauli contributions}
\begin{lemma}[Orthogonality of diagonal Pauli contributions]
\label{lem:pauli-orthogonality}
For an IQP $X$ program on $|0^n\rangle$, initialize all angles
independently and uniformly over a full period. Let $\mathsf G$
contain all generator supports as rows. If $\mathsf Gu\neq\mathsf Gv$,
\begin{equation}
    \E_{\theta}[L_uL_v]=0.
    \label{eq:pauli-crosszero}
\end{equation}
If $\mathsf Gu=\mathsf Gv$, then $L_u=L_v$ for every parameter vector.
In particular, $\rank_{\mathbb F_2}\mathsf G=n$ ensures
Eq.~\eqref{eq:pauli-crosszero} for all distinct $u,v$.
\end{lemma}
\begin{proof}
\emph{Step 1: find an angle active for exactly one observable.}
If $\mathsf Gu\neq\mathsf Gv$, some row $g_k$ satisfies
$g_k\cdot u\neq g_k\cdot v$. After interchanging $u$ and $v$ if
necessary, take $g_k\cdot u=1$ and $g_k\cdot v=0$. The angle
$\theta_k$ is active for $L_u$ but inactive for $L_v$. Independent
local rotations guarantee such a row whenever $u\neq v$: choose
a qubit $q$ where the masks differ and use $g_k=e_q$. More
generally, full column rank guarantees that
$\mathsf G(u\oplus v)\neq0$.

\emph{Step 2: average over that angle first.}
Hold every other angle fixed. Since the IQP generators commute,
the inactive $X^{g_k}$ rotation cancels from $U_X^\dagger Z^vU_X$.
Thus $L_v$ is independent of $\theta_k$. By
Eq.~\eqref{eq:Lu-xbasis}, every summand of $L_u$ contains the factor
$e^{-i\theta_k(-1)^{g_k\cdot b}}$, and all its other factors are independent
of $\theta_k$. For either value of $g_k\cdot b$,
\[
    \frac{1}{2\pi}\int_{-\pi}^{\pi}
       e^{-it(-1)^{g_k\cdot b}}\,dt=0.
\]
Consequently $\E_{\theta_k}[L_u]=0$ for every fixed choice of the
remaining angles. Independence of the parameters is important:
conditioning on those angles leaves $\theta_k$ uniform.

\emph{Step 3: average the product.}
Writing $\E_{\rm rest}$ for the average over the remaining angles,
iterated expectation gives
\[
    \E_{\theta}[L_uL_v]
    =\E_{\rm rest}\!\left[
        L_v\,\E_{\theta_k}[L_u]
      \right]=0.
\]
The factor $L_v$ can be taken outside the inner average because it
does not depend on $\theta_k$. Thus the two observables are
uncorrelated; they need not be statistically independent.

\emph{Step 4: identify what happens when the signatures coincide.}
If $\mathsf Gu=\mathsf Gv$, the active generator sets are identical.
Equation~\eqref{eq:Lu-xbasis} depends on the observable mask only
through that set, so $L_u=L_v$ pointwise. If their signature is
nonzero, their covariance is therefore their common variance,
$2^{-r^u}$, not zero. If the signature is zero, both are identically
one and their covariance is zero.
\end{proof}

The single-component rank law alone does not establish this
orthogonality. For example, if the only gate is
$U_X=e^{-i\theta X_1X_2/2}$, then
$L_{(1,0)}=L_{(0,1)}=\cos\theta$.
For uniform $\theta$ both individual variances equal $1/2$, but
their covariance is also $1/2$. Thus removing local rotations does
not invalidate the single-component rank identity, but can invalidate
the individual-term policy sum unless the remaining signatures
still distinguish the contributing observables.

\subsection{Proof of the IQP-policy part of Theorem~\ref{thm:main-trainability}}
\label{app:grouped-policy-variance}
\begin{proof}
Fix an observation and a classical decoder, so its real Pauli
coefficients $c_{a,u}$ do not depend on the initialized angles.

\emph{Step 1: separate the constant part.}
The Walsh--Pauli expansion gives
\[
    \pi(a)=c_{a,0}+\sum_{u\neq0}c_{a,u}L_u,
\]
because the identity expectation is $L_0=1$. Full column rank of
$\mathsf G$ implies $\mathsf Gu\neq0$ for each $u\neq0$, so at
least one generator is active. Lemma~\ref{lem:rank-law} yields
$\E_\theta[L_u]=0$ and $\E_\theta[L_u^2]=2^{-r^u}$.
It follows that $\E_\theta[\pi(a)]=c_{a,0}$.

\emph{Step 2: expand the centered square.}
By the definition of variance and linearity of expectation,
\[
\begin{aligned}
    \Var_\theta[\pi(a)]
    &=\E_\theta\!\left[
        \left(\sum_{u\neq0}c_{a,u}L_u\right)^2
      \right]\\
    &=\sum_{u\neq0}c_{a,u}^2\,\E_\theta[L_u^2]\\
    &\quad+
      \sum_{\substack{u,v\neq0\\u\neq v}}
      c_{a,u}c_{a,v}\,\E_\theta[L_uL_v].
\end{aligned}
\]
The second sum runs over ordered pairs, so no extra factor of two
is needed. This is the step at which cross covariances would enter;
a variance of a sum is not generally a sum of variances.

\emph{Step 3: remove cross terms and substitute the rank law.}
Lemma~\ref{lem:pauli-orthogonality} makes every term in the second
sum zero. The first sum therefore becomes
\[
    \Var_\theta[\pi(a)]
    =\sum_{u\neq0}c_{a,u}^2\,2^{-r^u},
\]
which is Eq.~\eqref{eq:main-policy-rank-variance}.
Every contribution is nonnegative because its coefficient is
squared. The decoder's Pauli terms may interfere in the probability
for a particular parameter vector, but not in this ensemble variance.

For local-and-edge graph IQPs, Lemma~\ref{lem:graph-lightcone}
identifies $r^u$ with the structural backward light-cone size,
giving Eq.~\eqref{eq:exact-policy-variance}.
The repeated-circuit part is established separately after
Corollary~\ref{cor:repeated-letch} in
Appendix~\ref{app:proofs-repeated-trainability}; it does not follow
by assuming that noncommuting repeated blocks satisfy the IQP
rank law.
\end{proof}

\paragraph*{Proof of Proposition~\ref{prop:grouped-policy-variance}.}
\begin{proof}
\emph{Step 1: group identical expectations.}
By Lemma~\ref{lem:pauli-orthogonality}, all masks in a signature
group have the same expectation, which we denote by $L_\sigma$.
The zero-signature group is constant, $L_0=1$. Thus
\[
    \pi(a)=C_{a,0}+\sum_{\sigma\neq0}C_{a,\sigma}L_\sigma,
\]
where the sum is over signatures realized by the generator matrix.

\emph{Step 2: average and remove cross terms between groups.}
Every nonzero-signature expectation has mean zero and variance
$2^{-r_\sigma}$ by Lemma~\ref{lem:rank-law}. Distinct signatures
have zero cross expectation by Lemma~\ref{lem:pauli-orthogonality}.
Taking the mean, then expanding the centered square as above,
gives Eq.~\eqref{eq:grouped-policy-rank-variance}.
\end{proof}

This identity shows exactly when grouping is unnecessary. If
$\rank_{\mathbb F_2}\mathsf G=n$, each group contains one mask,
recovering Theorem~\ref{thm:main-trainability} for every decoder.
For example, the three generators $X_1X_2$, $X_1X_3$, and
$X_1X_2X_3$ have
\[
    \mathsf G=\begin{pmatrix}1&1&0\\1&0&1\\1&1&1\end{pmatrix},
    \qquad \rank_{\mathbb F_2}\mathsf G=3.
\]
They contain no single-qubit rotation, yet the individual-term
policy sum holds. Even if $\mathsf G$ is rank deficient, it suffices
for a particular decoder that its nonconstant terms with nonzero
coefficients have distinct nonzero signatures. Any zero-signature
terms must be assigned to the constant part, not counted as
variance contributions. More generally, the exact difference from
the sum of individual nonconstant variances consists of the
within-group cross terms
$2\sum_{\sigma\neq0}2^{-r_\sigma}
\sum_{u<v:\,\mathsf Gu=\mathsf Gv=\sigma}c_{a,u}c_{a,v}$,
where $<$ is any fixed ordering of masks. Vanishing of this total
is the exact decoder-specific condition; distinct signatures are
a simple sufficient condition that avoids cancellations.

Without this separation, a protected component must be certified
by its \emph{grouped} coefficient: a nonzero signature with
$|C_{a,\sigma}|\geq1/\poly(n)$ and $r_\sigma=O(\log n)$ suffices.
An individual $c_{a,u}$ is not enough, because other coefficients
in the same group can cancel it.

\paragraph*{Pure-$R_{XX}$ rings and the policy-variance experiment.}
For any connected graph containing only independently parameterized
edge rotations, $\mathsf Gu=0$ means $u_i=u_j$ on every edge.
Connectivity then gives
$\ker\mathsf G=\{0,\mathbf1\}$ and
$\rank_{\mathbb F_2}\mathsf G=n-1$. Hence two masks have the
same signature precisely when they are equal or complementary.
Writing $\bar u=u\oplus\mathbf1$, choose one representative from
each pair $\{u,\bar u\}$ and omit the constant pair
$\{0,\mathbf1\}$; call this representative set $\mathcal R$.
Equation~\eqref{eq:grouped-policy-rank-variance} becomes
\begin{equation}
\begin{aligned}
    \E_\theta[\pi(a)]&=c_{a,0}+c_{a,\mathbf1},\\
    \Var_\theta[\pi(a)]
      &=\sum_{u\in\mathcal R}
        (c_{a,u}+c_{a,\bar u})^2\,2^{-r^u}.
\end{aligned}
\label{eq:pure-rxx-grouped-policy-variance}
\end{equation}
This is an exact policy-variance law for any decoder, including
the contiguous, recursive-parity, and modulo policies in
Fig.~\ref{fig:pure-rxx-policy-variance}. For binary recursive parity,
only the constant pair occurs, so the variance is exactly zero.
For a decoder reading a proper subset of the qubits, no two
nonzero Fourier coefficients have complementary masks; its
nonconstant policy variance therefore reduces to an individual-term
sum even though the circuit has no local rotations.

\subsection{Proof of Corollary~\ref{cor:main-protected-component}}
\begin{proof}
Every summand in Eq.~\eqref{eq:main-policy-rank-variance} is
nonnegative.  Hence the indicated nonidentity
component alone gives
\begin{equation}
    \Var_{\theta}[\pi(a)]
    \geq c_{a,u}^2 2^{-r^u}.
\end{equation}
There are constants $p,q,C_1,C_2>0$ such that, for all sufficiently
large $n$, $|c_{a,u}|\geq C_1n^{-p}$ and
$r^u\leq C_2\log_2n+q$.  Therefore
\begin{equation}
    c_{a,u}^2 2^{-r^u}
    \geq C_1^2 2^{-q}n^{-(2p+C_2)},
\end{equation}
which is inverse polynomial.
For a local-and-edge graph IQP,
Lemma~\ref{lem:graph-lightcone} gives
$r^u=\Delta_G^{\theta}(u)$, yielding the stated
light-cone specialization.

For the grouped extension, retain the specified nonzero-signature
term in Eq.~\eqref{eq:grouped-policy-rank-variance}. The same
calculation, with $C_{a,\sigma}$ and $r_\sigma$ in place of
$c_{a,u}$ and $r^u$, gives the stated inverse-polynomial bound.

For the repeated architecture, retain the $u$ contribution in
Eq.~\eqref{eq:main-repeated-policy-bound}. Since $4^{-x}$ decreases
with $x$, its ensemble average is bounded below by its smallest
value in the auxiliary Clifford ensemble:
\begin{equation}
    \begin{aligned}
        \Var_{\theta}[\pi_R(a)]
        &\geq c_{a,u}^2
          \E_{\theta}[4^{-\Delta_R^\theta(u)}]\\
        &\geq c_{a,u}^2
          4^{-\max_{\theta}\Delta_R^\theta(u)}.
    \end{aligned}
\end{equation}
The maximum is over $\theta\in\{0,\pi/2\}^{M}$. The last expression
is inverse polynomial under the coefficient and worst-case
propagated-weight assumptions in the corollary.
\end{proof}

\subsection{Proof of neural-embedding trainability}
\begin{proof}
\emph{Step 1: a uniform bias preserves a single angle.}
For a fixed scalar feature $x$, consider
\begin{equation}
    \varphi(x)=\lambda x+\theta,
    \qquad
    \lambda,\theta\overset{\mathrm{iid}}{\sim}\Unif[-\pi,\pi].
    \label{eq:single-feature-neural-angle-proof}
\end{equation}
Conditioning on $\lambda$ makes $\lambda x$ a constant shift.
For any measurable set $\mathcal I$ on the angle circle,
\begin{align}
    &\Pr\!\left(
       \varphi(x)\bmod 2\pi\in\mathcal I
       \mid x,\lambda
    \right)\nonumber\\
    &\qquad=\frac{1}{2\pi}\int_{-\pi}^{\pi}
      \mathbf1_{\mathcal I}\!\left((\theta+\lambda x)\bmod2\pi\right)\,d\theta
    =\frac{|\mathcal I|}{2\pi},
    \label{eq:single-feature-uniform-shift-proof}
\end{align}
where $|\mathcal I|$ is its arc length. A shift preserves uniform
measure on the circle, so the effective angle remains uniform.
The independent uniform bias is essential to this argument;
the weight $\lambda$ need not itself be uniform.
Policy probabilities are $2\pi$-periodic in each Pauli-rotation
angle, since $R_P(\theta+2\pi)=-R_P(\theta)$ changes only a global
phase. To verify the moments explicitly, fix the action, observation,
and all other effective circuit angles, and write $F(t)$ for the
policy probability as a function of this one gate angle.
For $r=1$ (the mean) and $r=2$ (the second moment),
\begin{equation}
    \begin{aligned}
        &\E_\theta\!\left[F(\lambda x+\theta)^r\mid\lambda\right]\\
        &\quad=\frac{1}{2\pi}\int_{-\pi}^{\pi}
            F(\lambda x+t)^r\,dt\\
        &\quad=\frac{1}{2\pi}\int_{-\pi+\lambda x}^{\pi+\lambda x}
            F(v)^r\,dv\\
        &\quad=\frac{1}{2\pi}\int_{-\pi}^{\pi}F(v)^r\,dv.
    \end{aligned}
    \label{eq:neural-shift-moments}
\end{equation}
The second equality substitutes $v=t+\lambda x$; the third uses
the $2\pi$-periodicity of $F^r$, whose integral is the same over
any interval of length $2\pi$. The result is independent of
$\lambda$. Thus, under the stated uniform initialization of
the weight,
\[
    \begin{aligned}
        \E_{\lambda,\theta}[F(\lambda x+\theta)^r]
        &=\frac{1}{(2\pi)^2}
          \int_{-\pi}^{\pi}\!\int_{-\pi}^{\pi}
          F(\lambda x+t)^r\,dt\,d\lambda\\
        &=\frac{1}{2\pi}\int_{-\pi}^{\pi}F(v)^r\,dv
         =\E_\theta[F(\theta)^r].
    \end{aligned}
\]
Subtracting the squared mean from the second moment gives
\[
    \begin{aligned}
        &\Var_{\lambda,\theta}[F(\lambda x+\theta)]\\
        &\quad=\frac{1}{2\pi}\int_{-\pi}^{\pi}F(v)^2\,dv
          -\left(\frac{1}{2\pi}\int_{-\pi}^{\pi}F(v)\,dv\right)^2\\
        &\quad=\Var_\theta[F(\theta)].
    \end{aligned}
\]
The same conclusion holds for any independent weight distribution,
because the conditional moments in
Eq.~\eqref{eq:neural-shift-moments} do not depend on $\lambda$.

\emph{Step 2: establish joint uniformity for the neural encoder.}
Fix $s_t$ and condition on all encoding weights in
Eq.~\eqref{eq:neural-angle-general}. Each weighted feature sum
$\sum_jw_j^{(\ell)}s_{t,j}$ is now a deterministic shift.
Assume the biases are mutually independent, uniform on
$[-\pi,\pi]$, and independent of all weights and remaining circuit
parameters. Applying Step~1 to each bias gives, for measurable
sets $\mathcal I_\ell$,
\begin{align}
    &\Pr\!\left(
       \phi_\ell(s_t)\bmod2\pi\in\mathcal I_\ell
       \text{ for every }\ell
       \mid s_t,\boldsymbol w
    \right)\nonumber\\
    &\qquad=\prod_{\ell=1}^{L_{\rm enc}}
        \frac{|\mathcal I_\ell|}{2\pi}.
    \label{eq:neural-angle-uniform-shift-proof}
\end{align}
This product law does not depend on the weights. Averaging over
them therefore preserves both uniformity and independence of the
effective angles, also jointly with the other independent circuit
angles. Equivalently, apply the change of variables in
Eq.~\eqref{eq:neural-shift-moments} successively to each bias
integral: both the policy and its square have the same expectation
as under independent uniform effective angles. Averaging also over
the remaining circuit parameters preserves these two moment
identities, and therefore preserves the full initialization variance
for the same generator architecture.

\emph{Step 3: apply the variance laws to the encoded IQP circuit.}
Every encoding generator is $Z$-type, so $E_{s_t}$ commutes with
$D_Z$ and Eq.~\eqref{eq:state-conditioned-iqp} remains IQP.
Step~2 supplies the independent uniform initialization required
by Theorem~\ref{thm:main-trainability}. With its other architectural
assumptions retained, the theorem therefore applies to the combined
circuit-and-encoding generator list.
Duplicate generators combine by adding their independent angles
modulo $2\pi$; the resulting angle is still uniform and independent
of the angles associated with other generators.
Such duplicates do not change the binary rank, whereas new
generators can increase it. Thus the variance identities are
preserved in form, but an inverse-polynomial lower bound still
requires a sufficiently weighted component with logarithmic rank
in the resulting circuit.

\emph{Step 4: retain independence under repeated encoding.}
For repeated IQP-like circuits, the same argument applies if every
encoding-gate occurrence, including occurrences in different layers,
has its own independent uniform bias. These biases must also be
independent of all weights and other circuit parameters. Shared
encoding weights are harmless under this condition, since Step~2
conditions on all weights.
Alternatively, an encoding rotation may combine with a rotation
of the same generator in the same commuting layer whose angle
provides an independent uniform offset for that occurrence.
With the other randomized-layer assumptions of
Corollary~\ref{cor:repeated-letch} retained, its bound then applies
using the propagated light cone of the encoded circuit.
Reusing the same neural angle across layers without such independent
offsets does not justify this conclusion: each angle may be uniform
while the joint angles are correlated. Parameter-tied repeated
encoders therefore require a separate verification of the bound's
assumptions.

All statements are pointwise at a fixed observation $s_t$.
They neither assert independence between different observations
nor directly establish a variance bound for a trajectory-averaged
RL objective.
\end{proof}

\section{Proofs for Repeated-IQP Trainability}
\label{app:proofs-repeated-trainability}

\subsection{Proof of Corollary~\ref{cor:repeated-letch}}
\begin{proof}
We verify each condition needed to specialize Theorem~1 of
Ref.~\cite{Letcher2024Tight}, using the same rotation convention
$R_P(\theta)=\exp(-i\theta P/2)$.  First, factor the first diagonal layer
as $D_Z^{(1)}=D_{Z,\mathrm{rest}}^{(1)}D_{Z,\mathrm{loc}}^{(1)}$,
where the second factor contains one local $Z_i$ rotation per qubit and
the first contains all remaining, possibly higher-body, $Z$-type
rotations.  All generators commute, so this reordering leaves the
unitary unchanged.  Applied from right to left, the circuit therefore
begins with the one-qubit $X$ rotations in $D_X^{(0)}$, followed by the
one-qubit $Z$ rotations in $D_{Z,\mathrm{loc}}^{(1)}$.  On each qubit
these two Pauli axes anticommute, so they are the two orthogonal local
rotation layers required by that theorem.  Every remaining gate in
$C_R$ is itself a parameterized Pauli rotation; identity separators may
be inserted between such rotations.  Finally, all angles
are mutually independent and uniform over a full period, as required.

For the product input $\rho_0=|0^n\rangle\langle0^n|$, the first local
axis is $X$ on every qubit.  Since
$\operatorname{Tr}(X|0\rangle\langle0|)=0$, the state-orthogonality
factor appearing in the lower bound of Ref.~\cite{Letcher2024Tight}
is one.  Specialize its Pauli-observable statement to $P_\alpha=Z^u$.
At a Clifford-restricted parameter vector
$\boldsymbol\theta\in\{0,\pi/2\}^M$, Pauli propagation maps $Z^u$ to a
Pauli word, and the number of nonidentity qubits in that propagated word
is $\Delta_R^{\theta}(u)$.  The loss-variance statement in that theorem
then gives, term by term,
\begin{equation}
    4^{-\Delta_R^{\theta}(u)}
    \quad\text{and}\quad
    2^{-\Delta_R^{\theta}(u)}
\end{equation}
as the lower and upper discrete contributions, respectively. Because
the discrete coordinates are independent and satisfy
$\theta_j\sim\operatorname{Unif}(\{0,\pi/2\})$, averaging these
contributions gives Eq.~\eqref{eq:repeated-letch-bound-paper}.

Finally, linearity gives
$\pi_R(a)=\sum_uc_{a,u}L_u^{(R)}$.  The first identity in Theorem~1 of
Ref.~\cite{Letcher2024Tight} states that distinct Pauli components
contribute additively to the loss variance; equivalently, all
coefficient-weighted cross covariances vanish after averaging over the
continuous initialization ensemble. Thus
\begin{equation}
    \Var_{\theta}[\pi_R(a)]
    =\sum_uc_{a,u}^2\Var_{\theta}[L_u^{(R)}].
\end{equation}
The $u=0$ component is deterministic because $L_0^{(R)}=1$, yielding
Eq.~\eqref{eq:repeated-policy-var-decomposition}.  Applying the lower
bound in Eq.~\eqref{eq:repeated-letch-bound-paper} to each remaining
nonnegative term yields
Eq.~\eqref{eq:repeated-policy-lightcone-bound-paper}.
\end{proof}

\subsection{Completion of the proof of Theorem~\ref{thm:main-trainability}}
\begin{proof}
The IQP equalities were proved in
Appendix~\ref{app:proofs-single-block}.  For the repeated circuit, the
preceding derivation gives
\begin{align}
    \Var_{\theta}[\pi_R(a)]
    &=\sum_{u\neq0}c_{a,u}^2\Var_{\theta}[L_u^{(R)}]
      \nonumber\\
    &\geq\sum_{u\neq0}c_{a,u}^2
      \E_{\theta}
      [4^{-\Delta_R^{\theta}(u)}],
\end{align}
which is Eq.~\eqref{eq:main-repeated-policy-bound}.  This proves the
remaining assertion of Theorem~\ref{thm:main-trainability}.
\end{proof}

\section{Proofs of the Classical-Simulability Results}
\label{app:proofs-classical}

\subsection{Proof of Proposition~\ref{prop:van-den-nest-iqp-estimator}}
\label{app:van-den-nest-estimator}
We give the IQP specialization of Van den Nest's estimator
\cite{VanDenNest2011}.
The rotation convention here is $e^{-i\theta_\ell P_\ell/2}$ rather
than $e^{i\theta_\ell P_\ell}$, so the phase below includes a factor
of $-1/2$. Circuit parameters are fixed throughout this proof:
all expectations and variances below concern the classical random
samples, not parameter initialization.

\begin{proof}
Write $P_\ell=X^{g_\ell}$, where $g_\ell\in\{0,1\}^n$ is the
generator's binary support. Conjugation by Hadamards gives the
diagonal circuit
\begin{equation}
    D_Z:=H^{\otimes n}U_XH^{\otimes n}
    =\prod_{\ell=1}^{m}e^{-i\theta_\ell Z^{g_\ell}/2}.
\end{equation}
Since $H^{\otimes n}|0^n\rangle=|+\rangle^{\otimes n}$ and
$H^{\otimes n}Z^uH^{\otimes n}=X^u$,
\begin{equation}
    L_u(\boldsymbol\theta)
    =\langle+|^{\otimes n}D_Z^\dagger X^uD_Z|+\rangle^{\otimes n}.
\end{equation}
Each computational-basis vector is an eigenvector of $D_Z$, with
an efficiently computable phase:
\begin{equation}
    D_Z|z\rangle=\lambda_z|z\rangle,
    \qquad
    \lambda_z=\exp\!\left[-\frac{i}{2}
        \sum_{\ell=1}^{m}\theta_\ell(-1)^{g_\ell\cdot z}\right].
\end{equation}
Expanding $|+\rangle^{\otimes n}=2^{-n/2}\sum_z|z\rangle$ and
using $X^u|z\rangle=|z\oplus u\rangle$ therefore yields
\begin{equation}
    \begin{aligned}
        L_u(\boldsymbol\theta)
        &=2^{-n}\sum_z\lambda_{z\oplus u}^*\lambda_z\\
        &=2^{-n}\sum_z\exp\!\left[-\frac{i}{2}
          \sum_{\ell=1}^{m}\theta_\ell(-1)^{g_\ell\cdot z}
          \bigl(1-(-1)^{g_\ell\cdot u}\bigr)\right].
    \end{aligned}
    \label{eq:classical-estimator-complex-form}
\end{equation}
Here binary inner products are evaluated modulo two. Because $L_u$
is real, taking the real part leaves the sum unchanged. The factor
$1-(-1)^{g_\ell\cdot u}$ equals zero for commuting generators and
two for anticommuting generators. Hence, for a uniformly sampled
$z\in\{0,1\}^n$, define
\begin{equation}
    Y_u(z):=\cos\!\left[
        \sum_{\ell:g_\ell\cdot u=1}
        \theta_\ell(-1)^{g_\ell\cdot z}\right].
    \label{eq:classical-iqp-estimator}
\end{equation}
The evenness of cosine removes the minus sign, giving
\begin{equation}
    L_u(\boldsymbol\theta)
    =2^{-n}\sum_zY_u(z)=\E_z[Y_u(z)],
    \qquad |Y_u(z)|\leq1.
    \label{eq:classical-estimator-unbiased-identity}
\end{equation}
The exponentially large sum need not be enumerated.
Draw independent uniformly random bit strings
$z^{(1)},\ldots,z^{(N)}\in\{0,1\}^n$ and define
\begin{equation}
    \widehat L_u
    :=
    \frac1N\sum_{r=1}^{N}Y_u(z^{(r)}),
\end{equation}
with $Y_u$ from Eq.~\eqref{eq:classical-iqp-estimator}.  By
Eq.~\eqref{eq:classical-estimator-unbiased-identity} and linearity,
\begin{equation}
    \E_{z}[\widehat L_u]
    =\frac1N\sum_{r=1}^N\E_{z}[Y_u(z^{(r)})]
    =L_u(\boldsymbol\theta),
\end{equation}
so the estimator is unbiased.  Independence of the samples gives
\begin{equation}
    \Var_{z}[\widehat L_u]
    =
    \frac1{N^2}\sum_{r=1}^N\Var_{z}[Y_u(z^{(r)})]
    =\frac{\Var_{z}[Y_u]}{N}\leq\frac1N,
\end{equation}
where $\Var_{z}[Y_u]\leq\E_{z}[Y_u^2]\leq1$.  Thus any
$N>\epsilon^{-2}$ ensures
$\operatorname{Std}(\widehat L_u)<\epsilon$, proving the proposition's
stated standard-deviation guarantee.

For completeness, the usual high-probability version follows from the
same bounded samples.  Since every $Y_u$ lies in $[-1,1]$, Hoeffding's
inequality gives
\begin{equation}
    \Pr\!\left(
        |\widehat L_u-L_u|>\epsilon
    \right)
    \leq 2e^{-N\epsilon^2/2}.
\end{equation}
Thus
$N\geq2\epsilon^{-2}\ln(2/\delta)$ suffices for additive error
$\epsilon$ with probability at least $1-\delta$.  Each realization
requires $m=\poly(n)$ binary inner products, sign evaluations, and phase
additions in Eq.~\eqref{eq:classical-iqp-estimator}.  Its time is
$O(mn)$ with a direct implementation and its working memory is
polynomial. Thus the total arithmetic cost is
$O(mn\epsilon^{-2}\log(2/\delta))$; samples can be processed
sequentially without storing the full batch. With efficiently specified
angles, evaluating each cosine to error at most $\epsilon/2$ and
using sampling tolerance $\epsilon/2$ preserves the additive-error
guarantee with polynomial bit complexity. The exact-arithmetic
estimator above is unbiased; finite numerical precision adds only
this controlled deterministic error. This proves polynomial classical
time and space for inverse-polynomial accuracy, without sampling
the circuit's Born distribution.
\end{proof}

\subsection{Sparse Boolean post-processing and explicit Pauli sums}
\label{app:sparse-postprocessing-explanation}

For an efficiently evaluable Boolean decoder
$h:\{0,1\}^n\to\{0,1\}$ and circuit output $X$, its Walsh expansion gives
\begin{equation}
    \Pr[h(X)=1]=\sum_u\widehat h(u)L_u,
    \qquad L_u=\E[(-1)^{u\cdot X}].
    \label{eq:appendix-postprocessing-pauli-sum}
\end{equation}
If the polynomially many nonzero terms are efficiently available,
one can estimate each $L_u$ and sum them with controlled error, as
proved in Proposition~\ref{prop:pauli-sparse-policy} below.

Takahashi \emph{et al.} remove the need to supply that expansion
when a polynomial upper bound $S$ on the exact Fourier sparsity is
known~\cite{TakahashiMiyamotoKunihiro2026}. Their
Kushilevitz--Mansour-based procedure uses evaluations of $h$ to identify
the coefficients needed at the requested accuracy and combines their
estimates with estimates of $L_u$. The relevant truncation bound is
elementary: omitting coefficients of magnitude at most $\tau$ changes
the sum by at most $S\tau$, since $|L_u|\leq1$. Taking
$\tau=O(\epsilon/S)$ and controlling the remaining estimation errors
gives inverse-polynomial additive accuracy in polynomial time under
their assumptions. This estimates the decoded probability without
sampling the full circuit output.

For our contiguous, basis-modulo, and recursive-parity decoders with
$A=2^k=\poly(n)$, the $A$ Pauli terms are explicitly known, so this
support-discovery procedure is unnecessary. General contiguous
intervals instead use the separate estimator in
Appendix~\ref{app:arbitrary-contiguous-sim}. Exponential exact Fourier
support places the majority families outside the sparse-postprocessing
guarantee, but does not establish classical hardness.

\subsection{Proof of Proposition~\ref{prop:pauli-sparse-policy}}
\begin{proposition}[Pauli-sparse policy evaluation]
\label{prop:pauli-sparse-policy}
Suppose an efficiently computable action projector has the expansion
\begin{equation}
    \Pi_a=\sum_{j=1}^{M}c_jZ^{u_j},
\end{equation}
where $M=\poly(n)$ and $\sum_j|c_j|=\poly(n)$. Then
$\pi(a)=\langle\Pi_a\rangle$ can be estimated to inverse-polynomial
additive precision with high probability in randomized classical
polynomial time.
\end{proposition}
\begin{proof}
Let $C_1:=\sum_{j=1}^M|c_j|$.  The case $C_1=0$ is trivial, so assume
$C_1>0$.  For each $j$, use independent copies of the bounded estimator
in Eq.~\eqref{eq:classical-iqp-estimator} to construct
$\widehat L_{u_j}$ satisfying
\begin{equation}
    |\widehat L_{u_j}-L_{u_j}|\leq\epsilon/C_1
    \label{eq:sparse-proof-individual-error}
\end{equation}
with failure probability at most $\delta/M$.  The preceding Hoeffding
calculation shows that
\begin{equation}
    N_j\geq
    2C_1^2\epsilon^{-2}\ln(2M/\delta)
\end{equation}
samples suffice for each term.  A union bound makes
Eq.~\eqref{eq:sparse-proof-individual-error} hold simultaneously for all
$M$ terms with probability at least $1-\delta$.  On this event, the
weighted estimator $\widehat\pi(a):=\sum_jc_j\widehat L_{u_j}$ obeys
\begin{align}
    |\widehat\pi(a)-\pi(a)|
    &\leq\sum_j|c_j|
      |\widehat L_{u_j}-L_{u_j}|\nonumber\\
    &\leq C_1(\epsilon/C_1)=\epsilon.
\end{align}
Because $M$, $C_1$, and the cost of computing every listed coefficient
and mask are polynomial, the total time and space are polynomial in
$n$, $\epsilon^{-1}$, and $\log\delta^{-1}$.
\end{proof}

\subsection{Proof of Corollary~\ref{cor:power-two-classical}}
\begin{proof}
For a contiguous or basis-modulo decoder with $A=2^k$, the action is
fixed by $k$ selected bits.  Expanding their $k$ one-bit projectors gives
$2^k=A$ distinct Pauli strings, each with coefficient magnitude $2^{-k}
=1/A$.  For recursive parity, expanding the $k$ commuting parity-bit
projectors gives the same count and magnitudes; distinct subsets of the
nested parity strings give distinct products because those strings are
linearly independent over $\mathbb F_2$.  In all three cases the masks
and signs are computable in polynomial time and
\begin{equation}
    \sum_{j=1}^{A}|c_j|=A(1/A)=1.
\end{equation}
Since $A=\poly(n)$, every hypothesis of
Proposition~\ref{prop:pauli-sparse-policy} holds.
\end{proof}

\subsection{Arbitrary contiguous action partitions}
\label{app:arbitrary-contiguous-sim}

The following result avoids enumerating the generally exponential Walsh
expansion of a contiguous interval projector.

\begin{theorem}[Classical simulation of arbitrary contiguous policies]
\label{thm:arbitrary-contiguous-sim}
Let the action bins be contiguous intervals with efficiently computable
boundaries. Every IQP action probability can be estimated
to additive error $\epsilon$ and failure probability $\delta$ in
randomized time
\begin{equation}
    \poly\!\left(n,\epsilon^{-1},\log\delta^{-1}\right).
\end{equation}
More explicitly, one action probability requires
$O(n^2\epsilon^{-2}\log\delta^{-1})$ samples of the bounded estimator
in Eq.~\eqref{eq:classical-iqp-estimator}.  The result does not require
$A$ to be a power of two.
\end{theorem}

\paragraph{Step 1: reduce an interval to two thresholds.}
Identify $x=x_1\cdots x_n$ with
$\operatorname{int}(x)=\sum_{i=1}^n2^{n-i}x_i$ and set
$F_\theta(t)=\Pr[\operatorname{int}(x)<t]$.  For
$I_a=[L_a,U_a)$,
\begin{equation}
    \pi(a)=F_\theta(U_a)-F_\theta(L_a).
    \label{eq:interval-cdf}
\end{equation}

\paragraph{Step 2: decompose a threshold into disjoint prefixes.}
If $t=t_1\cdots t_n$ is the binary threshold, then
\begin{equation}
    F_\theta(t)=\sum_{j:t_j=1}p_{j,t},
    \label{eq:prefix-decomp-paper}
\end{equation}
where the relevant prefix projector is
\begin{equation}
    Q_{j,t}
    =
    \left(\prod_{i=1}^{j-1}\frac{I+(-1)^{t_i}Z_i}{2}\right)
    \frac{I+Z_j}{2}.
\end{equation}

\paragraph{Step 3: sample, rather than enumerate, the Pauli expansion.}
Although $Q_{j,t}$ has $2^j$ Pauli terms, its probability can be written
as
\begin{equation}
    p_{j,t}
    =
    \E_{S\sim\Unif(2^{[j]})}
    \left[\chi_{j,t}(S)\langle Z_S\rangle\right].
    \label{eq:stochastic-prefix}
\end{equation}
\begin{proof}
The boundary cases $F_\theta(0)=0$ and $F_\theta(2^n)=1$ are known
exactly, so consider $1\leq t<2^n$ with binary digits
$t_1\cdots t_n$.  A string $x$ represents an integer below $t$ if and
only if, at the first position $j$ where $x$ differs from $t$, one has
$t_j=1$ and $x_j=0$.  The earlier bits must satisfy $x_i=t_i$ for
$i<j$.  These first-difference events are disjoint and are indexed by
the at most $n$ positions for which $t_j=1$, proving
Eq.~\eqref{eq:prefix-decomp-paper}.  The displayed operator $Q_{j,t}$ is
their projector because $(I+(-1)^{t_i}Z_i)/2$ projects onto bit $t_i$
and $(I+Z_j)/2$ projects onto bit zero.

Expanding this product gives
\begin{equation}
    Q_{j,t}=2^{-j}\sum_{S\subseteq[j]}\chi_{j,t}(S)Z_S,
\end{equation}
where, on defining $b_i=t_i$ for $i<j$ and $b_j=0$,
\begin{equation}
    \chi_{j,t}(S)
    =(-1)^{\sum_{i\in S}b_i}\in\{-1,+1\}.
\end{equation}
Thus the apparently exponential sum is an expectation over a uniformly
random subset $S\subseteq[j]$.  Taking the circuit expectation proves
Eq.~\eqref{eq:stochastic-prefix}.

An implementable single-sample estimator for $p_{j,t}$ is obtained by
drawing $S\sim\Unif(2^{[j]})$, independently drawing the bit string $z$
used in Eq.~\eqref{eq:classical-iqp-estimator}, and returning
\begin{equation}
    W_{j,t}:=\chi_{j,t}(S)Y_S(z).
\end{equation}
Iterated expectation together with
Eq.~\eqref{eq:classical-estimator-unbiased-identity} gives
$\E[W_{j,t}]=p_{j,t}$, while $|W_{j,t}|\leq1$ because
$|Y_S(z)|\leq1$.  This establishes an efficient bounded estimator for
one prefix probability without listing its $2^j$ Pauli terms.

\paragraph{Step 4: estimate the whole interval in one Monte Carlo
average.}
Collect all nonconstant prefix contributions in
$F_\theta(U_a)-F_\theta(L_a)$ into a signed list
$\mathcal R_a$.  A prefix from $U_a$ has sign $+1$, whereas a prefix
from $L_a$ has sign $-1$.  Endpoint values $F_\theta(0)$ and
$F_\theta(2^n)$ contribute a known constant $c_a$.  If
$K_a:=|\mathcal R_a|$, then $K_a\leq2n$ and
\begin{equation}
    \pi(a)
    =c_a+\sum_{r\in\mathcal R_a}\sigma_r p_r.
\end{equation}
When $K_a>0$, choose $r$ uniformly from $\mathcal R_a$, generate the
corresponding bounded prefix sample $W_r$, and return
\begin{equation}
    V_a:=c_a+K_a\sigma_rW_r.
\end{equation}
Uniform sampling of $r$ gives $\E[V_a]=\pi(a)$.  Moreover, the random
part lies in $[-K_a,K_a]$.  Hence Hoeffding's inequality shows that the
mean of
\begin{equation}
    N
    \geq
    2K_a^2\epsilon^{-2}\ln(2/\delta)
    =O\!\left(n^2\epsilon^{-2}\log\delta^{-1}\right)
    \label{eq:contiguous-estimator-sample-complexity}
\end{equation}
independent samples has additive error at most $\epsilon$ with
probability at least $1-\delta$.  If $K_a=0$, the probability is the
known constant $c_a$.  Generating $r$, $S$, and $z$ is polynomial-time,
and a direct evaluation of $Y_S(z)$ for an IQP circuit with $m$
rotations takes $O(mn)$ bit operations.  The resulting nonoptimized
classical runtime is therefore
$O(mn^3\epsilon^{-2}\log\delta^{-1})$, completing the proof.
\end{proof}

\subsection{Proof of Theorem~\ref{thm:main-simulability}}
\begin{proof}
Proposition~\ref{prop:van-den-nest-iqp-estimator}, including the explicit
construction in Eq.~\eqref{eq:classical-iqp-estimator}, proves the first
claim: every diagonal Pauli expectation of a polynomial-size single IQP
block has a randomized inverse-polynomial additive estimator using
polynomial resources.

For route~(i), write the action projector as
$\Pi_a=\sum_{j=1}^{M}c_jZ^{u_j}$.  The assumed polynomial term count,
polynomial coefficient $\ell_1$ norm, and efficient computability are
exactly the hypotheses of Proposition~\ref{prop:pauli-sparse-policy},
which constructs an additive estimator for
$\pi(a)=\sum_jc_jL_{u_j}$.

For route~(ii), let $B_a=[L_a,U_a)$ be a contiguous bin.  The proof of
Theorem~\ref{thm:arbitrary-contiguous-sim} constructs bounded unbiased
estimators for both threshold probabilities $F_\theta(L_a)$ and
$F_\theta(U_a)$ without enumerating the generally exponential Walsh
expansion of the interval projector.  Their difference is $\pi(a)$ and
has the requested additive accuracy and failure probability with
polynomial runtime.  The two routes establish all assertions of the
theorem for action probabilities.  Finally, if $\theta_j$ occurs in one
gate $e^{-i\theta_jP_j/2}$, then $P_j^2=I$ and the parameter-shift rule
gives
\begin{equation}
    \partial_{\theta_j}\pi(a)
    =\frac12\left[
      \pi_{\boldsymbol\theta+(\pi/2)e_j}(a)
      -\pi_{\boldsymbol\theta-(\pi/2)e_j}(a)
    \right].
\end{equation}
Both shifted circuits remain IQP circuits. Estimating each
probability to additive error $\epsilon$ therefore estimates the
derivative to additive error at most $\epsilon$ with only a constant
factor overhead.  Division by $\pi(a)$ is not controlled when that
probability is exponentially small, which proves the stated limitation.
\end{proof}

\section{Proofs of the Majority--Modulo Results}
\label{app:proofs-majority}

\subsection{Supplementary majority--modulo examples}
\label{app:supplementary-majority-examples}

\subsubsection{Complete five-qubit expansion}

Expanding the grouped expression in
Eq.~\eqref{eq:majmod-five-compact} term by term gives
\begin{align}
    \mathcal M_5
    ={}&\frac38(Z_1+Z_2+Z_3+Z_4+Z_5)\nonumber\\
      &-\frac18\bigl(
        Z_1Z_2Z_3+Z_1Z_2Z_4\nonumber\\
      &\hspace{0.8cm}+Z_1Z_2Z_5+Z_1Z_3Z_4\nonumber\\
      &\hspace{0.8cm}+Z_1Z_3Z_5+Z_1Z_4Z_5\nonumber\\
      &\hspace{0.8cm}+Z_2Z_3Z_4+Z_2Z_3Z_5\nonumber\\
      &\hspace{0.8cm}+Z_2Z_4Z_5+Z_3Z_4Z_5\bigr)\nonumber\\
      &+\frac38Z_1Z_2Z_3Z_4Z_5.
    \label{eq:majmod-five-explicit}
\end{align}

\subsubsection{Four actions without modulo collisions}

For $A=4$, use the two three-qubit blocks from
Eq.~\eqref{eq:majmod-three-bit-blocks}.  Here $k=2$, $C=4$, and $A=C$,
so modulo reduction is the identity on the four majority codes:
\begin{equation}
    00\mapsto0,\qquad 01\mapsto1,\qquad
    10\mapsto2,\qquad 11\mapsto3.
    \label{eq:majmod-four-action-code-map}
\end{equation}
Every action has exactly one codeword, and hence
\begin{align}
    \Pi_0^{\rm MM}&=Q_{00}
      =\tfrac14(I+\mathcal M_1)(I+\mathcal M_2),\nonumber\\
    \Pi_1^{\rm MM}&=Q_{01}
      =\tfrac14(I+\mathcal M_1)(I-\mathcal M_2),\nonumber\\
    \Pi_2^{\rm MM}&=Q_{10}
      =\tfrac14(I-\mathcal M_1)(I+\mathcal M_2),\nonumber\\
    \Pi_3^{\rm MM}&=Q_{11}
      =\tfrac14(I-\mathcal M_1)(I-\mathcal M_2).
    \label{eq:majmod-four-action-projectors}
\end{align}
If action $a$ has binary code $y_1y_2$, substituting the 1-local
sectors of $\mathcal M_1$ and $\mathcal M_2$ gives
\begin{align}
    \Pi_a^{\rm MM}
    ={}&\frac14 I
      +\frac18\left[
        (-1)^{y_1}\sum_{i\in V_1}Z_i
        +(-1)^{y_2}\sum_{j\in V_2}Z_j\right]\nonumber\\
      &+\frac{(-1)^{y_1+y_2}}{16}
        \sum_{\substack{i\in V_1\\j\in V_2}}Z_iZ_j
        +\text{higher-weight terms}.
    \label{eq:majmod-four-action-lowweight}
\end{align}
Unlike the $A=3$ collision in the main text, no standalone block
observable cancels.  Every action therefore retains 1-local terms from
both blocks, together with 2-local cross-block terms and higher-weight
majority correlations.

\subsubsection{Eight-qubit, three-action policy with even blocks}

Let $A=3$, $n=8$, $k=2$, and choose
$V_1=\{1,2,3,4\}$ and $V_2=\{5,6,7,8\}$.  Each block uses
$\mathcal M_4^{(+)}$ from Eq.~\eqref{eq:majmod-even-four-example}; its
six tied strings are assigned majority bit $0$.  The two majority bits
generate four codes, and modulo reduction gives
\begin{equation}
    00,11\mapsto0,
    \qquad 01\mapsto1,
    \qquad 10\mapsto2.
    \label{eq:eight-qubit-even-code-map}
\end{equation}
Thus $00$ and $11$ collide because $2^k>A$; this code collision is
separate from the ties within either four-qubit block.  In particular,
\begin{equation}
    \Pi_0^{\rm MM}
    =Q_{00}+Q_{11}
    =\frac12\left(I+\mathcal M_1^{(+)}\mathcal M_2^{(+)}\right).
    \label{eq:eight-qubit-even-collided-projector}
\end{equation}
The standalone block observables cancel, as they do for odd blocks.
However, writing their identity and 1-local coefficients as
$\mu_4=\eta_4^{(+)}=3/8$ gives
\begin{align}
    \Pi_0^{\rm MM}
    ={}&\left(\frac12+\frac{\mu_4^2}{2}\right)I
       +\frac{\mu_4\eta_4^{(+)}}{2}
        \sum_{i=1}^{8}Z_i
       +\cdots\nonumber\\
    ={}&\frac{73}{128}I+\frac{9}{128}\sum_{i=1}^{8}Z_i
       +\cdots.
    \label{eq:eight-qubit-even-lowweight}
\end{align}
The nonzero identity component created by deterministic ties therefore
restores physical 1-local terms inside the surviving product.  For a
uniform computational-basis distribution, the action masses are
$73/128$, $55/256$, and $55/256$.  Moreover, every Pauli coefficient of
each four-qubit block is nonzero, so all three action projectors in this
example have the complete $2^8=256$-string Pauli support.

\subsection{Numerical validation of block-majority Fourier coefficients}
\label{app:majority-fourier-validation}

\begin{figure*}[!t]
    \centering
    \includegraphics[width=\textwidth]{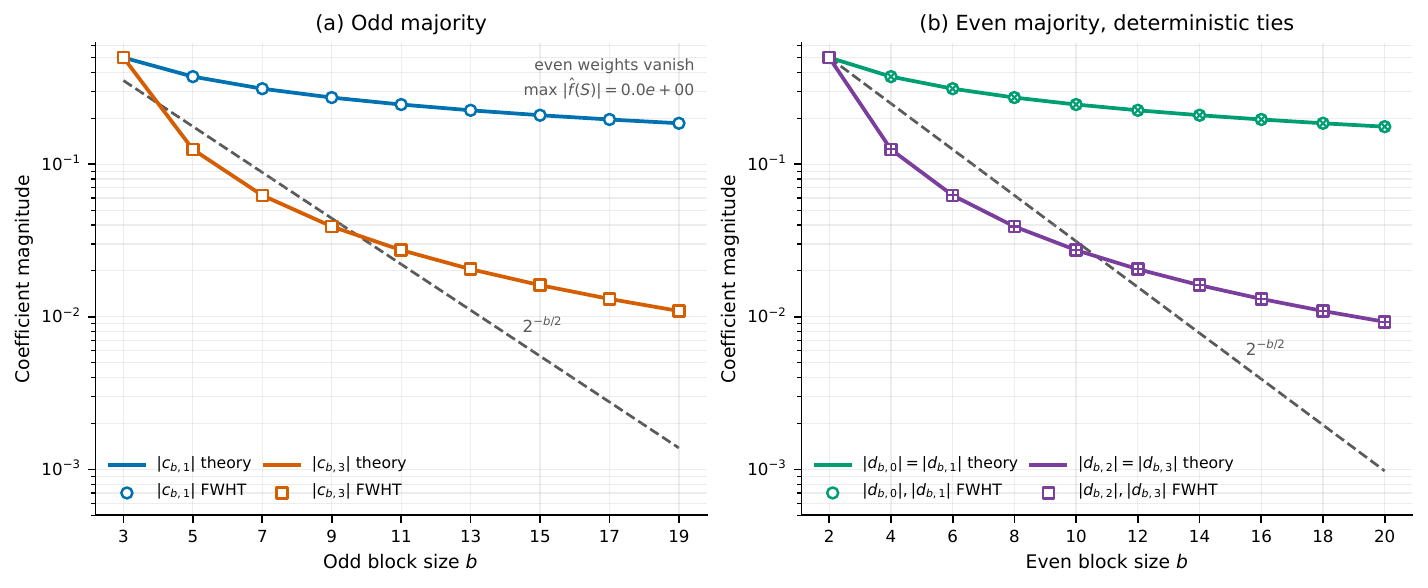}
    \caption{Exact numerical validation of the block-majority
    Walsh--Pauli coefficients.  Markers show complete truth-table
    enumeration followed by a normalized fast Walsh--Hadamard transform;
    solid lines show the analytical laws, and the dashed gray curve is
    the amplitude-scale exponential reference $2^{-b/2}$.  The ordinate
    is coefficient magnitude, so this reference corresponds after
    squaring to a $2^{-b}$ variance or power scale.  (a) For odd blocks, the
    1-local and 3-local sectors satisfy
    $c_{b,1}=\eta_b$ and $c_{b,3}=-\eta_b/(b-2)$, while every
    even-weight coefficient vanishes.  (b) For even blocks with ties
    assigned majority bit zero,
    $d_{b,0}=d_{b,1}=\binom b{b/2}/2^b$ and
    $d_{b,2}=d_{b,3}=-[\binom b{b/2}/2^b]/(b-1)$.
    Across all $2{,}097{,}148$ enumerated coefficients, the maximum
    complete-spectrum discrepancy was $1.0842\times10^{-19}$.}
    \label{fig:majority-block-fourier-validation}
\end{figure*}

We independently verified all branches of
Eq.~\eqref{eq:majority-general-weight-coefficient} by enumerating every
input of each majority function and applying a normalized fast
Walsh--Hadamard transform.  The calculation covers odd block sizes
$b=3,5,\ldots,19$ and even sizes $b=2,4,\ldots,20$, with ties in the
even case assigned majority sign $+1$.  All $2{,}097{,}148$ resulting
coefficients were compared with their closed forms, not only the local
sectors displayed in Fig.~\ref{fig:majority-block-fourier-validation}.
The maximum absolute discrepancy was
$1.0842\times10^{-19}$, below the prespecified tolerance $10^{-12}$,
and the maximum variation among coefficients of equal Pauli weight was
zero.  Figure~\ref{fig:majority-block-fourier-validation} also compares
the local coefficient magnitudes with $2^{-b/2}$.  Because the ordinate
is a coefficient amplitude, this dashed reference is the amplitude-scale
counterpart of a $2^{-b}$ variance or power law.  The analytical local
coefficients decay only polynomially and therefore remain parametrically
larger as $b$ grows.

\subsection{Proof of Lemma~\ref{lem:majmod-preimages-paper}}
\begin{lemma}[Majority--modulo preimages]
\label{lem:majmod-preimages-paper}
Let $R=C-A$. Then
\begin{equation}
    |\mathcal R_a|
    =
    \begin{cases}
        2, & 0\leq a<R,\\[1mm]
        1, & R\leq a<A.
    \end{cases}
\end{equation}
For a two-preimage action, the corresponding integers are $a$ and
$a+A$.
\end{lemma}
\begin{proof}
Every integer $r\in\{0,\ldots,C-1\}$ with residue $a$ has the form
$r=a+qA$ for an integer $q\geq0$.  Since
$C=2^{\lceil\log_2A\rceil}\leq2A$, one has $r<2A$ and hence
$q\in\{0,1\}$.  The value $q=0$ always gives the representative $a$.
The value $q=1$ gives a second representative $a+A$ exactly when
\begin{equation}
    a+A<C
    \quad\Longleftrightarrow\quad
    a<C-A=R.
\end{equation}
Thus actions $0\leq a<R$ have two preimages and actions
$R\leq a<A$ have one.  If $A$ is a power of two, then $C=A$ and
$R=0$, so the same formula correctly says that every action has one
preimage.
\end{proof}

\subsection{Proof of Theorem~\ref{thm:majmod-collision-paper}}
\begin{theorem}[Collided majority--modulo projector]
\label{thm:majmod-collision-paper}
For a majority codeword $y\in\{0,1\}^k$, define
$\mathcal M_T:=\prod_{r\in T}\mathcal M_r$ and
$\mathcal M_\varnothing=I$. Its projector is
\begin{equation}
    Q_y
    =\prod_{r=1}^{k}\frac{I+(-1)^{y_r}\mathcal M_r}{2}
    =\frac1C\sum_{T\subseteq[k]}(-1)^{T\cdot y}\mathcal M_T.
    \label{eq:majmod-micro-Walsh}
\end{equation}
For a two-preimage action with codewords $y,y'$ and changed-block set
$D=\{r:y_r\neq y'_r\}$,
\begin{equation}
    \Pi_a^{\mathrm{MM}}
    =
    \frac{2}{C}
    \sum_{\substack{T\subseteq[k]\\|T\cap D|\ {\rm even}}}
    (-1)^{T\cdot y}\mathcal M_T.
    \label{eq:majmod-collision-projector-paper}
\end{equation}
If $A$ is not a power of two, then $|D|\geq2$.
\end{theorem}
\begin{proof}
The action projector is the sum of the two mutually orthogonal code
projectors, $\Pi_a^{\mathrm{MM}}=Q_y+Q_{y'}$.  Applying
Eq.~\eqref{eq:majmod-micro-Walsh} to both gives
\begin{align}
    Q_y+Q_{y'}
    &=
    \frac1C
    \sum_T
    (-1)^{T\cdot y}
    \left[
        1+
        (-1)^{T\cdot(y\oplus y')}
    \right]
    \mathcal M_T.
\end{align}
Here all dot products in exponents are modulo two, and
\begin{equation}
    T\cdot(y\oplus y')
    =\sum_{r\in T}\mathbf1[y_r\neq y'_r]
    =|T\cap D|\pmod2.
\end{equation}
The bracket is consequently $2$ for even $|T\cap D|$ and zero for odd
$|T\cap D|$, proving the displayed projector formula.

The two integer representatives are $a$ and $a+A$, so their numerical
difference is $A$.  If $|D|=1$, their binary codewords differ in exactly
one bit, of place value $2^s$ for some $s\in\{0,\ldots,k-1\}$.  Their
integer difference would then be $2^s$, forcing $A=2^s$.  Therefore a
non-power-of-two $A$ cannot have $|D|=1$.  Since $y\neq y'$, $D$ is
nonempty, and hence $|D|\geq2$.
\end{proof}

\subsection{Proof of Proposition~\ref{prop:majmod-fourier-support-paper}}
\begin{proof}
Let $s_b:=2^{b-1}$ be the number of nonzero Pauli strings in one
odd-block majority observable, as established in
Eq.~\eqref{eq:majority-support-paper}.  Because the blocks are disjoint,
a choice of one nonidentity majority string on each selected block gives
a unique global Pauli support.  Terms belonging to different block
subsets or different within-block supports therefore cannot cancel.

For a one-preimage action, every $T\subseteq[k]$ occurs in
Eq.~\eqref{eq:majmod-micro-Walsh}.  Each block independently contributes
either the identity or one of its $s_b$ nonzero strings, so
\begin{equation}
    N_{\mathrm{Pauli}}^{(1)}=(1+s_b)^k.
    \label{eq:majmod-support-one-paper}
\end{equation}

For a collided action, let $h:=|D|$.  Among the $h$ changed blocks, an
even number $q$ must be selected; each selected block contributes one of
$s_b$ nonzero strings.  The other $k-h$ blocks are unrestricted.  Hence
\begin{align}
    N_{\mathrm{Pauli}}^{(2)}
    &=
    (1+s_b)^{k-h}
    \sum_{\substack{
        q=0\\
        q\ {\rm even}
    }}^h
    \binom hq s_b^q
    \\
    &=
    \frac12
    (1+s_b)^{k-h}
    \left[
        (1+s_b)^h+(1-s_b)^h
    \right].
    \label{eq:majmod-support-two-paper}
\end{align}
where the final equality follows by adding the binomial expansions of
$(1+s_b)^h$ and $(1-s_b)^h$, which removes every odd power.

It remains to turn these exact counts into the claimed asymptotic lower
bound.  Since $A=\poly(n)$, $k=\lceil\log_2A\rceil=O(\log n)$ and
$kb=n$.  In the one-preimage case,
\begin{equation}
    N_{\mathrm{Pauli}}^{(1)}
    \geq s_b^k=2^{k(b-1)}=2^{n-k}=2^{\Omega(n)}.
\end{equation}
For a collided action, Theorem~\ref{thm:majmod-collision-paper} gives
$h\geq2$.  If $h$ is even, the $q=h$ summand alone and the unrestricted
blocks give
\begin{equation}
    N_{\mathrm{Pauli}}^{(2)}
    \geq s_b^h s_b^{k-h}=s_b^k=2^{n-k}.
\end{equation}
If $h$ is odd, then $h\geq3$; retaining $q=h-1$ gives
\begin{equation}
    N_{\mathrm{Pauli}}^{(2)}
    \geq h s_b^{h-1}s_b^{k-h}
    =h\,2^{(k-1)(b-1)}.
\end{equation}
Here $k\geq3$, so
$(k-1)(b-1)=n-b-k+1\geq(2/3)n-O(\log n)=\Omega(n)$.
Thus every one- or two-preimage action has exponential exact Pauli
support.
\end{proof}

% ================================================================
% Exact simulation of the final quantum-inspired policy
% ================================================================

\section{Exact Tensor-Network Simulation of the Connected Majority-IQP Policy}
\label{app:connected-iqp-simulation}

We derive an exact tensor-network simulator for the circuit in
Sec.~\ref{subsec:connected-iqp-mlp}, explaining its efficient classical
majority readout despite its connected and entangled state. Fix the
observation $s$ and all parameters. There are $n=3m$ qubits in triples
$B_1,\ldots,B_m$, arbitrary diagonal gates within each triple,
and one bridge $R_{Z_{3j}Z_{3j+1}}(\beta_j)$ between successive
triples. The input and output layers are $H^{\otimes n}$.
The argument holds for every choice of these angles, not only at
initialization.

\subsection{One bridge gives at most two Schmidt components}

Cut the chain between $B_j$ and $B_{j+1}$, with sides $L$ and $R$.
All diagonal gates commute, so the crossing bridge can be applied
first to $|+\rangle^{\otimes n}$. All other gates, including the
final Hadamards, can be collected into unitaries $V_L$ and $V_R$
acting on their respective sides:
\[
 |\psi\rangle=(V_L\otimes V_R)
 R_{Z_aZ_b}(\beta_j)|+\rangle^{\otimes n},
 \qquad a=3j,\quad b=3j+1.
\]
These unitaries may entangle qubits within either side, but cannot
change entanglement across the cut. Since $(Z_aZ_b)^2=I$,
\[
 R_{Z_aZ_b}(\beta_j)
 =\cos(\beta_j/2)I-i\sin(\beta_j/2)Z_aZ_b.
\]
Define $|L_0\rangle=V_L|+\rangle^{\otimes|L|}$ and
$|L_1\rangle=V_LZ_a|+\rangle^{\otimes|L|}$, and define
$|R_0\rangle,|R_1\rangle$ analogously using $Z_b$. Then
\[
 |\psi\rangle
 =\cos(\beta_j/2)|L_0\rangle|R_0\rangle
 -i\sin(\beta_j/2)|L_1\rangle|R_1\rangle.
\]
The two states on each side are orthogonal, because
\[
 \langle L_0|L_1\rangle
 =\langle+|^{\otimes|L|}Z_a|+\rangle^{\otimes|L|}
 =0,
\]
and similarly on $R$. Thus this is a Schmidt decomposition up to
phases. The reduced-state eigenvalues are
$\cos^2(\beta_j/2)$ and $\sin^2(\beta_j/2)$, giving
\begin{equation}
 \operatorname{Tr}(\rho_L^2)
 =\cos^4(\beta_j/2)+\sin^4(\beta_j/2)
 =1-\tfrac12\sin^2\beta_j.
 \label{eq:connected-cut-purity}
\end{equation}
The cut is entangled precisely when $\beta_j\notin\pi\mathbb Z$,
but its Schmidt rank never exceeds two. Observation-dependent
diagonal encodings enter $V_L,V_R$ and do not change this conclusion.

\subsection{Compact amplitudes and the doubled probability network}

Write $x=(x_1,\ldots,x_m)$ with $x_j\in\{0,1\}^3$.
Expanding each bridge introduces one binary index $t_j$.
Let $|v_j\rangle$ be the state of triple $j$ after its initial
Hadamards and all its internal diagonal gates, including the neural
encoding. Set
$a_j(0)=\cos(\beta_j/2)$ and
$a_j(1)=-i\sin(\beta_j/2)$ for $j<m$, and
$a_m(0)=1$, $a_m(1)=0$. A concrete block tensor is
\[
 [A_j(x_j)]_{l,r}
 =a_j(r)\langle x_j|H^{\otimes3}
 Z_1^{\,l}Z_3^{\,r}|v_j\rangle,
 \qquad l,r\in\{0,1\},
\]
where $Z_1,Z_3$ act on the first and third qubits of that triple.
With boundary indices $t_0=t_m=0$, the exact amplitude is
\begin{equation}
 \langle x|\psi\rangle
 =\sum_{t_1,\ldots,t_{m-1}}
   \prod_{j=1}^{m}[A_j(x_j)]_{t_{j-1},t_j}.
 \label{eq:connected-mps-amplitude}
\end{equation}
This is a matrix-product representation with bond dimension two:
the sum is evaluated by multiplying $2\times2$ matrices, not by
enumerating the $2^{m-1}$ bridge-index assignments.
Each block tensor has only eight physical outputs.

For a probability, an amplitude is multiplied by its complex
conjugate. Each bridge then carries a pair $(t_j,t'_j)$ with four
possible values. Consequently the doubled network has bond
dimension at most four. This retains quantum interference and
inter-block correlations; it is not a product of independent
block-output distributions.

\subsection{Majority probabilities from a generating polynomial}

Let $q_w(s)=\Pr_\theta[|X|=w\mid s]$. Majority requires only
these $n+1$ Hamming-weight probabilities:
\[
 \pi_\theta(0\mid s)=\sum_{w=0}^{\lfloor n/2\rfloor}q_w(s),
 \qquad \pi_\theta(1\mid s)=1-\pi_\theta(0\mid s).
\]
For even $n$, this convention assigns ties to action 0.
The polynomial $F_s(z)$ in
Eq.~\eqref{eq:connected-majority-generating-function} collects the
$q_w(s)$ as its coefficients. Since $|x|=\sum_j|x_j|$, its weight
factor separates as $z^{|x|}=\prod_j z^{|x_j|}$.
Define
\begin{equation}
 \begin{aligned}
 [T_j(z)]_{(l,l'),(r,r')}
 &=\sum_{x_j\in\{0,1\}^3}z^{|x_j|}
 [A_j(x_j)]_{l,r}\,
 \overline{[A_j(x_j)]_{l',r'}} .
 \end{aligned}
 \label{eq:connected-majority-transfer}
\end{equation}
Substituting Eq.~\eqref{eq:connected-mps-amplitude} into $F_s(z)$
and exchanging the finite sums gives
\begin{equation}
 F_s(z)=e_{00}^{\mathsf T}T_1(z)\cdots T_m(z)e_{00},
 \label{eq:connected-majority-contraction}
\end{equation}
where $e_{00}$ selects the boundary pair $(0,0)$.
Each $T_j(z)$ is $4\times4$ and sums only eight local outcomes.
For fixed $z$, constructing these matrices and multiplying them
successively into a boundary vector therefore costs $O(n)$.
The matrices can be complex; they need not be stochastic matrices.

To recover all coefficients, take $N=n+1$ and
$z_r=e^{-2\pi i r/N}$ for $r=0,\ldots,N-1$. Then
\begin{equation}
 q_w(s)=\frac1N\sum_{r=0}^{N-1}F_s(z_r)e^{2\pi i rw/N}.
 \label{eq:connected-majority-inverse-fourier}
\end{equation}
Indeed, substituting $F_s(z_r)=\sum_{v=0}^{n}q_v(s)e^{-2\pi i rv/N}$
reduces the right-hand side to
\[
 \sum_{v=0}^{n}q_v(s)
 \underbrace{\frac1N\sum_{r=0}^{N-1}
 e^{2\pi i r(w-v)/N}}_{=\,1\text{ if }w=v,\;0\text{ otherwise}}.
\]
There is no aliasing because the polynomial degree is less than $N$.
Thus $n+1$ contractions cost $O(n^2)$ operations, the inverse FFT
costs $O(n\log n)$, and the majority sum costs $O(n)$.
For 21 qubits, this uses 22 evaluations of a seven-block network
instead of enumerating $2^{21}$ outputs or the decoder's Pauli terms.
Sampling the resulting binary policy is a Bernoulli draw, so the
calculation also supplies a classical weak simulator. An additive
error $\epsilon$ in the normalized action-0 probability gives TV
error at most $\epsilon$ for that binary action distribution.

\subsection{Exactness and limits of the argument}

These are exact algebraic contractions with no bond truncation.
The complexity above counts arithmetic operations after constructing
the local angles; computing the neural encoder adds its classical
evaluation cost. Floating-point implementations incur numerical
error. The training implementation additionally floors action
probabilities at $10^{-8}$ and renormalizes them for stability.
The reported statevector comparison checks the underlying
Hamming-weight probabilities, not the asymptotic complexity proof.
Differentiating the contraction also has polynomial arithmetic cost;
this does not guarantee successful optimization at arbitrary scale.

The generating-polynomial identity holds for every majority policy,
but efficient evaluation of $F_s$ is circuit-dependent.
If $c$ two-qubit diagonal bridges cross a block cut, their
two-term expansions give upper bounds $2^c$ and $4^c$ on the ket
and doubled bond dimensions, rather than constant bounds.
These bounds can be loose, and another contraction order may help.
General tensor-network methods depend on contraction width, not
merely on the number of gates incident on an individual
qubit~\cite{MarkovShi2008}. Bounded degree or IQP commutativity
alone therefore does not establish polynomial contraction cost.
Nor does failure of this contraction approach prove classical
hardness or exclude an efficient approximate algorithm.

Finally, Schmidt rank and contraction width are distinct from
the active-generator binary rank $r^u$ controlling initialization
variance. A protected Pauli component need not make the entire
majority readout tractable. Here tractability follows from the
specific chain structure together with additive Hamming-weight
aggregation: exponential exact Pauli support does not obstruct
this alternative classical simulation.

% Supplementary studies retained from the author's current manuscript.
\section{Supplementary CartPole Studies and Gradient Diagnostics}
\label{app:supplementary-numerics}

These studies supply the detailed protocols and secondary evidence
summarized in Sec.~\ref{sec:experiments}. They use distinct ensembles
and sample sizes and should not be pooled into a single comparison.

\subsection{Matched CartPole policy optimization}
\label{subsec:experiment-cartpole}

We next ask whether the readout-induced trainability distinction is
visible in an end-to-end control task.  We begin with recursive parity
because it was proposed by Meyer \emph{et al.} as an effective CartPole
decoder and was reported to improve on the earlier softmax-based quantum
policies of Jerbi \emph{et al.}~\cite{Meyer2023ActionDecoding,
Jerbi2021Parametrized}.  We train binary parity and
majority--modulo Born policies in the CartPole environment using the same five-qubit
graph-IQP circuit on a ring and $L\in\{1,2,3\}$ repeated blocks. The
majority decoder uses one tie-free block containing all five measured
qubits. For each depth and seed in $\{3,7,11,23,37\}$, both policies
receive elementwise-identical initial circuit and input-scaling tensors;
the environment and action-sampling streams are also seeded identically.
Only the final action projector differs at initialization.  The
five-qubit trainable tensors are initialized independently from a
unit-variance zero-mean normal distribution, as implemented by QASP's
\texttt{normal\_0\_1} option.

At every environment time step, the four CartPole observation features
$s_t\in\mathbb R^4$ are encoded using the classical neural state
embedding of Sec.~\ref{subsec:neural-state-embeddings}. Specifically,
we set $L_{\rm enc}=n$ and $P_\ell=Z_\ell$ in
Eq.~\eqref{eq:neural-state-embedding}, giving
\begin{equation}
    \begin{aligned}
    E_{s_t}&=\prod_{\ell=1}^{n}R_{Z_\ell}\!\left(\phi_\ell(s_t)\right),\\
    \phi_\ell(s_t)&=\sum_{j=1}^{4}w_j^{(\ell)}s_{t,j}+b^{(\ell)}.
    \end{aligned}
    \label{eq:cartpole-neural-encoding}
\end{equation}
Each qubit therefore receives a learned affine combination of all four
features through a single-qubit $R_Z$ rotation. In the QASP
implementation, the feature-dependent rotation and the trainable local
$R_Z$ gate are drawn separately; their angles add, so the latter
supplies the bias $b^{(\ell)}$ without introducing an additional
parameter. This is parallel data encoding across the $n$ qubits: the
encoding rotations act on distinct wires and can be applied
simultaneously. For $L=1$, they remain inside the commuting diagonal
IQP program. For $L>1$, the same observation $s_t$ is also reuploaded
serially across the $L$ layers, with independent trainable weights and
biases in each layer. The intervening Hadamard layers make the full
architecture repeated IQP-like rather than globally commuting IQP.
Thus $L_{\rm enc}=n$ counts encoding gates per layer, while the circuit
contains $nL$ data-encoding rotations in total. This encoding strategy
is shared by the CartPole studies below, with initialization specified
separately for each experiment.

Both policies are trained for 500 episodes with REINFORCE, discount
$\gamma=0.99$, learning rate $3\times10^{-2}$, batches of five episodes,
an exponential-moving-average baseline, normalized returns, entropy
coefficient $5\times10^{-3}$, and unit gradient-norm clipping. Episodes
are limited to 200 environment steps. The shared circuit template is summarized in
Fig.~\ref{fig:cartpole-circuit-schematic}, and
Fig.~\ref{fig:cartpole-ring-results} reports 20-episode moving averages
as mean $\pm$ standard error over the five matched seeds.

\begin{table}[t]
\caption{CartPole mean return over the final 50 episodes}
\label{tab:cartpole-final-return}
\centering
\begin{tabular}{c@{\hspace{1.2em}}cc}
\toprule
Depth $L$ & Parity & Majority--modulo \\
\midrule
1 & $157.86\pm32.85$ & $191.28\pm3.85$ \\
2 & $188.50\pm 5.51$ & $193.66\pm2.19$ \\
3 & $172.12\pm 8.21$ & $194.52\pm1.89$ \\
\bottomrule
\end{tabular}
\vspace{0.4ex}

\footnotesize Mean $\pm$ standard error over five matched seeds.
\end{table}

As summarized in Table~\ref{tab:cartpole-final-return}, majority--modulo
has the larger mean late-training return at every tested depth, with
mean differences of $33.42$, $5.15$, and $22.40$ for $L=1,2,3$,
respectively. Its between-seed standard error is also smaller in all
three comparisons. This result is consistent with the low-weight
majority sector providing a more stable optimization signal than the
global binary-parity projector. The experiment is nevertheless a small
statevector study with five matched seeds and one control environment;
it establishes neither an asymptotic learning separation nor a quantum
advantage. Rather, it supplies an end-to-end check that the spectral
choice of action decoder can affect learning even when the circuit,
parameter count, initialization, optimizer, and environment are held
fixed.

\subsection{Ten-qubit cubic-graph CartPole stress test}
\label{subsec:experiment-expander-cartpole}

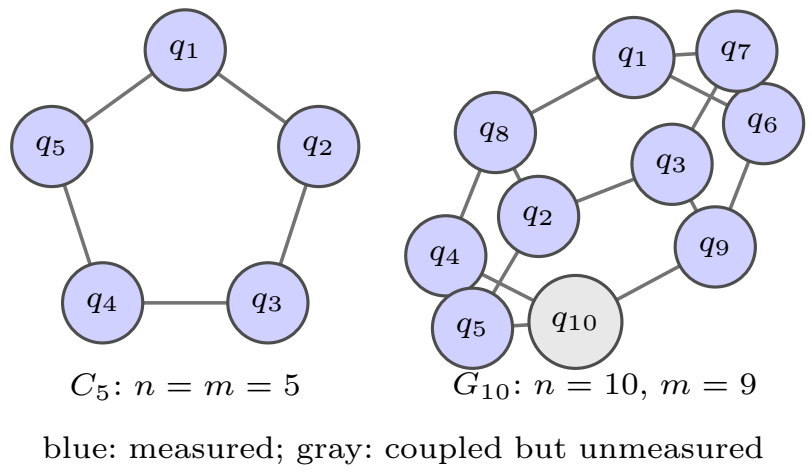
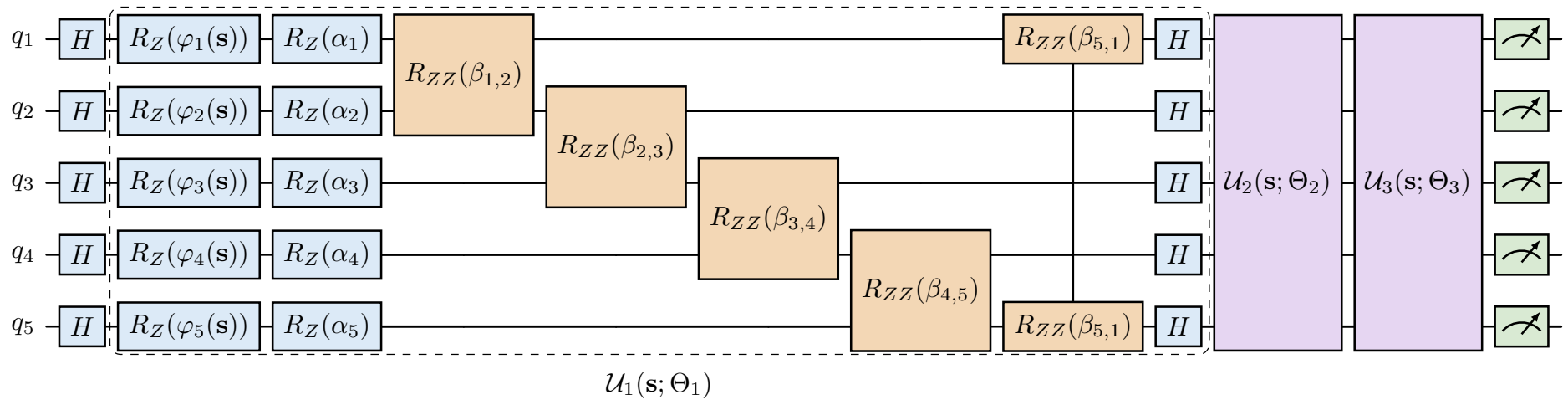
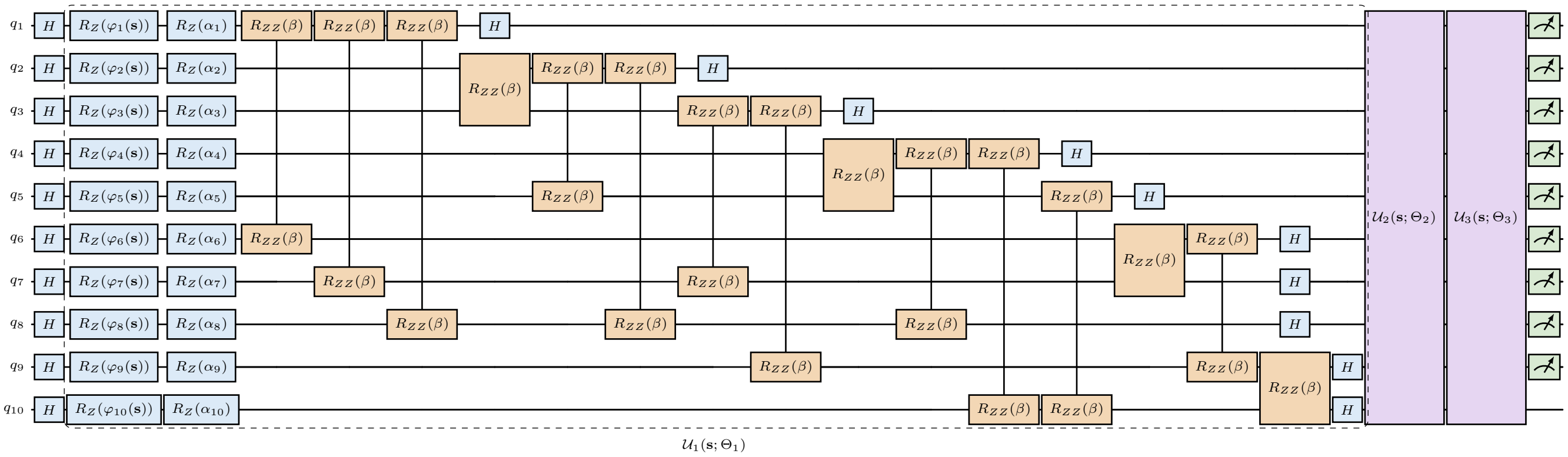
\begin{figure*}[!htbp]
    \centering
    \begin{subfigure}[t]{0.29\textwidth}
        \centering
        \resizebox{\linewidth}{!}{%
        \begin{tikzpicture}[
            topoedge/.style={line width=0.8pt, draw=black!55},
            expnode/.style={circle, draw=black!70, line width=0.7pt,
                           minimum size=5.8mm, font=\scriptsize},
            note/.style={font=\scriptsize, align=center}
        ]
            \coordinate (c1) at (1.40,1.10);
            \coordinate (c2) at (2.45,0.34);
            \coordinate (c3) at (2.05,-0.89);
            \coordinate (c4) at (0.75,-0.89);
            \coordinate (c5) at (0.35,0.34);
            \foreach \u/\v in {1/2,2/3,3/4,4/5,5/1}
                \draw[topoedge] (c\u) -- (c\v);
            \foreach \i in {1,...,5}
                \node[expnode, fill=blue!18] at (c\i) {$q_{\i}$};
            \node[note] at (1.40,-1.55) {$C_5$: $n=m=5$};

            \coordinate (q1p) at (4.93,1.04);
            \coordinate (q2p) at (4.18,-0.21);
            \coordinate (q3p) at (5.23,0.20);
            \coordinate (q4p) at (3.45,-0.52);
            \coordinate (q5p) at (3.66,-1.09);
            \coordinate (q6p) at (5.95,0.52);
            \coordinate (q7p) at (5.74,1.09);
            \coordinate (q8p) at (3.84,0.45);
            \coordinate (q9p) at (5.57,-0.45);
            \coordinate (q10p) at (4.47,-1.04);
            \foreach \u/\v in {1/6,1/7,1/8,2/3,2/5,2/8,3/7,3/9,
                                 4/5,4/8,4/10,5/10,6/7,6/9,9/10}
                \draw[topoedge] (q\u p) -- (q\v p);
            \foreach \i in {1,...,9}
                \node[expnode, fill=blue!18] at (q\i p) {$q_{\i}$};
            \node[expnode, fill=gray!18] at (q10p) {$q_{10}$};
            \node[note] at (4.70,-1.55) {$G_{10}$: $n=10$, $m=9$};
            \node[note, anchor=west] at (0.15,-2.05)
                {blue: measured; gray: coupled but unmeasured};
        \end{tikzpicture}}
        \caption{Interaction graphs.}
    \end{subfigure}\hfill
    \begin{subfigure}[t]{0.69\textwidth}
        \centering
        \includegraphics[width=\linewidth]{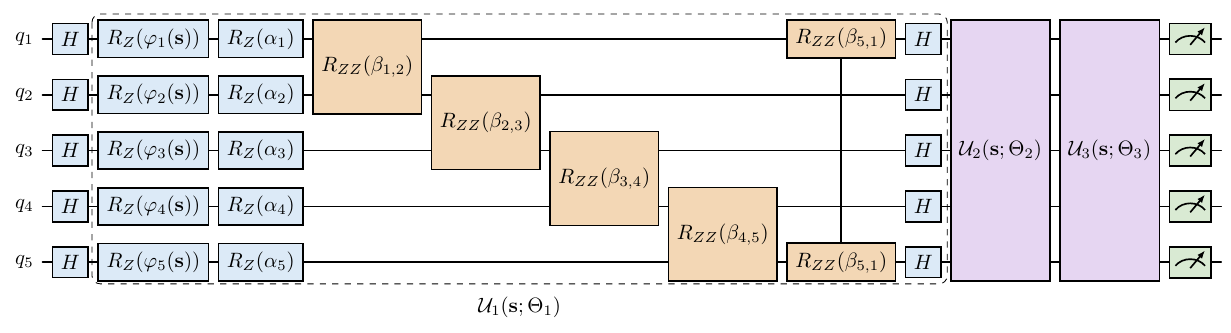}
        \caption{Five-qubit ring circuit.}
    \end{subfigure}

    \vspace{0.8em}
    \begin{subfigure}[t]{0.98\textwidth}
        \centering
        \includegraphics[width=\linewidth]{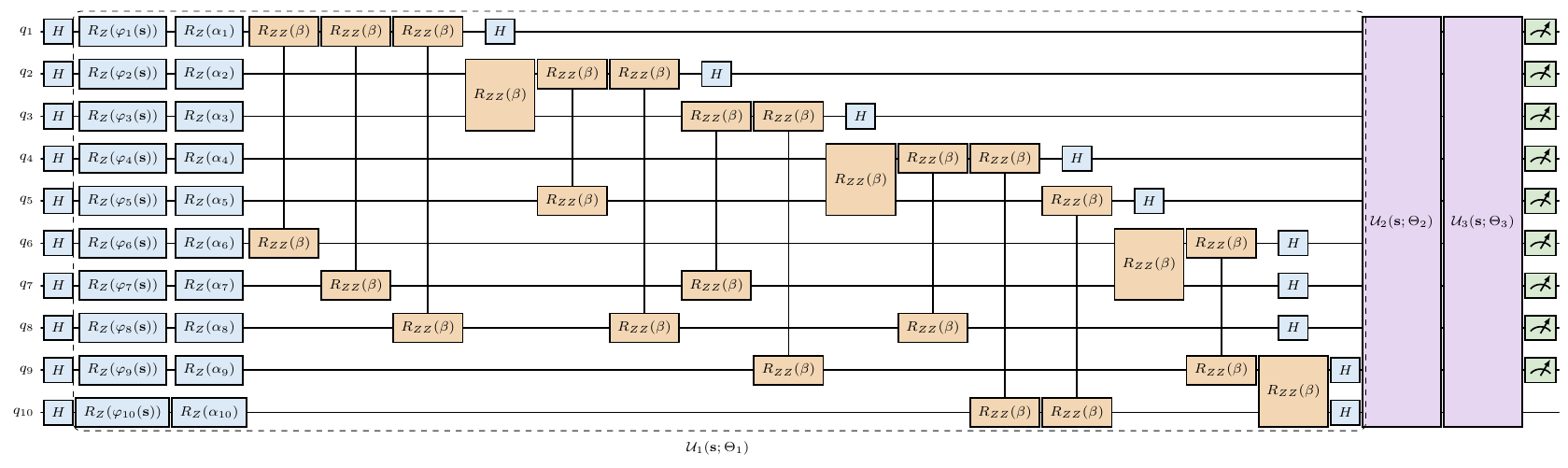}
        \caption{Ten-qubit cubic-graph circuit.}
    \end{subfigure}
    \caption{CartPole graph-IQP circuits used in both training studies.
    (a) The five-qubit baseline instantiates $(n,m,G_n)=(5,5,C_5)$; the
    ten-qubit stress test uses $(n,m,G_n)=(10,9,G_{10})$, where $G_{10}$
    is the displayed connected 3-regular expander and $q_{10}$ is coupled
    but unmeasured. Panels (b) and (c) are exported directly from the
    QASP models. The first state-reuploading block $\mathcal U_1$ is
    expanded gate by gate; $\mathcal U_2$ and $\mathcal U_3$ compactly
    represent the same layer at depths $L=2$ and $3$ with independent parameters per layer. Each layer contains
    $n$ parallel neural encoding rotations with $L_{\rm enc}=n$ and
    $P_\ell=Z_\ell$, as in Eq.~\eqref{eq:cartpole-neural-encoding},
    with the affine biases shown as separate trainable local $R_Z$ gates,
    graph-edge $R_{ZZ}$ gates, and terminal Hadamards. Parity and
    majority--modulo share the complete quantum circuit and differ only
    in their classical decoder. The terminal meter symbols denote
    computational-basis measurement on all five qubits in (b) and on
    $q_1,\ldots,q_9$ in (c).}
    \label{fig:cartpole-circuit-schematic}
\end{figure*}

The five-qubit ring isolates the readout while retaining a
quasi-one-dimensional circuit topology. To test whether the same
trainability mechanism survives in a structurally more demanding model,
we repeat the matched comparison on a ten-qubit connected cubic graph.
The seeded graph is 3-regular, has 15 edges, and has adjacency spectral
gap $d-\lambda_2=0.853351$.  It was selected from NetworkX random
3-regular graphs using master seed $31415$ and accepted graph seed
$1039606416$ subject to a gap threshold of $0.5$; the complete edge set
is displayed in Fig.~\ref{fig:cartpole-circuit-schematic}. It is a finite sparse,
non-one-dimensional instance rather than an asymptotic expander-family
construction: bounded degree preserves local majority sectors, whereas
its connectivity removes the path or ring geometry that enables
especially simple contraction orderings. Since the experiment is
performed at ten qubits by statevector simulation, it is a
model-complexity stress test, not evidence of classical hardness or
quantum advantage.

Figure~\ref{fig:cartpole-circuit-schematic} consolidates the two CartPole
circuit instances. Both use the parallel single-qubit neural encoder in
Eq.~\eqref{eq:cartpole-neural-encoding}, with serial reuploading for
$L>1$. All ten qubits of the larger circuit are encoded, including the
unmeasured qubit. The five-qubit
baseline uses the cycle $C_5$ and measures all five qubits. The expander
instance uses ten local $R_Z$ rotations and 15 graph-edge $R_{ZZ}$
rotations per layer; nine qubits are measured so that the majority block
is odd and tie-free, while the tenth remains coupled to the readout. For
every depth, parity and majority--modulo use identical circuit and
input-scaling tensors and differ only in the classical partition of the
same Born distribution.

The parity observable on the measured register is
\begin{equation}
    Z^{\mathrm P}=Z_1Z_2\cdots Z_9.
\end{equation}
Although it has weight nine, its active-generator light cone includes
all ten qubits because $q_{10}$ is connected across the support boundary.
Consequently, the exact IQP law gives
\begin{equation}
    \Var_{\theta}[\langle Z^{\mathrm P}\rangle]
    =2^{-10}=0.0009765625.
    \label{eq:expander-parity-observable-variance}
\end{equation}
For the binary action probability
$\pi^{\mathrm P}(0|s)=(1+\langle Z^{\mathrm P}\rangle)/2$, the projector
normalization yields
\begin{equation}
    \Var_{\theta}[\pi^{\mathrm P}(0|s)]
    =2^{-12}=0.000244140625.
    \label{eq:expander-parity-policy-variance}
\end{equation}
By contrast, each 1-local component of the nine-bit majority projector
has light-cone size four. The 1-local sector alone therefore gives
\begin{equation}
    \Var_{\theta}[\pi^{\mathrm{MM}}(0|s)]
    \geq
    9\left(\frac{\binom{8}{4}}{2^9}\right)^2 2^{-4}
    =0.01051426,
    \label{eq:expander-majority-lower-bound}
\end{equation}
which is $43.07$ times the parity action-probability variance. Equations
\eqref{eq:expander-parity-observable-variance}--
\eqref{eq:expander-majority-lower-bound} are one-layer initialization
statements; for $L>1$, reuploading enlarges propagated light cones, so the
comparison below tests whether the predicted signal hierarchy persists
under end-to-end optimization rather than assuming the one-layer formula
remains exact.

\begin{figure*}[!htbp]
    \centering
    \begin{subfigure}[t]{0.49\textwidth}
        \centering
        \includegraphics[width=\linewidth]{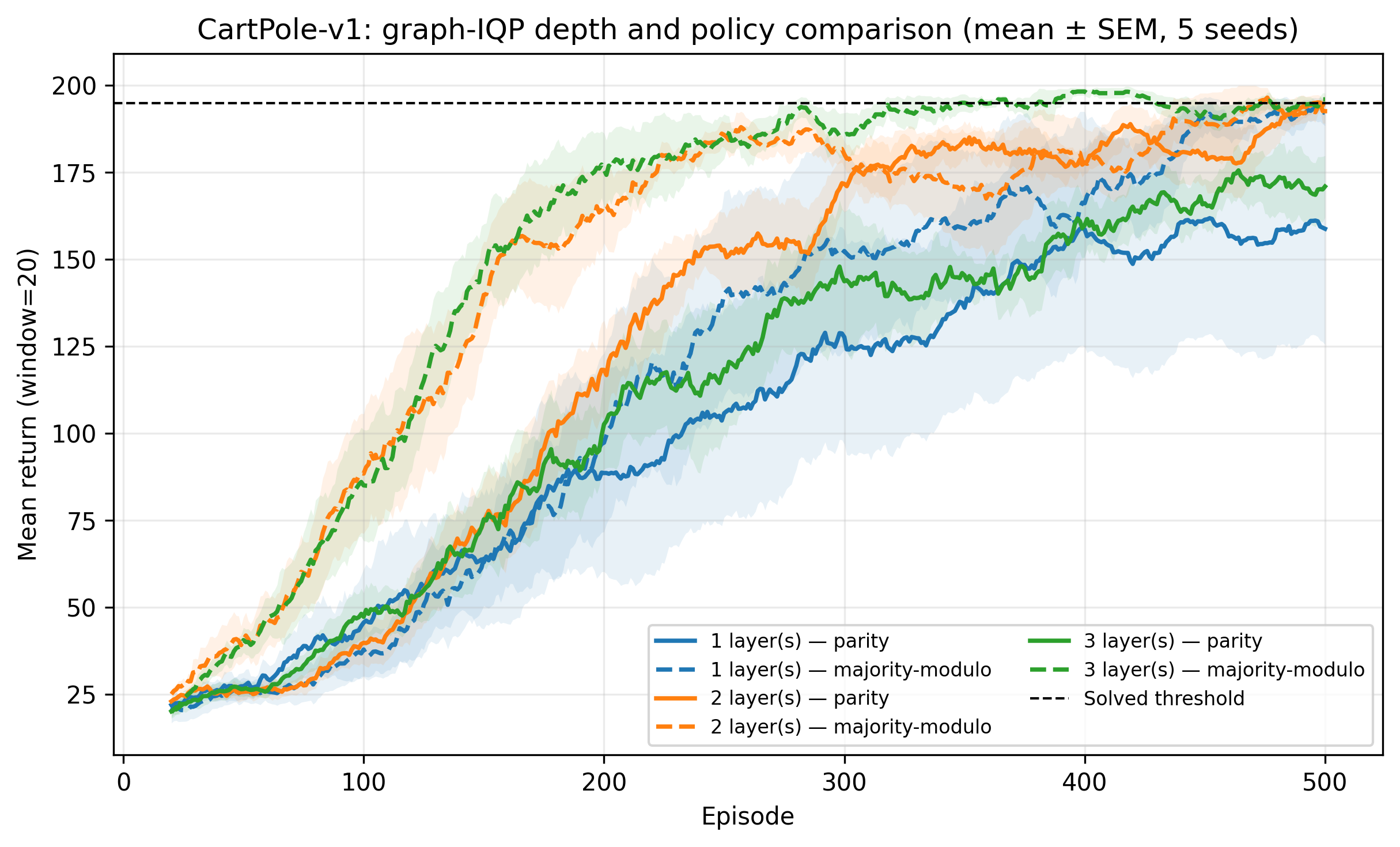}
        \caption{Five-qubit ring, five matched seeds.}
        \label{fig:cartpole-ring-results}
    \end{subfigure}\hfill
    \begin{subfigure}[t]{0.49\textwidth}
        \centering
        \includegraphics[width=\linewidth]{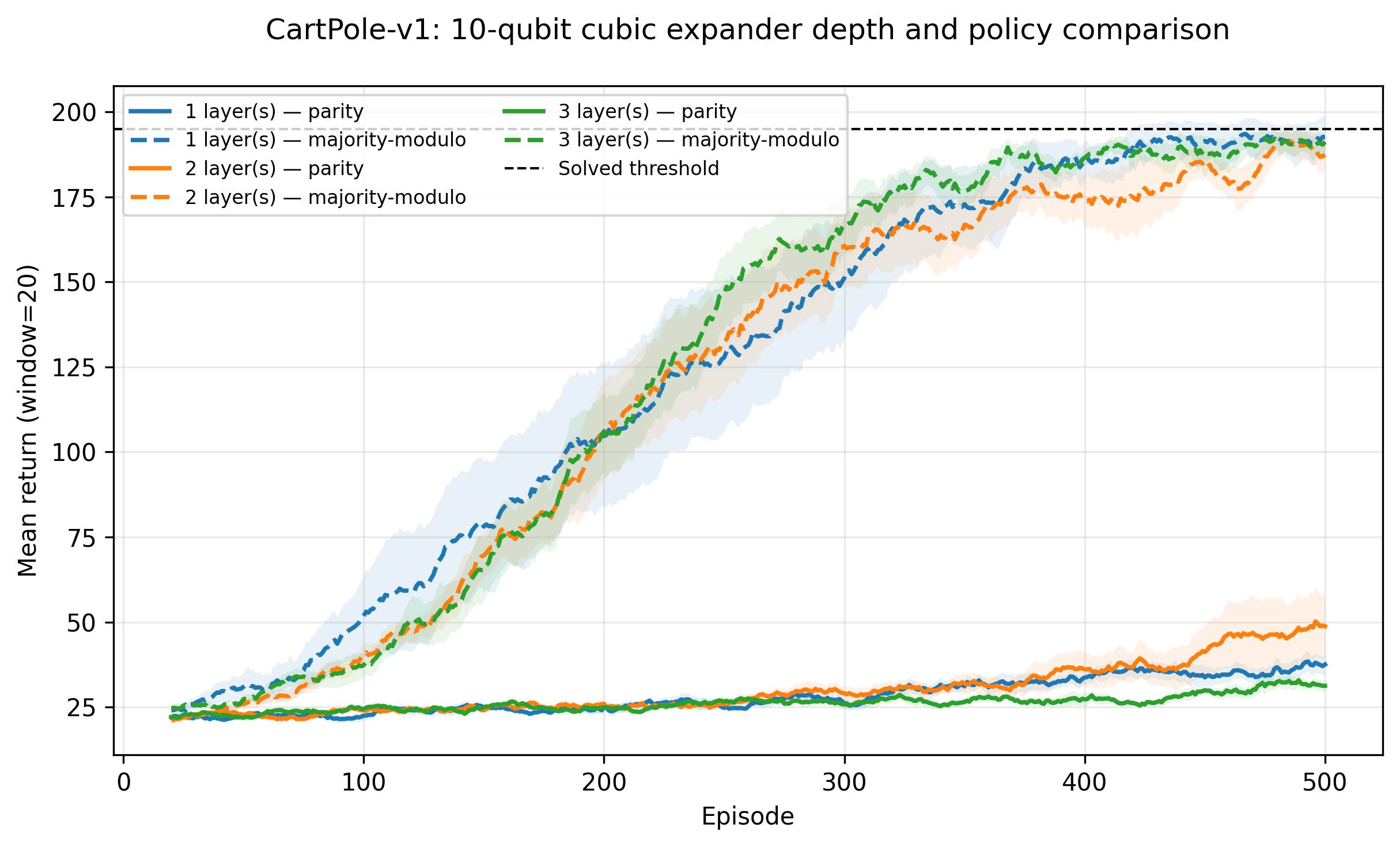}
        \caption{Ten-qubit cubic expander, ten matched seeds.}
        \label{fig:cartpole-expander-results}
    \end{subfigure}
    \caption{Matched CartPole learning curves for the two graph-IQP
    instances in Fig.~\ref{fig:cartpole-circuit-schematic}. Solid lines
    denote parity and dashed lines denote majority--modulo; color denotes
    graph-IQP depth. Curves show the mean 20-episode moving return and
    bands show the standard error over the matched seeds. The horizontal
    line marks the study-specific rolling-100 success threshold of 195;
    the maximum return is 200.}
    \label{fig:cartpole-combined-results}
\end{figure*}

We use the REINFORCE protocol specified above, now with ten matched seeds
$\{3,7,11,23,37,41,43,47,53,59\}$ and full-period uniform parameter
initialization. Within each depth and seed, the two policies receive
elementwise-identical trainable tensors and identical environment and
action-sampling seeds. Figure~\ref{fig:cartpole-combined-results} places
the five-qubit baseline and ten-qubit stress test side by side;
Fig.~\ref{fig:cartpole-expander-results} reports the latter's mean
20-episode moving return and its standard error over the ten runs.

\begin{table}[t]
\caption{Ten-qubit expander: mean return over the final 50 episodes}
\label{tab:expander-cartpole-final-return}
\centering
\begin{tabular}{c@{\hspace{1.2em}}cc}
\toprule
Depth $L$ & Parity & Majority--modulo \\
\midrule
1 & $36.63\pm2.41$ & $192.13\pm5.23$ \\
2 & $47.76\pm9.96$ & $186.03\pm4.30$ \\
3 & $31.05\pm1.67$ & $190.90\pm2.36$ \\
\bottomrule
\end{tabular}
\vspace{0.4ex}

\footnotesize Mean $\pm$ standard error over ten matched seeds.
\end{table}

The separation is large at every depth. Majority--modulo improves from
the initial-return regime to mean final-50 returns between $186.03$ and
$192.13$, whereas parity remains between $31.05$ and $47.76$. The mean
late-training advantages are $155.50$, $138.27$, and $159.86$ returns for
$L=1,2,3$, respectively. Using the study-specific rolling-100 criterion,
majority--modulo meets the threshold in 7 of 10, 3 of 10, and 2 of 10
runs at the three depths; no parity run meets it. The result is therefore
consistent with the low-weight majority sector retaining appreciable
initial policy variance while the global-parity sector is already strongly
concentrated at ten qubits. Because this is a finite matched experiment,
it supports the proposed mechanism but does not by itself establish that
initialization variance is the unique cause of the learning separation.
All reported terminal comparisons use the same prespecified final-50
episode statistic, with the seed as the independent experimental unit.

\subsection{Graph-local three-body rotations in CartPole}
\label{subsec:cartpole-graph-local-three-body}

We next keep the majority decoder fixed and test whether adding
three-body interactions improves learning on the same ten-qubit cubic
graph. Both policies use $L=1$, the neural encoder in
Eq.~\eqref{eq:cartpole-neural-encoding}, and majority on
$q_1,\ldots,q_9$, with $q_{10}$ coupled but unmeasured. The baseline
has 65 trainable parameters: 40 encoding weights, ten local rotation
angles, and 15 edge rotation angles. The extended circuit inserts ten
independently parameterized $R_{Z_iZ_jZ_k}$ rotations into the commuting
diagonal block, giving 75 parameters while remaining an IQP circuit.
Each triple contains a length-two path of the underlying graph; together
the triples cover all 15 edges and include each qubit exactly three times.
With the one-based wire numbering of
Fig.~\ref{fig:cartpole-circuit-schematic}, their supports are
\[
\begin{gathered}
\{1,6,7\},\ \{2,5,8\},\ \{2,3,7\},\ \{4,5,8\},\ \{4,5,10\},\\
\{1,6,9\},\ \{3,6,7\},\ \{1,2,8\},\ \{3,9,10\},\ \{4,9,10\}.
\end{gathered}
\]

This controlled comparison uses 20 matched seeds, distinct from the
preceding ten-seed majority--parity study. For each seed, the 65 shared
parameters are initialized to bitwise-identical values from
$\operatorname{Unif}[0,2\pi)$; the ten additional angles use an
independent reproducible draw from the same distribution. Both models
use the preceding REINFORCE settings, including Adam with learning rate
$0.03$ for every parameter, 500 training episodes with a 200-step limit,
and 100 held-out evaluation episodes per seed. Thus the comparison
changes the three-body sector and parameter count, but not the decoder,
shared initialization, or optimization protocol.

\begin{figure*}[!t]
    \centering
    \includegraphics[width=\textwidth]{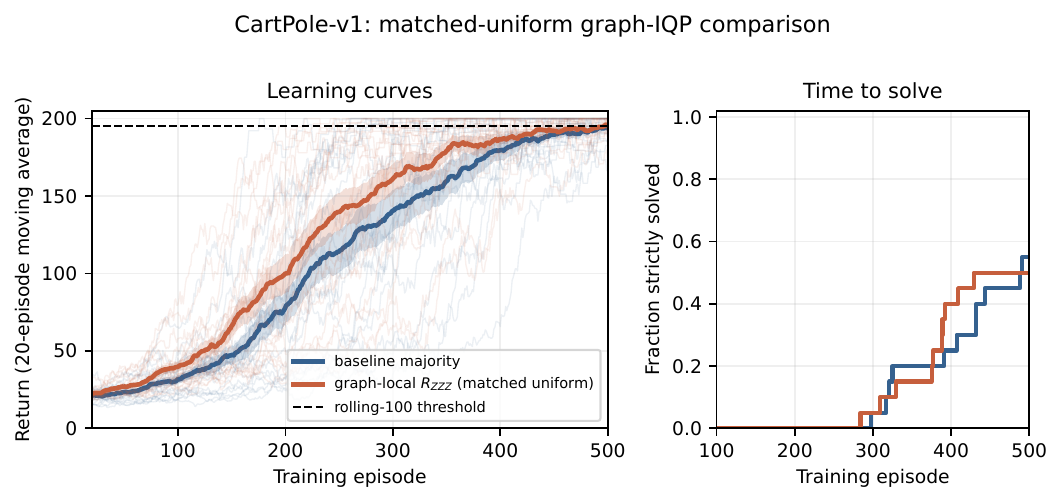}
    \caption{Matched-uniform CartPole comparison of the ten-qubit
    majority policy with and without ten graph-local $R_{ZZZ}$ rotations,
    over 20 matched seeds. Left: mean 20-episode moving return, with
    standard-error bands and faint individual-seed trajectories. The
    dashed line marks return 195; solving is assessed using the
    rolling-100 mean, not the displayed rolling-20 curves. Right:
    fraction of seeds whose rolling-100 mean has reached at least 195
    by each episode. The extended circuit has higher normalized
    training-return AUC, but the difference in restricted solve time
    is not statistically significant.}
    \label{fig:cartpole-graph-local-three-body}
\end{figure*}

The primary endpoint is normalized training-return AUC, computed as
the sum of the 500 episode returns divided by $500\times200$.
Figure~\ref{fig:cartpole-graph-local-three-body} shows the corresponding
learning curves. Mean AUC increases from $0.55507\pm0.03061$ to
$0.61675\pm0.02571$ (mean $\pm$ standard error across seeds).
The paired increase is $0.06168$, with a 95\% paired-bootstrap
confidence interval $[0.00396,0.12309]$ and one-sided paired
sign-randomization $p=0.0303$; 13 of 20 seed pairs favor the extended
circuit. Intervals use 50,000 bootstrap resamples and tests use
200,000 sign randomizations, with the seed as the independent unit.
This supports improved cumulative training performance over the fixed
episode budget.

The improvement does not establish significantly earlier threshold
solving or higher terminal returns. The fractions reaching a
rolling-100 mean of at least 195 are $0.55$ and $0.50$ for the baseline
and extended circuit, respectively. Assigning unsolved runs episode
501 gives mean restricted solve episodes of $442.65$ and $434.55$;
the $8.10$-episode advantage has 95\% confidence interval
$[-27.50,44.45]$ and one-sided $p=0.340$. Likewise, the paired
final-50 return gain is $1.66$ (95\% interval $[-2.37,6.35]$,
$p=0.253$), and the held-out gain is $1.97$
($[-1.38,6.27]$, $p=0.204$). The evidence therefore concerns the
learning trajectory rather than a resolved improvement in final policy
quality.

\subsection{Finite-shot CartPole policy optimization}
\label{subsec:finite-shot-cartpole}

The preceding majority--parity comparisons use exact probabilities and
analytic gradients. To test whether the decoder contrast persists when
action probabilities must be estimated from measurements, we repeat the
five-qubit $L=3$ and ten-qubit $L=1$ experiments with a finite-shot SPSA
estimator. A two-direction pilot compares 1,024 and 4,096 shots; the final
experiment retains 4,096 shots and increases the number of directions
according to the model size. These are the strongest five-qubit and least
costly ten-qubit contrasts in the depth sweeps. We use the same five matched
seeds $\{3,7,11,23,37\}$ for every condition. For binary actions, the
worst-case standard error of a sampled probability is $1/(2\sqrt{S})$,
equal to $0.015625$ and $0.0078125$ at $S=1024$ and $4096$, respectively.
The latter is half the initialization standard deviation
$\sqrt{2^{-12}}=0.015625$ of the ten-qubit parity action probability in
Eq.~\eqref{eq:expander-parity-policy-variance}; the majority lower bound
in Eq.~\eqref{eq:expander-majority-lower-bound} has standard deviation
at least $0.1025$.

Both interaction and optimization use sampled probabilities.  If $N_a$
counts action $a$ among $S$ circuit shots, symmetric Jeffreys smoothing
gives
\begin{equation}
    \widetilde\pi(a|s)=\frac{N_a+1/2}{S+1}.
    \label{eq:finite-shot-jeffreys}
\end{equation}
This prevents undefined logarithms without privileging either action.
Native parameter-shift would require 91 circuit evaluations per visited
state for the five-qubit circuit and 71 for the ten-qubit circuit.
To improve convergence without incurring the full parameter-shift cost,
we test two SPSA direction counts, $K=2$ and
$K=\lceil\log_2 p\rceil$, where $p$ is the total number of scalar
trainable parameters in the model. Both settings use the same two-sided
estimator of the batched REINFORCE loss,
\begin{equation}
    \begin{aligned}
    \widehat g_k
    &=\frac{1}{K}\sum_{j=1}^{K}
      \frac{\widehat{\mathcal L}(\theta+c_k\Delta_j)
            -\widehat{\mathcal L}(\theta-c_k\Delta_j)}{2c_k}\,
      \Delta_j,\\
    K&\in\left\{2,\left\lceil\log_2 p\right\rceil\right\}.
    \end{aligned}
    \label{eq:finite-shot-spsa}
\end{equation}
The Rademacher directions $\Delta_j$ and finite-shot loss evaluations
on the two sides are independent. The five- and ten-qubit models have
$p=90$ and $p=65$, respectively, so the logarithmic setting uses seven
directions in both cases. Averaging more directions aims to reduce
direction noise: the two settings require $2K=4$ and $14$ perturbed
circuit evaluations per visited state, respectively, below the
parameter-shift counts above. We test $K=2$ at both 1,024 and 4,096 shots,
and $K=7$ only at 4,096 shots. The parameter counts refer to the
weights perturbed by SPSA, rather than the smaller number of effective
gate angles after feature contraction. We set
$c_k=0.1(k+1)^{-0.101}$. All other environment, optimizer, batching,
initialization, and stopping settings are unchanged. Because the
finite-shot condition changes both probability access and gradient
estimation relative to the analytic baseline, it should be interpreted
as a shot-based mechanism study rather than a controlled shot-noise
comparison or a claim of near-term hardware practicality.

\begin{figure*}[!t]
    \centering
    \includegraphics[width=0.98\textwidth]{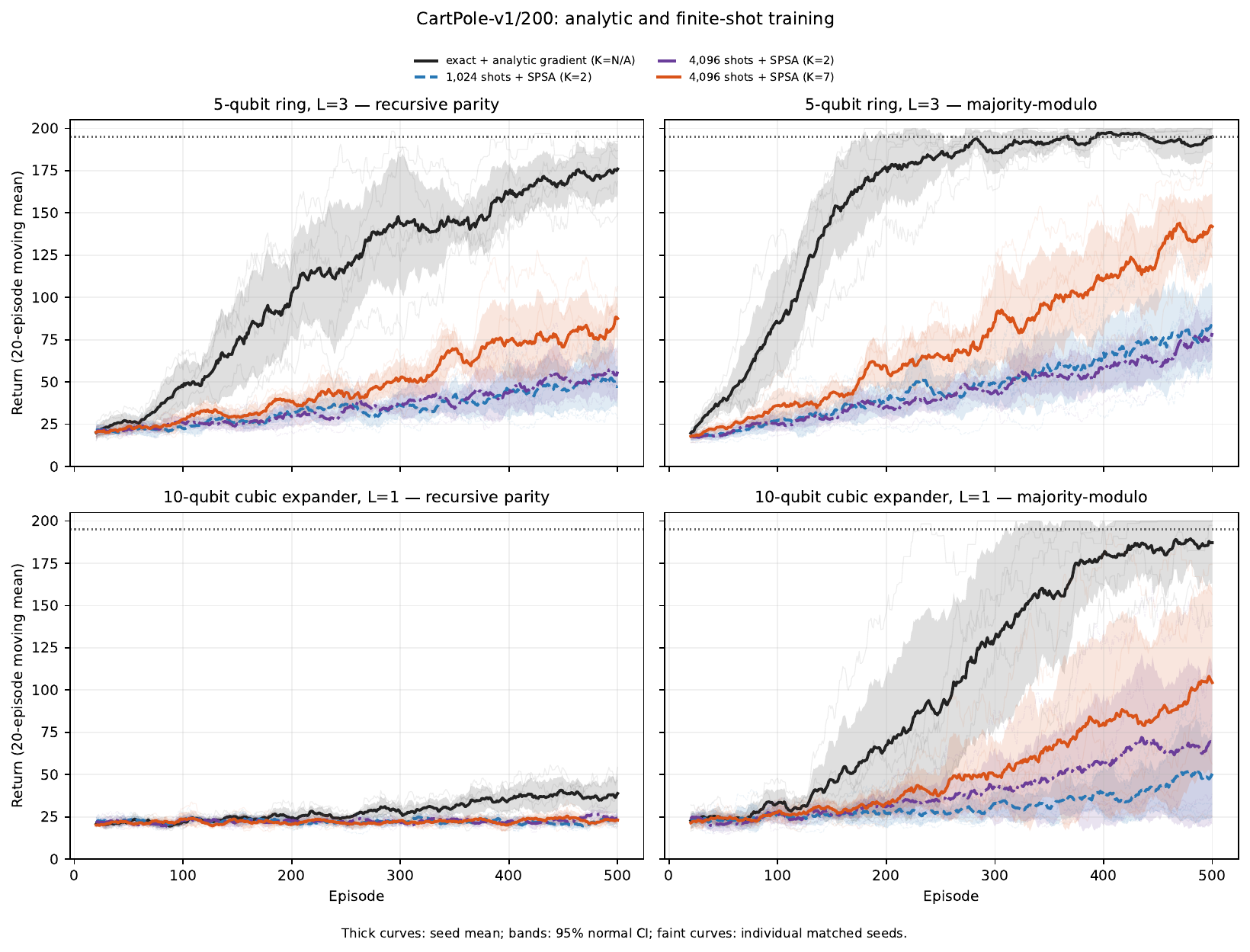}
    \caption{Analytic-gradient and finite-shot SPSA training for the
    selected five-qubit $L=3$ ring and ten-qubit $L=1$ cubic graph on
    the CartPole environment. The curves compare the analytic baseline, $K=2$ at
    1,024 and 4,096 shots, and $K=7$ at 4,096 shots, making the effects
    of shot count and direction count separately visible.
    Columns show recursive parity and majority--modulo; rows show the two
    circuit instances.  Thick curves are seed means of the 20-episode
    moving return, bands are 95\% normal intervals across five matched
    seeds, and faint curves are individual seeds.  The dotted line marks
    the study-specific rolling-100 threshold of 195; the maximum episode
    return is 200.}
    \label{fig:finite-shot-cartpole}
\end{figure*}

Paired uncertainty is reported using a 95\% confidence interval (CI).

\begin{table*}[t]
\caption{Finite-shot CartPole terminal performance}
\label{tab:finite-shot-cartpole}
\centering
\scriptsize
\setlength{\tabcolsep}{4.5pt}
\begin{tabular}{llccc}
\toprule
Circuit & Training condition & Parity & Majority--modulo
& Paired difference $\Delta$ (95\% CI) \\
\midrule
5 qubits, $L=3$ & Exact + analytic gradient
& $173.89\pm7.46$ & $192.60\pm4.86$ & $18.71\ [8.05,32.90]$ \\
& 1,024 shots + SPSA ($K=2$)
& $50.04\pm7.62$ & $78.56\pm12.26$ & $28.52\ [13.04,49.58]$ \\
& 4,096 shots + SPSA ($K=2$)
& $53.36\pm5.87$ & $74.68\pm5.99$ & $21.33\ [9.80,32.85]$ \\
& 4,096 shots + SPSA ($K=7$)
& $83.79\pm8.18$ & $138.83\pm9.65$ & $55.04\ [50.16,59.92]$ \\
\midrule
10 qubits, $L=1$ & Exact + analytic gradient
& $37.65\pm4.77$ & $187.17\pm10.37$ & $149.52\ [136.99,158.12]$ \\
& 1,024 shots + SPSA ($K=2$)
& $21.96\pm0.71$ & $49.67\pm13.30$ & $27.70\ [4.76,50.90]$ \\
& 4,096 shots + SPSA ($K=2$)
& $24.66\pm0.75$ & $67.17\pm23.78$ & $42.51\ [4.49,86.66]$ \\
& 4,096 shots + SPSA ($K=7$)
& $22.88\pm0.91$ & $98.80\pm28.41$ & $75.91\ [25.96,125.87]$ \\
\bottomrule
\end{tabular}
\vspace{0.4ex}

\footnotesize Returns are final-50 means $\pm$ standard errors over five
matched seeds.  $\Delta$ is majority minus parity; intervals are paired
seed-bootstrap 95\% intervals.  With five seeds, they are descriptive
and should not be read as precise population-level coverage statements.
\end{table*}

Figure~\ref{fig:finite-shot-cartpole} and
Table~\ref{tab:finite-shot-cartpole} show that the decoder ordering
survives finite sampling. With $K=7$, the paired majority advantage is
$55.04$ returns on the five-qubit ring (95\% CI
$[50.16,59.92]$) and $75.91$ on the ten-qubit graph
($[25.96,125.87]$). Neither finite-shot policy meets the strict
rolling-100 success criterion; the exact analytic majority condition
meets it in three of five seeds for each circuit.

The preserved $K=2$ runs separate shot count from direction count.
Increasing the shots from 1,024 to 4,096 changes the paired final-50
return by $+3.32$ and $-3.88$ for five-qubit parity and majority, and by
$+2.70$ and $+17.50$ for their ten-qubit counterparts. Three of the four
paired intervals include zero; the sole exception is the small
ten-qubit parity change of $2.70$ ($[1.03,4.42]$). Thus four times as
many shots yields no consistent material improvement at fixed $K=2$.

Increasing the number of SPSA directions at 4,096 shots has a much
larger effect. Relative to $K=2$, $K=7$ raises the five-qubit final-50
return by $30.43$ for parity ($[11.82,48.94]$) and $64.15$ for
majority--modulo ($[48.58,80.60]$). On ten qubits it changes parity by
$-1.78$ ($[-4.26,0.53]$) but raises majority--modulo by $31.63$
($[14.62,46.02]$). The 20 new $K=7$ runs use 7,378,845
state-conditioned circuit evaluations and 30,223,749,120 shots. The
finite-shot returns remain below analytic-gradient training, so the
experiment demonstrates a robust relative decoder advantage and the
benefit of reducing SPSA direction noise, not recovery of the analytic
learning rate.  The approximately $3.0\times10^{10}$-shot budget is
prohibitive for a near-term end-to-end implementation and is reported to
make that resource limitation explicit.

\subsection{Gradient participation by Pauli locality}
\label{subsec:majority-gradient-locality-training}

\subsubsection{Five-qubit ring}

We instrument the valid full majority policy on the five-qubit $L=3$
ring throughout 500 CartPole episodes.  The circuit, initialization,
state encoding, optimizer, five-episode batching, and ten seeds are the
same as in the preceding controlled comparison.  The only protocol change is diagnostic:
the episode at which the source solve rule would freeze optimization is
recorded, but optimization continues through episode 500 so that post-solve
zero gradients cannot be mistaken for disappearance of a Fourier sector.

For each update $t$, automatic differentiation resolves every Pauli string
in $\mathcal M_5$ and groups the resulting vectors by locality.  With
$\mathcal W_5=\{1,3,5\}$, the quantum-policy gradient is
\begin{equation}
    g_t=\sum_{w\in\mathcal W_5}g_t^{(w)}.
    \label{eq:training-gradient-sector-decomposition}
\end{equation}
Here $g_t^{(w)}$ is the sum of the parameter-gradient vectors generated by
all weight-$w$ strings, including the selected-action log probability,
normalized advantage, entropy term, and the common clipping factor used by
the optimizer.  For either circuit size, we quantify the relative gradient
magnitude of locality $w$ by
\begin{equation}
    q_t^{(w)}
    :=\frac{\lVert g_t^{(w)}\rVert_2}
            {\sum_{v\in\mathcal W_m}\lVert g_t^{(v)}\rVert_2},
    \qquad
    \sum_{w\in\mathcal W_m}q_t^{(w)}=1,
    \label{eq:gradient-sector-norm-share-training}
\end{equation}
where $\mathcal W_m$ is the set of nonzero Pauli weights for the measured
$m$-bit majority observable.  We average $q_t^{(w)}$ within each 50-episode
bin for every seed and then across the ten seeds.  This is an exact
sector-level gradient-norm share: it is not an average of the individual
string-gradient norms.  Weights two and four vanish identically for odd
five-bit majority.  Across all 1,000 updates, the 16 string vectors
reconstruct the optimizer gradient to maximum relative error
$1.4\times10^{-6}$.

\begin{figure*}[t]
    \centering
    \includegraphics[width=0.98\textwidth]{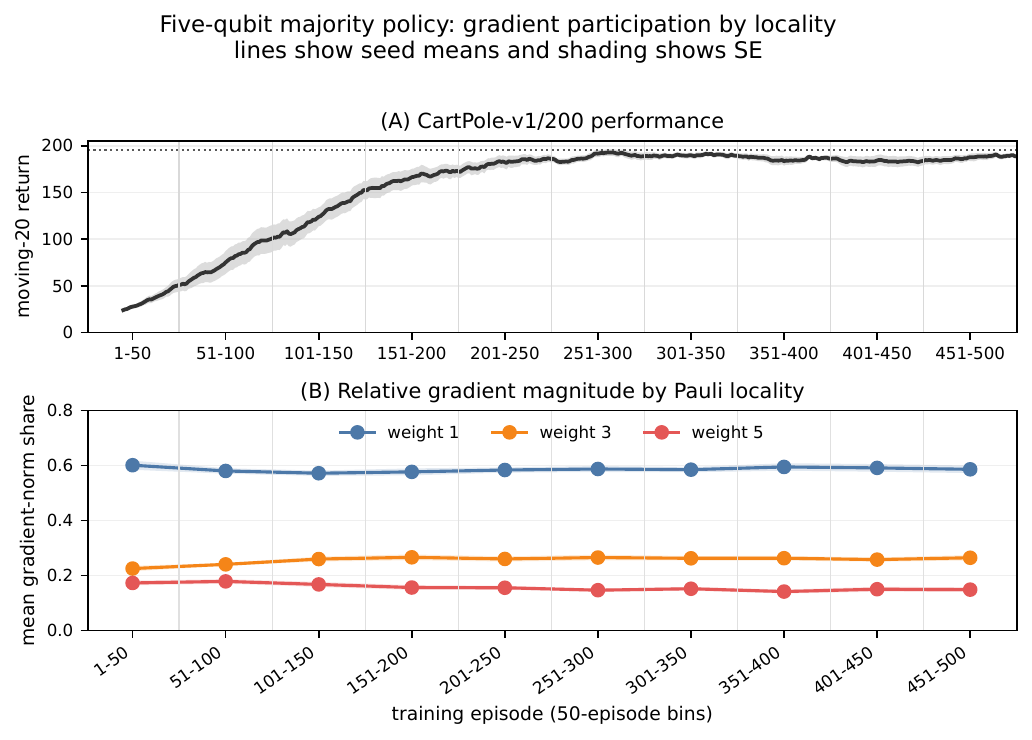}
    \caption{Time-resolved gradient participation by Pauli locality for the valid full
    five-qubit majority policy on the $L=3$ ring.  (A) Mean moving-20
    CartPole return with standard-error shading over ten seeds; the
    dotted line marks return 195.  (B) Mean relative gradient magnitude
    $q_t^{(w)}$ of each Pauli-weight sector in 50-episode bins, with
    standard-error shading.  Weights two and four vanish exactly.  The 16
    Pauli strings are grouped into three exact aggregate sector gradients;
    the curves are not averages of per-string gradient norms.  Optimization
    continues after the recorded solve threshold solely to preserve
    diagnostic late gradients.}
    \label{fig:majority-gradient-norm-share-training-5q}
\end{figure*}

Figure~\ref{fig:majority-gradient-norm-share-training-5q} and
Table~\ref{tab:majority-gradient-norm-share-training}(a) show that the
higher-locality sectors are substantial throughout the five-qubit training
trajectory.  Between episodes 1--50 and 451--500, the 1-local share changes
from $0.6012$ to $0.5863$, the weight-three share grows from $0.2255$ to
$0.2648$, and the weight-five share decreases from $0.1732$ to $0.1489$.
The combined share
$q_t^{(>1)}:=q_t^{(3)}+q_t^{(5)}$ changes from $0.3988$ to $0.4137$, a paired
increase of $0.0149$ with bootstrap 95\% interval $[-0.0279,0.0566]$.
Hence ten seeds do not resolve a net late-training increase: weight three
grows while weight five falls, leaving about $40\%$ of the summed
sector-gradient magnitudes in higher-locality terms across the trajectory.
The mean return rises from $36.98\pm4.13$ to $187.73\pm2.71$ over the same
bins.  This temporal comparison measures gradient participation, not a
counterfactual performance advantage of the higher-locality sectors.

\subsubsection{Ten-qubit cubic graph}

We next repeat the diagnostic on the connected ten-qubit cubic graph used
in the policy comparison, now with one IQP layer and 65 trainable quantum
parameters.  The majority decoder acts on measured qubits one through nine;
the tenth qubit remains coupled to the graph but is not measured.  We train
the valid deterministic nine-bit majority policy for 500 CartPole
episodes using the same ten seeds, five-episode updates, exact statevector
gradients, and continued post-threshold optimization as above.

The observable $\mathcal M_9$ contains all 256 nonzero odd-weight Pauli
strings at weights
$\mathcal W_9=\{1,3,5,7,9\}$.  To make attribution tractable without an
approximation, automatic differentiation first sums all strings of a fixed
weight and then evaluates one exact vector--Jacobian product per sector:
\begin{equation}
    g_t=\sum_{w\in\mathcal W_9}g_t^{(w)}.
    \label{eq:training-gradient-sector-decomposition-10q}
\end{equation}
We use the relative gradient magnitude in
Eq.~\eqref{eq:gradient-sector-norm-share-training}, now with
$\mathcal W_m=\mathcal W_9$.  This procedure resolves five exact sector
vectors, but not the 256 individual string gradients; the plotted values
are therefore sector-level gradient-norm shares, not averages of per-string
gradient norms.  Across all 1,000 updates, the five vectors reconstruct the
full optimizer gradient to maximum relative error $4.6\times10^{-6}$.

\begin{figure*}[!t]
    \centering
    \includegraphics[width=0.98\textwidth]{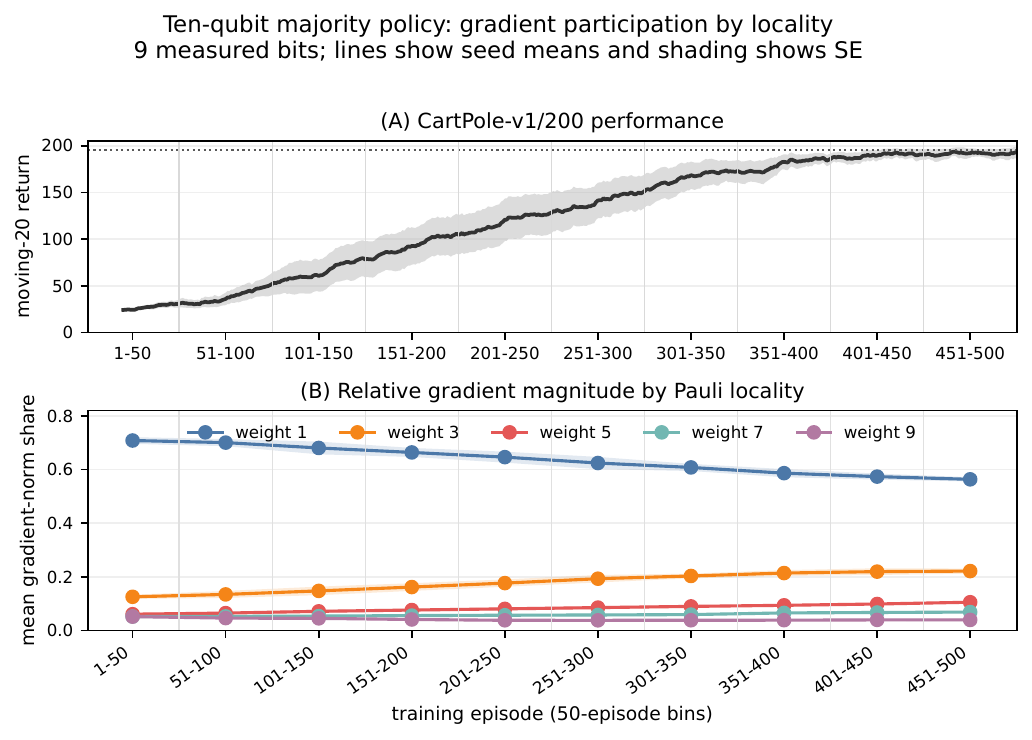}
    \caption{Time-resolved gradient participation by Pauli locality for the valid full
    nine-bit majority decoder on the one-layer ten-qubit cubic graph.  The
    tenth qubit is coupled but unmeasured.  (A) Mean moving-20
    CartPole return with standard-error shading over ten seeds; the
    dotted line marks return 195.  (B) Mean relative gradient magnitude
    $q_t^{(w)}$ of each Pauli-weight sector in 50-episode bins, with
    standard-error shading.  Even weights vanish exactly.  The 256 Pauli
    strings are grouped by weight into five exact aggregate sector gradients;
    the curves are not averages of per-string gradient norms.  Optimization
    continues after the recorded solve threshold solely to preserve
    diagnostic late gradients.}
    \label{fig:majority-gradient-norm-share-training-10q}
\end{figure*}

Figure~\ref{fig:majority-gradient-norm-share-training-10q} and
Table~\ref{tab:majority-gradient-norm-share-training}(b) show that the
higher-locality sectors become progressively more prominent as training
advances.  Let
$q_t^{(>1)}:=\sum_{w\in\{3,5,7,9\}}q_t^{(w)}$.  Between episodes 1--50 and
451--500, this combined share increases from $0.292$ to $0.436$, a paired
change of $0.145$ with bootstrap 95\% interval $[0.116,0.172]$.
Individually, weight three rises from $0.126$ to $0.222$, weight five from
$0.061$ to $0.106$, and weight seven from $0.053$ to $0.069$, whereas
weight nine changes from $0.052$ to $0.040$.  The 1-local share correspondingly
decreases from $0.708$ to $0.564$.  The final-bin return is
$192.14\pm5.23$, and six of ten seeds exceed the final-50 threshold of 195.
The redistribution of gradient magnitude provides direct evidence that
higher-locality sectors become increasingly engaged during learning.  The
concurrent return increase supplies temporal context only; it does not show
that these sectors cause the improvement or confer a performance advantage.

As a secondary directional diagnostic, the combined signed alignment of
weights three, five, seven, and nine with the full optimizer update changes
from $0.055$ to $-0.042$ (paired change $-0.097$, interval
$[-0.160,-0.039]$).  Thus the larger higher-locality sector vectors are not
collinear with the net update and partly cancel one another.  This caveat
does not make the policy equivalent to a local decoder: deleting those
sectors changes its probabilities, trajectories, and objective, and the
deployed decoder remains full majority.

\begin{table*}[!b]
    \centering
    \caption{Mean gradient-norm share by Pauli locality for the experiments in
    Figs.~\ref{fig:majority-gradient-norm-share-training-5q} and
    \ref{fig:majority-gradient-norm-share-training-10q}: (a) the five-qubit
    ring and (b) the ten-qubit cubic graph, for which majority acts on nine
    measured qubits.  Entries are means over ten seeds after averaging the
    updates within each 50-episode bin.  Each row sums to one up to rounding.}
    \label{tab:majority-gradient-norm-share-training}
    \footnotesize
    \begin{minipage}[t]{0.39\textwidth}
        \centering
        \textbf{(a) Five-qubit ring}\\[2pt]
        \setlength{\tabcolsep}{5pt}
        \begin{tabular}{lccc}
            \hline\hline
            Episode bin & $q^{(1)}$ & $q^{(3)}$ & $q^{(5)}$ \\
            \hline
            1--50    & 0.601 & 0.226 & 0.173 \\
            51--100  & 0.580 & 0.241 & 0.179 \\
            101--150 & 0.572 & 0.260 & 0.168 \\
            151--200 & 0.577 & 0.266 & 0.157 \\
            201--250 & 0.584 & 0.260 & 0.156 \\
            251--300 & 0.587 & 0.266 & 0.147 \\
            301--350 & 0.585 & 0.263 & 0.152 \\
            351--400 & 0.595 & 0.263 & 0.142 \\
            401--450 & 0.591 & 0.258 & 0.150 \\
            451--500 & 0.586 & 0.265 & 0.149 \\
            \hline\hline
        \end{tabular}
    \end{minipage}\hfill
    \begin{minipage}[t]{0.59\textwidth}
        \centering
        \textbf{(b) Ten-qubit cubic graph}\\[2pt]
        \setlength{\tabcolsep}{5pt}
        \begin{tabular}{lccccc}
            \hline\hline
            Episode bin & $q^{(1)}$ & $q^{(3)}$ & $q^{(5)}$ & $q^{(7)}$ & $q^{(9)}$ \\
            \hline
            1--50    & 0.708 & 0.126 & 0.061 & 0.053 & 0.052 \\
            51--100  & 0.700 & 0.135 & 0.065 & 0.053 & 0.047 \\
            101--150 & 0.681 & 0.148 & 0.072 & 0.054 & 0.046 \\
            151--200 & 0.664 & 0.162 & 0.077 & 0.056 & 0.041 \\
            201--250 & 0.646 & 0.177 & 0.081 & 0.057 & 0.038 \\
            251--300 & 0.624 & 0.193 & 0.086 & 0.058 & 0.038 \\
            301--350 & 0.608 & 0.204 & 0.090 & 0.060 & 0.039 \\
            351--400 & 0.587 & 0.214 & 0.094 & 0.066 & 0.039 \\
            401--450 & 0.574 & 0.220 & 0.099 & 0.067 & 0.040 \\
            451--500 & 0.564 & 0.222 & 0.106 & 0.069 & 0.040 \\
            \hline\hline
        \end{tabular}
    \end{minipage}
\end{table*}

% ================================================================

% ================================================================
% Supplementary circuit diagrams
% ================================================================

\clearpage
\onecolumn
\section{Neurally Encoded Random-Cubic IQP Circuits}
\label{app:neural-cubic-circuits}

\begin{figure}[H]
    \centering
    \begin{subfigure}[t]{0.48\textwidth}
        \vspace{0pt}
        \centering
        \includegraphics[origin=c,angle=-90,totalheight=0.69\textheight]{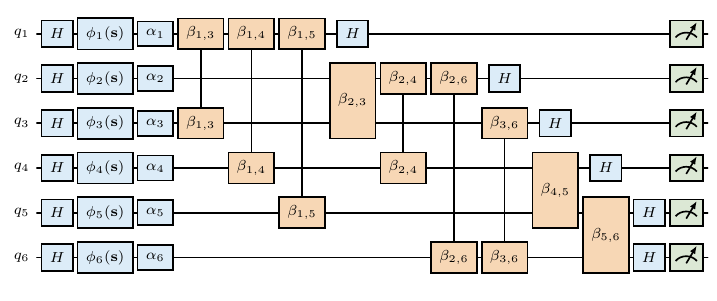}
        \caption{Base cubic-graph IQP policy.}
    \end{subfigure}\hfill
    \begin{subfigure}[t]{0.48\textwidth}
        \vspace{0pt}
        \centering
        \includegraphics[origin=c,angle=-90,totalheight=0.69\textheight]{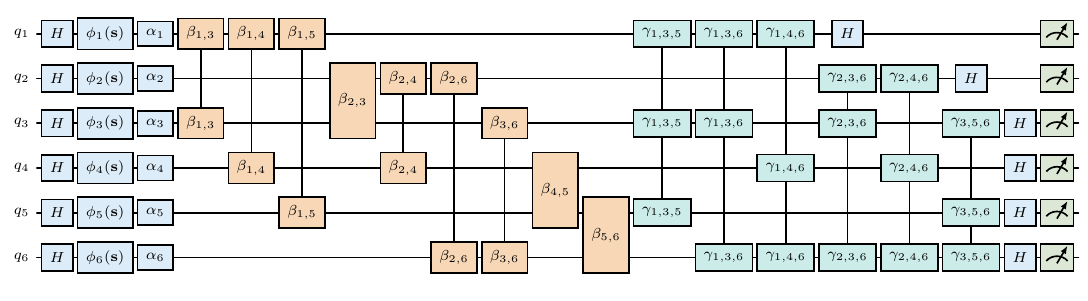}
        \caption{Paired three-body augmentation.}
    \end{subfigure}
    \caption{Unbroken QASP circuit exports, rotated so circuit time
    runs from top to bottom. Both circuits start in
    $|0\rangle^{\otimes6}$ and measure every wire in the computational
    basis. Compact gate labels $\phi_i(s)$ and $\alpha_i$ denote
    $R_{Z_i}$ rotations, $\beta_{ij}$ denotes $R_{Z_iZ_j}$, and
    $\gamma_{ijk}$ denotes $R_{Z_iZ_jZ_k}$; $H$ denotes a Hadamard.
    Panel (b) shares the neural weights and local/edge angles of (a)
    and adds the six recorded three-body rotations. Connected boxes
    with the same label represent one multi-qubit rotation on the
    marked wires, excluding intermediate crossed wires. Disjoint
    gates are scheduled in parallel by QASP.}
    \label{fig:neural-cubic-circuits}
\end{figure}

Figure~\ref{fig:neural-cubic-circuits} gives the unbroken QASP exports
for recorded topology 0 at $n=6$ in
Sec.~\ref{subsec:neural-cubic-variance}. The circuits implement
Eq.~\eqref{eq:neural-cubic-circuits}; only their orientation and gate
label formatting have been changed for presentation.

\clearpage

% ================================================================
% Connected-IQP circuit diagram
% ================================================================

\onecolumn
\section{Connected Majority-IQP Circuit}
\label{app:connected-majority-circuit}

\clearpage
\twocolumn

\section{Analytical Runtime and Implementation Controls}
\label{app:analytical-runtime}

The connected policy is polynomially contractible, but its complex-valued
transfer matrices and reverse-mode differentiation have larger constants
than a small real-valued MLP. We therefore distinguish asymptotic
tractability from measured training time.

A three-seed audit (2001--2003) reruns all 500 episodes on an Apple M1
CPU, with one PyTorch thread and PyTorch 2.7.0. It measures training
only; compilation warm-up is recorded separately. Four conditions
were run serially in rotated order; removal of discarded post-freeze
MLP calculations was a subsequent matched follow-up. The desktop was
shared rather than reserved exclusively, so these are descriptive
timings, not precision hardware benchmarks.

The first IQP optimization compiles the no-gradient rollout path and
caches parameter-dependent quantities without changing differentiable
training. A second optimization omits loss calculations whose results
would be discarded after the prescribed learning freeze. Neither
changes the circuit, decoder, precision, initialization, objective,
optimizer, random streams, batching, or rollout budget. Across the
three seeds, return sequences and final parameters are bitwise equal
to their corresponding reference runs, with identical freeze episodes
and update counts. The analogous post-freeze optimization is also
applied to the MLP for the final comparison.

\begin{table}[tbp]
\centering
\caption{Analytical training time over 500 episodes, averaged over
three matched seeds. Warm-up and held-out evaluation are excluded.}
\label{tab:analytical-runtime}
\small
\begin{tabular}{lr}
\toprule
Implementation & Mean seconds\\
\midrule
IQP, reference contraction & 43.58\\
IQP, compiled rollout & 38.45\\
IQP, compiled and post-freeze optimized & 34.03\\
MLP, reference & 8.98\\
MLP, post-freeze optimized & 7.90\\
\bottomrule
\end{tabular}
\end{table}

The optimized IQP is about $22\%$ faster than its reference
implementation, but remains about $4.3$ times slower than the
correspondingly optimized MLP on this audit. Thus the original
episode-AUC gain is not a classical runtime advantage. The comparison
uses each model's actual trajectory and freeze time, not an equal
number of environment steps, and should not be interpreted as a
per-update kernel benchmark. Per-run times, warm-up records, and
identity checks are included in the public companion experiment repository.

% ================================================================
% References
% ================================================================

\def\IEEEbibitemsep{-1.5pt plus 0.2pt}
\bibliographystyle{IEEEtran}
\bibliography{references}

\end{document}